\documentclass[prb,superscriptaddress,amsfonts,amssymb,amsmath,floats,twocolumn,aps,footinbib,shownopacs,longbibliography]{revtex4-2}

\usepackage{amsmath}
\usepackage{amsfonts}
\usepackage{graphicx}
\usepackage{color}
\usepackage{xcolor}
\usepackage{float}
\usepackage{hyperref}
\usepackage{soul}
\usepackage{braket}

\begin{document}
\title {Poor man's Majorana bound states in quantum dot based Kitaev chain coupled to a photonic cavity}
\author{Francesco Buonemani}
\affiliation{CPHT, CNRS, École polytechnique, Institut Polytechnique de Paris, 91120 Palaiseau, France}
\author{\'Alvaro G\'omez-Le\'on}
\affiliation{Institute of Fundamental Physics IFF-CSIC, Calle Serrano 113b, 28006 Madrid, Spain}
\author{Marco Schir\`o}
\affiliation{JEIP, UAR 3573 CNRS, Coll\`ege de France,   PSL  Research  University, 11,  place  Marcelin  Berthelot,75231 Paris Cedex 05, France}
\author{Olesia Dmytruk}
\affiliation{CPHT, CNRS, École polytechnique, Institut Polytechnique de Paris, 91120 Palaiseau, France}

\date{\today}

\begin{abstract} 
Quantum dot based platforms offer a promising route towards realizing the Kitaev chain Hamiltonian hosting Majorana bound states (MBSs). Poor man's MBSs arise in a two-site Kitaev chain when the parameters of the system are fine-tuned to the sweet spot. Based on our previous work~\cite{gomez2025majorana}, we consider a microscopic model for the Kitaev chain based on quantum dots with proximity effect embedded in a photonic cavity. 
We find that the photon coupling in the microscopic model yields an effective Hamiltonian where the cavity affects the pairing term. However, we demonstrate that even in this case, it is possible to screen particle interactions and reach the sweet spot condition for the emergence of the poor man's MBSs.
In particular, we find that attractive particle interactions can be canceled for the cavity prepared in the zero-photon state, while repulsive ones can be screened with a cavity prepared in the one-photon state. Furthermore, in case of a large number of photons in the cavity, we find that the hopping amplitudes are suppressed resulting in a degenerate spectrum. This motivates the use of quantum light for engineering poor man's MBSs with cavity embedding.
\end{abstract}

\maketitle

\section{Introduction}
Majorana bound states (MBSs) are zero-energy quasiparticles that emerge at the boundaries of topological superconductors~\cite{kitaev2001unpaired}. MBSs are robust against perturbations, making them promising building blocks for topological quantum computation~\cite{Kitaev2003quantumcomputationanyons}. One of the most successful theoretical proposals for the realization of the topological superconductor is the superconductor-semiconductor nanowire model~\cite{Oreg2010HelicalLiquids,Lutchyn2010Majorana,Alicea2012newdirections}, on which many experimental efforts have focused ~\cite{Mourik2012Signatures,Microsoft2023,Microsoft2025}. However, the observed signatures of the MBSs in superconductor-semiconductor devices could originate from other physical mechanisms~\cite{liu2012zero,kells2012near,prada2012transport,liu2017andreev,Reeg2018ZeroAndreev,penaranda2018quantifying,prada2020,hess2021local,hess2023trivial,sahu2023effect,prem2026distinguishing}.

Recently, a new way of engineering MBSs based on quantum dot chains came into research focus~\cite{SeoaneSouto2024}. The first proposals~\cite{Leijnse2012Parityqubits,Sau2012RealizingrobustMajorana,Fulga2013AdaptiveMajorana} introduced a simple model made by two quantum dots coupled to a conventional superconductor, 
in which each of the dots is regarded as a site of the Kitaev chain.
Although it is a very promising system, it presents some drawbacks given its simplicity.  Firstly, the MBSs that emerge do not have the topological protection of the original Kitaev chain model~\cite{kitaev2001unpaired} and due to this
they are referred to as "poor man’s" MBSs. Second, the system has to be tuned to an exact configuration of parameters, called the sweet spot, in which the electron-electron interactions have to be screened. Recent theoretical works have aimed to develop a more realistic quantum dot-based setup~\cite{Tsintzis2022,LiuQDABS,souto2023probing,samuelson2024minimal,Svensson2024QuantumdotKitaev,Miles2024dotAbs,liu2024enhancing,pino2024minimal,Miles2024dotAbs,luna2024flux,liu2024coupling,samuelson2024minimal,Svensson2024QuantumdotKitaev,Luethi2024from,Luethi2025fate,Luethi2025properties,sanches2025spin,sanches2026revisiting,zhang2026sensitivedependencepoormans,alvarado2026optimalmajoranasmesoscopickitaev}, whose low-energy effective model reduces to the minimal Kitaev chain model~\cite{Leijnse2012Parityqubits}. Two main investigated routes are based on hybrid Andreev bound states - quantum dot arrays~\cite{LiuQDABS,souto2023probing,Miles2024dotAbs,luna2024flux,liu2024enhancing,liu2024coupling,Luethi2024from,Luethi2025fate,alvarado2026optimalmajoranasmesoscopickitaev} and quantum dot chains with proximity-induced superconductivity~\cite{Tsintzis2022,samuelson2024minimal,Svensson2024QuantumdotKitaev,Luethi2024from,zhang2026sensitivedependencepoormans}.   
Several recent experiments have 
realized two~\cite{Dvir2023exp2site,tenHaaf2024exp2site,Zatelli2024exp2site,vanLoo2026single} and three-site Kitaev chains~\cite{Bordin2025exp3site,ten2025exp3site}.
However, the presence of electron-electron interactions, that is responsible for the hybridization of poor man's MBSs and deterioration of their quality, remains an open question in theoretical works.

Coupling quantum matter to photonic cavities offers a promising route for modifying properties of materials~\cite{garciavidal2021manipulating,schlawin2022cavity,bretscher2026fluctuation}. Recent experiments have demonstrated cavity control of quantum Hall effect~\cite{appugliese2022breakdown,enkner2025}, metal-to-insulator transition~\cite{jarcNature2023}, and cavity-altered superconductivity~\cite{Keren2026cavity}. Multiple theoretical works studied cavities coupled to various electronic systems, such as topological superconductors hosting MBSs~\cite{trif2012resonantly,cottet2013squeezing,dmytruk2015cavity,dmytruk2016josephson,dartiailh2017direct,trif2019braiding,mendez2020renyi,contamin2021topological,dmytruk2023microwave,bacciconi2023topological,dmytruk2024hybrid,becerra2025fermion,prem2026distinguishing,kobialka2026topology}, two-dimensional superconductors~\cite{sentef2018,schlawin2019cavity,curtis2019cavity,alloca2019,kozin2025cavity}, interacting electronic systems~\cite{mazza2019,passetti2023cavity,kass2024manybody,Fadler2024}, moir\'e  materials~\cite{nguyen2023electron}, quantum Hall systems~\cite{ciuti2021cavity,winter2025fractional,borici2025}, and the Su-Schrieffer-Heeger chains~\cite{dmytruk2022controlling,perez2023light,vlasiuk2023cavity,nguyen2024electron,shaffer2024entanglement,Sueiro25,ritzzwilling2026}. Moreover, for a two-site interacting Kitaev chain it was theoretically demonstrated that coupling to photons can cancel the Coulomb interactions in the system and allows for better tunability to the sweet spot~\cite{gomez2025majorana}. 

In this work, we propose to use cavity embedding to realize poor man's MBSs in an interacting quantum-dot based array. We focus on the more realistic model of two quantum dots with local superconducting proximity effect coupled to a single-mode cavity~\cite{samuelson2024minimal}. On the experimental side, coupling between the quantum dots and a microwave cavity has been successfully realized~\cite{delbecq2011coupling,petersson2012coupling,frey2012dipole,basset2013single,delbecq2013engineering,viennot2014out,viennot2015coheret,stockklauser2015microwave,mi2016strong,cottet2017cavity,mi2018electron,samkharadze2018strong,landig2018electron,scarlino2022in,gu2023probing}. Extending the previous study~\cite{gomez2025majorana}, we demonstrate the emergence of poor man's MBSs and find that a more realistic platform for poor man's MBSs coupled to photons offers more knobs for tuning to the sweet spot and screening the inter-dot interactions.

The paper is organized as follows. We introduce the quantum dot based microscopic model coupled to a single mode cavity in Section~\ref{ModelHamiltonian}.  Section~\ref{EffectiveHamiltonian} is dedicated to the derivation of the effective electronic Hamiltonian and finding the condition for the emergence of poor man's MBSs. Subsequently, in Section~\ref{Mainresults} we study the sweet spot condition for the cavity prepared in a state with different number of photons. Finally, in Section~\ref{Conclusion} we summarize our findings and provide an outlook for future work.

\section{Model Hamiltonian}\label{ModelHamiltonian}

\begin{figure}[t]
    \centering
    \includegraphics[width=\linewidth]{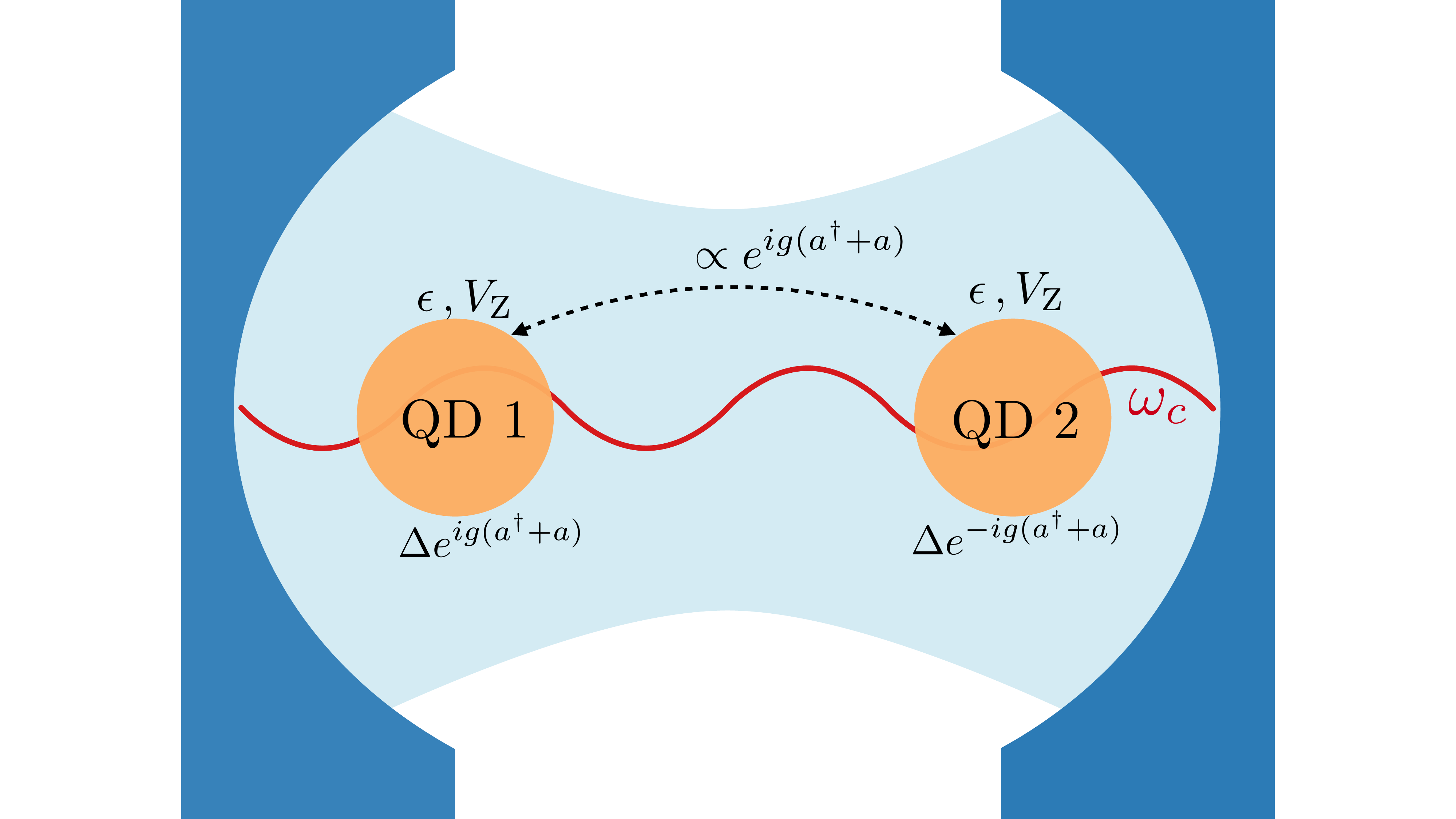}
    \caption{Scheme of the microscopic setup for poor man's Majorana bound states consists of two spinful quantum dots (QDs) ($1,2$) (in orange) with on-site energy $\epsilon$, Zeeman energy $V_Z$ and an $s$-wave superconducting pairing $\Delta$ embedded in a cavity (in blue). The QDs are coupled by both spin-conserving tunneling $t$ and spin-flipping tunneling $t_\text{so}$. The microscopic setup is coupled to a single mode cavity with frequency $\omega_\text{c}$ (in red), with the light-matter coupling strength $g$. }
    \label{fig:two_dot_cavity_micro}
\end{figure}

We consider a model of two coupled quantum dots with proximity-induced pairing and spin-orbit coupling, which effectively realizes a Kitaev chain with poor man's MBSs, embedded in a photonic cavity (see Fig.~\ref{fig:two_dot_cavity_micro}). This system is described by the total Hamiltonian
\begin{align}\label{eq:Htot}
H_\text{tot} = \omega_{\text{c}} a^\dag a + H_\text{QD}+H_\text{C},
\end{align}
where the first term describes the cavity mode, while $H_{\text{QD}}$ and $H_\text{C}$ read respectively as
\begin{align}  
&H_{\text{QD}} = \sum_{i=1}^2\left[\left(\epsilon+V_\text{Z}\right)n_{i\uparrow}+\left(\epsilon-V_\text{Z}\right)n_{i\downarrow}\right]
\notag\\
&+\Delta \left(e^{ig\left(a^\dag+a\right)}d^\dag_{1\uparrow}d^\dag_{1\downarrow}+ e^{-ig\left(a^\dag+a\right)}d^\dag_{2\uparrow}d^\dag_{2\downarrow}+\text{h.c.}\right),
\label{eq:two_dot_ligt_micro_2}\\
&H_\text{C} = \sum_{\sigma\,\in\left\{\uparrow,\downarrow\right\}} t \left(e^{ig\left(a^\dag+a\right)}d^\dag_{1\sigma}d_{2\sigma}+\text{h.c.}\right)\nonumber\\
&+t_\text{so}\left[e^{ig\left(a^\dag+a\right)}\left(d^\dag_{1\uparrow}d_{2\downarrow}-d^\dag_{1\downarrow}d_{2\uparrow}\right)+\text{h.c.}\right].
\label{eq:two_dot_ligt_micro_3}
\end{align}
Here, $a^\dag$ ($a$) are the photonic creation (annihilation) operators, $d^\dag_{i,\sigma}$ ($d_{i,\sigma}$) are the fermionic creation (annihilation) operators with spin $\sigma = \uparrow,\downarrow$ on QD $i = 1,2$, and $n_{i \sigma}=d^\dag_{i\sigma}d_{i\sigma}$ is the number operator. Moreover,  $\omega_\text{c}$ is the frequency of the cavity mode, $g$ the light-matter coupling strength, $\epsilon$ is the on-site energy of the QD, $V_\text{Z}$ is the Zeeman energy due to an externally applied magnetic field, $t$ is the spin-conserving tunneling amplitude, $t_\text{so}$ is the spin flipping amplitude, and $\Delta$ is an $s$-wave superconducting pairing. Since we are interested in tuning the sweet spot through the cavity coupling, we note that compared to the microscopic model in~\cite{samuelson2024minimal}, we take the on-site energy on the QDs to be equal and the superconducting phase is assumed to be zero.  The light-matter coupling is included through the Peierls substitution~\cite{dmytruk2021gauge,dmytruk2022controlling,perez2023light,passetti2023cavity,bacciconi2023topological,dmytruk2024hybrid,nguyen2024electron} introduced as $d_{j\sigma} \rightarrow d_{j\sigma} e^{ig(a+a^\dag) (j-j_0)}$, where $j_0 \equiv 3/2$ is fixed such that the phase on the superconducting pairing term $\Delta e^{2 i g (j-j_0)(a+a^\dag)}$ has the opposite sign for $j=1,2$~\cite{perez2022topology,dmytruk2024hybrid}.  
In the following, we will compare the spectrum of the full system with the effective Kitaev-like Hamiltonian obtained by projecting out high-energy degrees of freedom in the electron-photon system. In the next section we describe the derivation of this effective Hamiltonian, leaving the technical details to Appendix~\ref{Kitaevlimitnocavity}.  To derive the effective electronic Hamiltonian, we neglect the inter-dot and intra-dot particle interactions~\cite{samuelson2024minimal}. However, in Section~\ref{Mainresults} we will add the electron interactions term phenomenologically to the effective Hamiltonian to account for the particle interactions.

\section{Effective Hamiltonian}\label{EffectiveHamiltonian}

To find the sweet spot condition for poor man's MBSs, we derive a low-energy, effective electronic Hamiltonian. The path to do so is the following:

we express the total Hamiltonian in the photon number basis $\ket{n}$; we impose the Kitaev limit for the electronic system, $V_\text{Z},\Delta\gg  t,t_\text{so}$; and we derive the effective electronic Hamiltonian coupled to the cavity photons. 
Then we perform the adiabatic elimination of the photon subspace in the large detuning regime, $\omega_\text{c}\gg\Delta_{\text{bw}}(n,g)$, where $\Delta_{\text{bw}}(n,g)$ is the renormalized bandwidth of $n$-th photonic band of the effective Kitaev chain Hamiltonian.  
This procedure allows us to obtain the low-energy electronic effective Hamiltonian that depends on the parameters of the cavity.

\subsection{Kitaev limit for electronic Hamiltonian in the absence of the cavity}

We start by considering the Hamiltonian $H_{\text{tot}}$~\eqref{eq:Htot} in the absence of the cavity coupling, i.e. $g=0$. To derive the effective low-energy description we use the projector method~\cite{FESHBACH1958357} in the regime where $V_\text{Z}.\Delta\gg\,t,\,t_\text{so}$
known as the Kitaev limit. Treating the coupling Hamiltonian~\eqref{eq:two_dot_ligt_micro_3} as a perturbation and eliminating the high-energy electronic degree of freedom, we arrive at the effective Kitaev chain Hamiltonian
\begin{equation}
    H_\text{eff} =  \sum_{i=1}^2 E_\alpha^{\text{eff}} \alpha_i^\dag \alpha_i + \left(t_{\alpha}\alpha_1^\dag \alpha_{2}+\Delta_{\alpha}\alpha_1\alpha_{2}+\text{h.c.}\right).
    \label{eq:effective_two_dot_hamiltonianMain}
\end{equation}
Here, $\alpha_i = \left(\sqrt{\gamma-\epsilon}\,d^\dag_{i\downarrow}+\sqrt{\gamma+\epsilon}\,d_{i\uparrow}\right)/\sqrt{2\gamma}$, $\gamma=\sqrt{\Delta^2 + \epsilon^2}$, $t_{\alpha}= -\epsilon t/\gamma$, $\Delta_{\alpha}=-t_\text{so}\Delta/\gamma$, and $E_\alpha^{\text{eff}} = E_\alpha+2\left[(t_\text{so}\,\epsilon)^2-(\Delta\,t)^2\right]/\left[\gamma^2\left(\gamma+V_\text{Z}\right)\right]$ (see Appendix~\ref{Kitaevlimitnocavity} for more details). 

Noting that the parity is conserved, we separate the many-body electronic Hamiltonian into  even $\{ |0_{1} 0_{2} \rangle  , |1_{1} 1_{2} \rangle \} = \{ |0_{1} 0_{2} \rangle  , \alpha_1^\dag \alpha_2^\dag|0_{1} 0_{2} \rangle \} $ and odd $\{  |1_{1} 0_{2} \rangle ,|0_{1} 1_{2} \rangle  \}  = \{ \alpha_1^\dag |0_{1} 0_{2} \rangle  , \alpha_2^\dag|0_{1} 0_{2} \rangle \}$ parity sectors, where $| 0_10_2 \rangle$ is  the empty electronic state. The many-body energies corresponding to the even (odd) parity sector eigenvalues $E_{\pm}^{\text{even}}$ ($E_{\pm}^{\text{odd}}$) read
\begin{align}
&E_{\pm}^{\text{even}} = E_\alpha^\text{eff} \pm \sqrt{\left(E_\alpha^{\text{eff}}
\right)^2 + \Delta^2 t_{so}^2/\gamma^2},\\
&E_{\pm}^{\text{odd}} = E_\alpha^\text{eff} \pm \dfrac{\epsilon t}{\gamma}.
\end{align}

The sweet spot condition for the isolated effective Kitaev chain model is $E_\alpha = 0$ and $\epsilon\, t = \Delta\, t_{so}$. We note that the ground state is degenerate at the sweet spot, $E_{-}^{\text{even}} = E_{-}^{\text{odd}}$.

\subsection{Kitaev limit for electronic Hamiltonian coupled to cavity}
In this section, we take the Kitaev limit for the coupled electron-photon Hamiltonian given by Eq.~\eqref{eq:Htot} and derive the effective light-matter Hamiltonian. To do so, we express $H_\text{tot}$ in the photon number basis $\ket{n}$ and then identify a diagonal $H_{\text{QD}}^\text{D}$ and off-diagonal part $H_{\text{QD}}^\text{OD}$ in the photon Fock subspace for the $H_\text{QD}$. This allows us to diagonalize exactly $H_{\text{QD}}^\text{D}$ and treat the $H_{\text{QD}}^\text{OD}$ and $H_\text{C}$ 
as a perturbation. Then, we can derive the effective low-energy electronic Hamiltonian coupled to the cavity photon using the projectors method~\cite{FESHBACH1958357,gomez2025majorana}.

More details are provided in Appendix~\ref{KitaevLimit}. Thus, we
arrive at the effective electron-photon Hamiltonian expressed in the photon number basis:

\begin{align}
H_{\text{QD}}=\sum_n\Bigg[\sum _{i=1}^2E^\text{eff}_\alpha(n)\alpha_i^\dag\alpha_i+2\left(\epsilon-\gamma(n)\right)+n\omega_{\text{c}}\Bigg]Y^{n,n},
\label{eq:HCdiag}
\end{align}

\begin{align}
H_\text{C}^\text{even}=-\sum_{m,n}\frac{t_\text{so} \left[\Delta_{m,n}(g)+\Delta_{m,n}(-g)\right]}{2\gamma(n)}\left(\alpha_2\alpha_1+\text{h.c.}\right)Y^{m,n},
\label{eq:HCeven}
\end{align}

\begin{align}
H_\text{C}^\text{odd}=\sum_{m,n}\left[\left(\frac{t_{m,n}(g)}{2\gamma(n)}-\frac{t_{m,n}(-g)}{2\gamma(n)}\right)\alpha_1^\dag\alpha_2\,+\text{h.c.}\right]\,Y^{n,m},%
    \label{eq:HCodd}
\end{align}
where we separated two contributions to the coupling Hamiltonian $H_\text{C}$ given by Eq.~\eqref{eq:HCeven} and \eqref{eq:HCodd} corresponding to the even and odd parity sectors, respectively. Here, $Y^{n,m}=\ket{n}\bra{m}$ are the bosonic Hubbard operators,
$\alpha_i^\dag \left(\alpha_i\right)$ are fermionic creation (annihilation) operators of the Bogoliubov quasiparticles that diagonalize $H_\text{QD}^\text{D}$~\ref{eq:HCdiag}, $E^\text{eff}_\alpha(n) =\gamma(n)-V_\text{Z} +2\left[(t_\text{so}\,\epsilon)^2-(\Delta(n)\,t)^2\right]/\left[\gamma(n)^2\left(\gamma(n)+V_\text{Z}\right)\right]$, $\gamma(n) = \sqrt{\epsilon^2 + \Delta(n)^2}$,   $\Delta(n)=\Delta e^{-g^2/2}L_n(g^2)$.

The Hamiltonian given by Eqs.~\eqref{eq:HCdiag}-\ref{eq:HCodd} shows that the cavity couples nonlinearly to the system and affects all the electronic processes. The superconducting pairing and the tunneling amplitudes become photon dependent and acquire a non-linear structure, shown by the functions $\Delta(n)$ and $t_{m,n}(g)$. 
We note in particular that the cavity couples to low-energy electrons in both the even and odd subspaces. In the odd parity sector the cavity generates photon-assisted hopping while in the even parity sector the cavity induces an effective pairing proportional to the product of spin-orbit and superconducting terms. We emphasize that this last term would be absent if the cavity-coupling is performed 
directly on the effective Kitaev model, as in Ref.~\cite{gomez2025majorana}, and represents therefore a novelty arising from our fully microscopic description.

\subsection{Adiabatic elimination of photons in large cavity detuning regime}
\label{subsection:adiabatic_elimination}
In this section, we consider the high frequency regime, $\omega_\text{c}\gg\Delta_{\text{bw}}$, and derive the effective electronic Hamiltonian using the projector's method~\cite{FESHBACH1958357,gomez2025majorana} (see Appendix~\ref{ProjectorsMethod} for more details). Using the mapping $\alpha_1\rightarrow c_1$ and $\alpha_2\rightarrow c_2$, we arrive at the electronic effective Hamiltonian:
\begin{align}
&\tilde{H}\left(n\right) =  \tilde{U}\left(n\right) c_1^\dag c_2^\dag c_2c_1-\tilde{\mu}\left(n\right)\left(c_1^\dag c_1+c_2^\dag c_2\right)  +\tilde {C}\left(n\right)\notag\\
&+\tilde{\Delta}\left(n\right)\left(c_1c_2+c_2^\dag c_1^\dag\right)-\tilde{t}\left(n\right)\left(c_1^\dag c_2+c_2^\dag c_1\right),
\label{eq:hamiltonian_two_dot_final0}
\end{align}
where the chemical potential, superconducting pairing and hopping terms are modified by photon coupling and now depend on the parameters of the cavity. Moreover, $\tilde{H}\left(n\right)$ contains a cavity-induced electron-electron interaction term. The coefficients in Eq.~\eqref{eq:hamiltonian_two_dot_final0} are given by
\begin{align}
    \tilde{U}\left(n\right) &= -2\left[\frac{n\tilde{\kappa}_n^2\omega_c}{\omega_c^2-\tilde{\omega}_n^2}-\frac{(n+1)\tilde{\kappa}^2_{n+1}\omega_c}{\omega_c^2-\tilde{\omega}_{n+1}^2}\right],\\
    \tilde{\mu}\left(n\right) &= -\Bigg(E_\alpha(n)+\frac{n\tilde{\kappa}_n^2\omega_c}{\omega_c^2-\tilde{\omega}_n^2}-\frac{(n+1)\tilde{\kappa}^2_{n+1}\omega_c}{\omega_c^2-\tilde{\omega}_{n+1}^2}\Bigg),\\
    \tilde{C}(n)&=n\omega_c+2\left(\epsilon-\gamma(n)\right),\\
\tilde{\Delta}\left(n\right) &= \frac{t_{so}\Delta(n) e^{-g^2/2}}{\gamma(n)}L_n\left(g^2\right),\\
\tilde{t}\left(n\right)&=-\Big(\frac{n\tilde{\kappa}_n^2\tilde{\omega}_n}{\omega_c^2-\tilde{\omega}_n^2}-\frac{(n+1)\tilde{\kappa}^2_{n+1}\tilde{\omega}_{n+1}}{\omega_c^2-\tilde{\omega}_{n+1}^2}\notag\\
&-\frac{\epsilon t}{\gamma(n)}e^{-g^2/2}L_n\left(g^2\right)\Big),
\end{align}
where we introduced the parameters $\tilde{\kappa}_n$ $= $ $gte^{-g^2/2}{}_1F_1\left(1-n;2;g^2\right)$ 
and $\tilde{\omega}_n $$=$$~\epsilon t\left[L_n(g^2)+L_{n-1}(g^2)\right]/\gamma(n)$. Here,  $L_{n}\left(g^2\right)$ is the $n$th Laguerre polynomial and ${}_1F_1\left(1-n;2;g^2\right)$ is the Kummer confluent hypergeometric function. We formally recover the effective Hamiltonian obtained for a two-site Kitaev chain embedded in a cavity~\cite{gomez2025majorana}. The main difference is that the superconducting pairing term is now modified and depends on the parameters of the cavity. 

Nevertheless, we find that even though both the even and odd parity sectors are dressed by photons, it is possible to find a sweet spot condition formally similar to the one found in~\cite{gomez2025majorana}.

To find the sweet spot condition for poor man's MBSs, we rewrite the Hamiltonian~\eqref{eq:hamiltonian_two_dot_final0} in the Majorana basis: $c_j = \left(\gamma_{2j-1}+i\gamma_{2j}\right)/2$, $c_j^\dag =\left(\gamma_{2j-1}-i\gamma_{2j}\right)/2$, with $\{\gamma_i,\gamma_j\}=2\delta_{ij}$, and we arrive at
\begin{align}
\tilde{H}\left(n\right) &= \frac{\tilde{U}\left(n\right)}{4}-\tilde{\mu}\left(n\right)-\frac{\tilde{U}\left(n\right)}{4}\gamma_1\gamma_2\gamma_3\gamma_4\notag\\
 &-\frac{i}{2}\left[\tilde{\mu}\left(n\right)-\frac{\tilde{U}\left(n\right)}{2}\right]\left(\gamma_1\gamma_2+\gamma_3\gamma_4\right)\notag\\
 &+i\frac{\tilde{\Delta}\left(n\right)-\tilde{t}\left(n\right)}{2}\gamma_1\gamma_4+i\frac{\tilde{\Delta}\left(n\right)+\tilde{t}\left(n\right)}{2}\gamma_2\gamma_3.
\label{eq:Hamiltonian_majorana_two_dot_most}
\end{align}
To decouple $\gamma_1$ and $\gamma_4$ operators from the Hamiltonian given by Eq.~\eqref{eq:Hamiltonian_majorana_two_dot_most}, a sufficient condition for the appearance of the MBSs~\cite{kitaev2001unpaired,gomez2025majorana}, we have to impose the following conditions:
\begin{align}
&\tilde{U}\left(n\right)=0,\label{eq:Ueff}\\
&\tilde{\mu}\left(n\right) = \dfrac{\tilde{U}\left(n\right)}{2},\label{eq:muEff}\\
&\tilde{\Delta}\left(n\right)=\tilde{t}\left(n\right)\label{eq:tEff}.
\end{align}

Furthermore, we find the many-body eigenvalues of the effective Hamiltonian  \eqref{eq:hamiltonian_two_dot_final0} corresponding to the even and odd parity sectors for the cavity prepared in a state with $n$ photons:
\begin{align}
\tilde{E}_{\pm}^{\text{even}}(n) &= \tilde{C}\left(n\right)+\dfrac{\tilde{U}(n)}{2} - \tilde{\mu}(n) \notag\\
&\pm \sqrt{\left(\dfrac{\tilde{U}(n)}{2} - \tilde{\mu}(n)\right)^2 + \tilde{\Delta}^2(n)}\label{eq:E0even},\\
\tilde{E}_{\pm}^{\text{odd}}(n) &=\tilde{C}\left(n\right) -\tilde{\mu}(n) \pm \tilde{t}(n) \label{eq:E0odd},
\end{align}
We note that at the sweet spot the even and odd parity ground state energies are degenerate, $\tilde{E}_{-}^{\text{even}}(n) = \tilde{E}_{-}^{\text{odd}}(n)$. Moreover, the energy spectrum is symmetric with respect to the constant $\tilde{C}\left(n\right)$ at the sweet spot.

\section{Main results}\label{Mainresults}
In this section, we present the main findings of our work. 
\begin{figure}[b]
    \centering
    \includegraphics[width=\columnwidth]{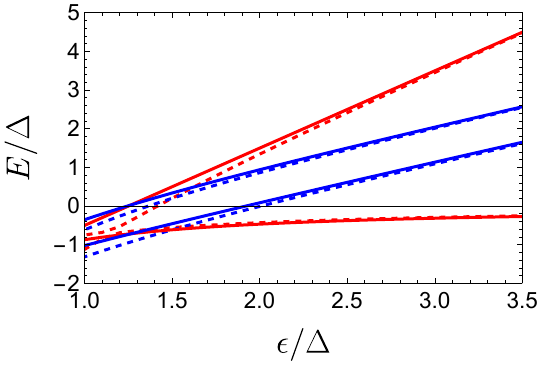}
    \caption{
    Many-body spectrum of the Hamiltonian for two quantum dot chain with superconductivity embedded in a cavity as a function of $\epsilon/\Delta$. 
 Solid red (blue) lines correspond to the even (odd) parity eigenvalues of obtained from the exact diagonalization of the original Hamiltonian $H_{\text{tot}}$ given by Eqs.~\eqref{eq:two_dot_ligt_micro_2} and \eqref{eq:two_dot_ligt_micro_3}. 
    Dashed red (blue) lines correspond to the even (odd) parity eigenvalues of the effective Hamiltonian obtained after eliminating high-energy and photonic degrees of freedom given by Eq.~\eqref{eq:hamiltonian_two_dot_final0}. There is a good agreement between the energy spectrum obtained for the effective Hamiltonian and the original Hamiltonian for large $\epsilon$.
    The parameters are chosen as $t/\Delta =0.5$, $t_{so}/\Delta = 0.1$ and $V_Z/\Delta = 1.25$, $g = 0.3$ and  $\omega_c/\Delta = 5.0$.
    For the exact diagonalisation is imposed a photonic $\text{cutoff} = 10$.
   }
    \label{fig:SpectrumEpsilon}
\end{figure}
To first demonstrate the effect of the cavity on the many-body energy spectrum of the microscopic model and to support our analytical derivation of the effective Hamiltonian, we start by comparing the many-body energy spectrum in the absence of the particle interactions obtained by exact diagonalization of the microscopic model of two quantum dots with superconductivity coupled to a single mode cavity given by Eqs.~\eqref{eq:Htot}, and the effective Hamiltonian obtained by eliminating high energy electronic degrees of freedom and adiabatic elimination of photons given by Eq.~\eqref{eq:hamiltonian_two_dot_final0} [see Fig.~\ref{fig:SpectrumEpsilon}].
 The spectrum of the microscopic Hamiltonian in the photonic subspace with zero photons consists of eight energy levels. In the case of the effective Hamiltonian, the number of electronic degrees of freedom is reduced by a factor of two since we eliminate high energy degrees of freedom. In Fig.~\ref{fig:SpectrumEpsilon} we plot the four lowest energy states for the total Hamiltonian~\eqref{eq:Htot} and compare with the effective model given by Eq.~\eqref{eq:hamiltonian_two_dot_final0}. There is a crossing between even and odd parity ground state energies. However, this crossing does not correspond to a sweet spot for realizing poor man's MBSs since the sweet spot conditions~\eqref{eq:Ueff}-\eqref{eq:tEff} are not fulfilled.
We find a good agreement between the two spectra for values of $\epsilon > \sqrt{V_{\text{Z}}^2-\Delta^2}$, 
the same range of $\epsilon/\Delta$ for which the effective model in the Kitaev limit without the cavity agrees with the microscopic model (see Appendix~\ref{Kitaevlimitnocavity} for more details).

Next, we discuss the sweet spot condition for the cavity prepared in a state with different number of photons: $n=0$, $n=1$, and $g\sqrt{n} = \text{const}$.  We demonstrate that embedding two QDs with a proximity-induced superconductivity in a cavity prepared in a state with small number of photons could be used to screen electron interactions and to realize poor man's MBSs.

\subsection{Sweet spot for $n=0$}
In this section, we express the set of conditions given by Eqs.~\eqref{eq:Ueff}-\eqref{eq:tEff} for the sweet spot for the cavity prepared in the ground state ($n = 0$) in terms of the on-site energy $\epsilon$.
As we ignored the presence of particle interactions $U$ in the microscopic Hamiltonian, we can add them phenomenologically in the effective Hamiltonian by shifting $\tilde{U}(n)\rightarrow\tilde{U}(n)+U$.  Since we consider the setup consisting of two QDs, we can only get the particle interactions of the form $ U n_1 n_2$.

In what follows, we demonstrate that cavity embedding induces a cavity-mediated electron-electron interaction term that cancels the inter-dot Coulomb interaction term in the effective Hamiltonian $U$, allowing us to obtain fully localized isolated poor man's MBS.

In the high frequency limit (taking the limit $\omega_c\rightarrow\infty$), the sweet spot conditions Eqs.~\eqref{eq:Ueff}-\eqref{eq:tEff}  for $n=0$ take the form
\begin{align}
    &U \xrightarrow{\omega_c\rightarrow \infty} -\dfrac{2e^{-g^2}g^2t^2}{\omega_c }\label{eq:Ueffn=0} \\
    &t\xrightarrow{\omega_c\rightarrow \infty}\sqrt{\frac{\Delta^2E_\beta\left(U/2-E_\beta\right)}{\Delta^2-e^{g^2}\epsilon^2}}\\
    &t_\text{so}\xrightarrow{\omega_c\rightarrow \infty}\dfrac{\epsilon\,t\,e^{g^2/2}}{\Delta},
\end{align}
where $\gamma(0)=\sqrt{e^{-g^2/2}\Delta^2+\epsilon^2}$ and $E_\beta = V_\text{Z}+\gamma(0)$. 
\begin{figure}[h]
    \centering
    \includegraphics[width=\columnwidth]{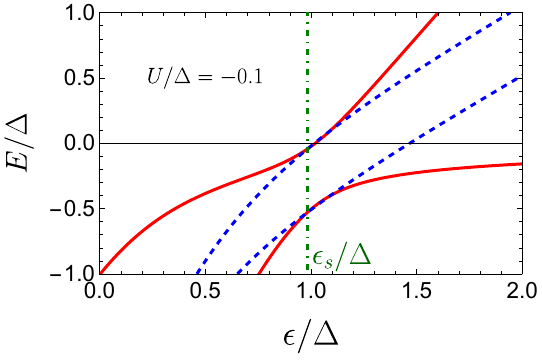}
    \caption{ Many-body spectrum of the interacting effective Hamiltonian for two quantum dot chain with superconductivity embedded in a cavity with $n=0$ as a function of $\epsilon/\Delta$. We consider attractive electron-electron interactions with $U/\Delta = -0.1$. 
    Red solid (blue dashed) lines correspond to the even (odd) parity eigenvalues of the effective Hamiltonian~\eqref{eq:hamiltonian_two_dot_final0}. The green dashed vertical line depicts the sweet spot value  $\epsilon_s/\Delta = 0.98$ obtained from Eqs.~\eqref{eq:Ueff}-\eqref{eq:tEff}. The parameters are chosen as $t/\Delta = 0.5$, $t_{so}/\Delta = 0.9$ and $V_Z/\Delta = 1.5$, and solving the sweet spot conditions we find that $g = 1.09$,  $\omega_c/\Delta =1.81$. The bandwidth at the sweet spot is given by 
   $\Delta^\text{s}_\text{bw}(n=0,g)/\Delta=0.5 < \omega_c/\Delta$.
        }\label{fig:SpectrumEpsilonU}
\end{figure}

From Eq.~\eqref{eq:Ueffn=0} we note that cavity embedding allows us to cancel only the attractive electron-electron interactions (with $U<0$). This was also found to be the case in the minimal interacting two-site Kitaev chain model embedded in a cavity in its ground state~\cite{gomez2025majorana}. Fig.~\ref{fig:SpectrumEpsilonU} shows the energy spectrum of the effective Hamiltonian given by Eqs.~\eqref{eq:E0even} and \eqref{eq:E0odd}. 
The spectrum consists of two ground states $\tilde{E}_{-}^{\text{even}}(0)$ and $\tilde{E}_{-}^{\text{odd}}(0)$ that cross at the value of the on-site energy corresponding to the sweet spot, and two excited states $\tilde{E}_{+}^{\text{even}}(0)$ and $\tilde{E}_{+}^{\text{odd}}(0)$.  Moreover, at the sweet spot
the excited states become also degenerate, making the spectrum symmetric with respect to the constant $\tilde{C}(n=0)$, resulting in additional signature that the sweet spot condition is fulfilled.

\begin{figure}[h!]
    \centering
    \includegraphics[width=\columnwidth]{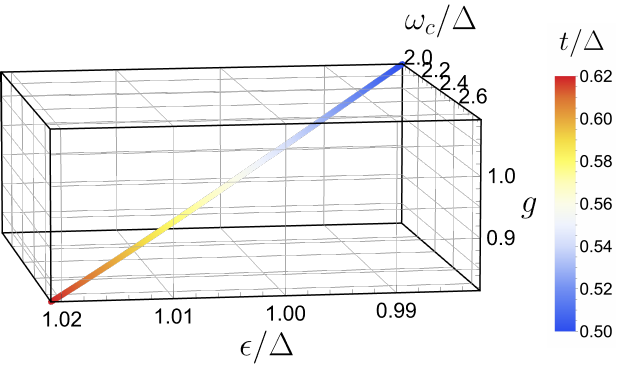}
    \caption{Density plot of the hopping $t/\Delta$ corresponding to the solution of the sweet spot conditions~\eqref{eq:Ueff}-\eqref{eq:tEff} as a function of  $\epsilon/\Delta$, $\omega_c/\Delta$ and  $g$ for the cavity in its ground state, $n=0$. 
    The other parameters are fixed as $U/\Delta=-0.1$, $V_Z/\Delta=1.5$, and $t_\text{so}/\Delta=0.9$.
    }
    \label{fig:PhaseDiagramAttractiveU}
\end{figure}
We note that the solution of Eqs.~\eqref{eq:Ueff}-\eqref{eq:tEff} corresponding to the sweet spot exists for different combinations of the parameters $\omega_c$, $t$,  $\epsilon$, and takes the form of the continuous curve in this parameter space (see Fig.~\ref{fig:PhaseDiagramAttractiveU}). The color coding further indicates that the sweet spot can be accessed over a wide range of inter-dot hopping amplitudes $t$ by tuning the on-site energy $\epsilon$, the cavity frequency $\omega_c$, and the light–matter coupling strength $g$. Therefore, coupling to cavity photons broadens the sweet spot into a continuous line
compared to the isolated electronic chain, also canceling the attractive effective interaction $U<0$.

\subsection{Sweet spot for $n=1$}

Next, we consider the cavity prepared in a state with one photon $n=1$. Taking the high frequency limit $\omega_c\rightarrow \infty$ in Eq.~\eqref{eq:Ueff}, we arrive at the condition for canceling the Coulomb interaction term:
\begin{equation}
    U \xrightarrow{\omega_c\rightarrow \infty} -\dfrac{e^{-g^2}\left(2-4g^2+g^4\right)g^2t^2}{\omega_c }.
\end{equation}

\begin{figure}[b]
    \centering
    \includegraphics[width=\columnwidth]{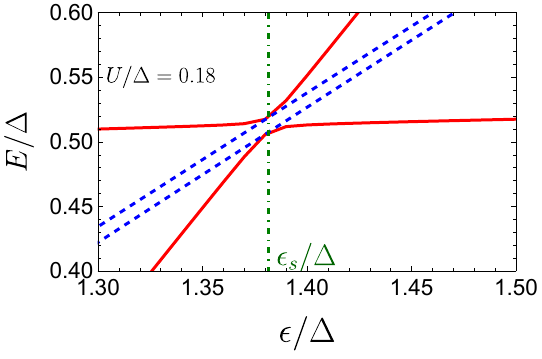}
    \caption{Many-body spectrum of the interacting effective Hamiltonian for two quantum dot chain with superconductivity embedded in a cavity as a function of $\epsilon/\Delta$ for $n=1$. We consider repulsive electron-electron interactions with $U/\Delta = 0.18$.
     The green dashed vertical line depicts the sweet spot value  $\epsilon_s/\Delta = 1.38$ obtained from Eqs.~\eqref{eq:Ueff}-\eqref{eq:tEff}. The parameters are chosen as $t/\Delta = 0.5$, $t_{so}/\Delta = 0.1$ and $V_Z/\Delta = 1.5$, and solving the sweet spot conditions we find that $g = 0.78$,  $\omega_c/\Delta = 0.57$. The bandwidth at the sweet spot is given by 
    $\Delta^\text{s}_\text{bw}(n=1,g)/\Delta=0.011 < \omega_c/\Delta$}. 
    \label{fig:SpectrumEpsilonUrepulsive1}
\end{figure}

We note that in the high frequency regime, we can screen repulsive interactions for $\sqrt{2-\sqrt{2}}\,\leq g \leq\, \sqrt{2+\sqrt{2}}$. Numerically solving the conditions for obtaining isolated poor man's MBSs in the $n=1$ photonic subspace, we find that we can screen the repulsive (with $U>0$) Coulomb interaction Fig.~\ref{fig:SpectrumEpsilonUrepulsive1}.

We further study the many-body energy spectrum of the effective model, see Fig.~\ref{fig:SpectrumEpsilonUrepulsive1}. For the cavity prepared in a state with one photon there is a degeneracy between even and odd parity ground states $\tilde{E}_{-}^{\text{even}}(1)=\tilde{E}_{-}^{\text{odd}}(1)$ at the value of the parameter $\epsilon$ that is a solution of Eqs.~\eqref{eq:Ueff}-\eqref{eq:tEff} that define the sweet spot. 

For a two-site Kitaev chain coupled to a cavity~\cite{gomez2025majorana}, it was found that for the cavity prepared in a state $n=1$ the degeneracy occurs between the even and odd parity ground states corresponding to different photonic subspaces.

This is due to the form of the many-body eigenvalues and $n\omega_c$ term that entered only in the odd subspace since the even parity sector did not include cavity coupling. However, now the even parity sector is also coupled to a cavity leading to a many-body eigenvalues that contain the photon energy shift, $n\omega_c$, both in the even and the odd parity sectors. Therefore, the degeneracy between even and odd parity sectors arises within the same photonic subspace, contrary to the result obtained in Ref.~\cite{gomez2025majorana}. 
We note that the excited states also cross $\tilde{E}_{+}^{\text{even}}(1)=\tilde{E}_{+}^{\text{odd}}(1)$ at the sweet spot, and excited and ground states become symmetric  with respect to $\tilde{C}(1)$.

\begin{figure}[t]
    \centering
    \includegraphics[width=\columnwidth]{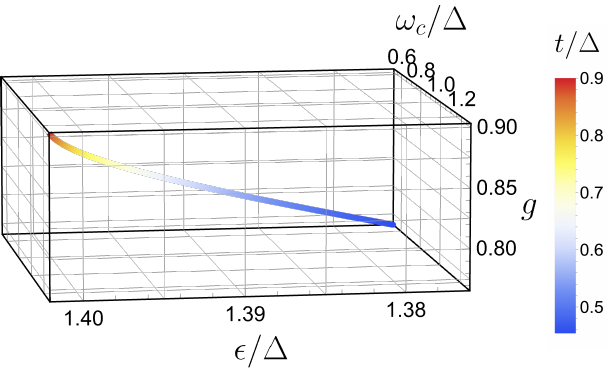}
    \caption{Density plot of the hopping $t/\Delta$ corresponding to the solution of the sweet spot conditions~\eqref{eq:Ueff}-\eqref{eq:tEff} as a function of  $\epsilon/\Delta$, $\omega_c/\Delta$ and  $g$ for $n=1$. The other parameters are fixed as $U/\Delta=0.18$, $V_z/\Delta=1.5$, $\Delta=1.0$ and $t_\text{so}/\Delta=0.1$.
    }\label{fig:PhaseDiagramRepulsiveU}
\end{figure}

Fig.~\ref{fig:PhaseDiagramRepulsiveU} shows that the sweet spot solution of Eqs.~\eqref{eq:Ueff}-\eqref{eq:tEff} for the cavity prepared in an excited state with one photon ($n=1$) is a curve in the parameter space of $\epsilon$, $\omega_c$, and $g$. Photon coupling enlarges the isolated sweet-spot condition into a continuous line, with the possibility to screen the Coulomb interactions. The color map shows that the sweet spot remains accessible over a wide range of inter-dot hopping amplitudes by tuning $\epsilon$, $\omega_c$, and the light–matter coupling $g$.


\subsection{Sweet spot for $g\sqrt{n} = \text{const}$}

In this section, we consider the limit of a large number of photons in the cavity, recovering the classical light limit.
Following Ref.~\cite{sentef2020quantum}, we fix $g\sqrt{n} = \text{const}$ that corresponds to a large photon number $n\rightarrow\infty$ and small light-matter coupling $g\rightarrow 0$.  
Starting from the light-matter Hamiltonian~\eqref{eq:Htot}, we perform a unitary transformation $\hat{U}(t) = \exp{\left(-i\omega_c\,a^\dag a\,t\right)}$. This transformation moves the system description to the interaction picture with respect to $\omega_c\,a^\dag a$. The transformed Hamiltonian $H_\text{el-ph}^\text{sc}\left(t\right) = \hat{U}^\dag(t)\left(H_\text{el-ph}-i\partial_t\right)\hat{U}(t)$ is then periodic in time, and one can perform a high-energy frequency expansion.  Therefore, considering the limit $n\to\infty$ at the lowest order in the expansion with $2g\sqrt{n} \equiv \lambda =\text{const}$,  we find the renormalized parameters (see Appendix~\ref{largeNlimit} for more details)
\begin{align}
    &t\to tJ_0\!\left(\lambda\right)e^{-g^2/2},\\
    &\Delta \to \Delta J_0\!\left(\lambda\right)e^{-g^2/2},\\
    &t_\text{so}\to t_\text{so}J_0\!\left(\lambda\right)e^{-g^2/2},
\end{align}
where $J_0\left(x\right)$ is the zero-th Bessel function of the first kind.
We observe that hopping and pairing amplitudes are renormalized in the same way. 
The sweet spot conditions become
\begin{align}
    &E_\alpha(\lambda) = 0,\label{eq:floquet_sweet_spot_1}\\
    &t_\text{so}\Delta\,e^{-g^2/2}J_0(\lambda) = t\epsilon,\label{eq:floquet_sweet_spot_2}
\end{align}
 where  
 \begin{align}
 &\gamma(\lambda)=\sqrt{\Delta^2e^{-g^2}J_0^2(\lambda)+\epsilon^2},\\
 &E_\alpha(\lambda)= \gamma(\lambda)-V_{\text{Z}} +\frac{2}{\gamma^2(\lambda)\left(\gamma(\lambda)+V_{\text{Z}}\right)}\notag\\
 &\times\left[(t_\text{so}\epsilon)^2-(t\,\Delta e^{-g^2/2}J_0(\lambda))^2\right]. 
 \end{align}

 We find that even in the limit of a large number of photons, the cavity embedding can influence the sweet spot for poor man's MBSs. However, in this regime, the zero-order Bessel function goes to zero, suppressing the hopping amplitudes and making the energy spectrum degenerate.  We stress that the degeneracy of the energy levels originates from the dressing of the superconducting pairing, $\Delta$, with the photonic operators through the Peierls phase. Indeed, without the term $\Delta\,e^{-g^2/2}J_0(\lambda)$ in Eq.~\eqref{eq:floquet_sweet_spot_2}, we recover the sweet spot conditions for the isolated effective Kitaev chain.  Therefore, we conclude that it is beneficial to work with a cavity prepared in a state with a small number of photons to realize isolated poor man's MBSs without suppressing the hopping amplitudes in the quantum dot-based platform.


\section{Conclusions}\label{Conclusion}
In this work, we theoretically demonstrated the emergence of the poor man's MBSs in an interacting quantum dot-superconductor array from cavity embedding. Starting with a microscopic model for two spinful quantum dots with local superconducting pairing coupled to a single mode cavity, we derived the effective Hamiltonian description for the problem and a set of sweet spot conditions for poor man's MBSs that also depend on the parameters of the cavity. Moreover, we found that cavity embedding engineers effective electron interactions that could be used to compensate for the intrinsic particle interactions in the quantum dot-superconductor array. In particular, the cavity prepared in its ground state generates repulsive electron interactions, while cavity prepared in the state with one photon gives rise to attractive electron interactions. Finally, a large number of photons in the cavity suppresses the electron hopping highlighting the importance of working with quantum light for controlling properties of quantum system.

A possible extension of this work is to study the effect of the photon losses and quasiparticle poisoning on poor man's MBSs.

\section*{Acknowledgments}
F.B. and O.D. acknowledge helpful discussions with Paul Fadler.
This work is supported by ERC grant (Q-Light-Topo, Grant Agreement No. 101116525) (F.B. and O.D.). A.G.L acknowledges support by MICIU/AEI/10.13039/501100011033 and by ERDF/EU to
project PID2023-146531NA-I00. Also acknowledges support from CSIC Interdisciplinary Thematic Platform (PTI+) on Quantum Technologies (PTI-QTEP+) and by FEDER Una Manera de Hacer Europa Project from the QUANTERA project
MOLAR with reference PCI2024-153449 funded by
Grant No. MICIU/AEI/10.13039/501100011033. M.S. acknowledges funding from the European Research Council (ERC) under the European Union's Horizon 2020 research and innovation program (Grant agreement No. 101002955 -- CONQUER).

\appendix

\section{Derivation of the Kitaev chain Hamiltonian in the absence of the cavity}
\label{Kitaevlimitnocavity}

In this section, we discuss the details of the derivation of the low-energy electronic Hamiltonian corresponding to the microscopic quantum dot chain with the local superconducting proximity effect following Ref.~\cite{samuelson2024minimal}. 
The Hamiltonian describing the system reads
\begin{equation}
    \tilde{H}_\text{tot} = \tilde{H}_\text{QD} + \tilde{H}_\text{C},
\end{equation}
where $\tilde{H}_\text{QD}$ is the Hamiltonian of the quantum dots with proximity-induced $s$-wave superconductivity and $\tilde{H}_\text{C}$ is the \textit{coupling Hamiltonian} accounting for the interactions between QDs.
The $\tilde{H}_\text{QD}$ and $\tilde{H}_\text{C}$ are give by
\begin{align}
    &\tilde{H}_{QD} = \sum_{i=1}^2\left(\epsilon_i-V_\text{Z}\right)n_{i\downarrow}+\left(\epsilon_i+V_\text{Z}\right)n_{i\uparrow}\notag\\&\hspace{2em}+\left(\Delta_id^\dag_{i\uparrow}d^\dag_{i\downarrow}+\text{h.c.}\right)+U^{\text{loc}}_i\, n_{i\uparrow}\,n_{i\downarrow}\label{eq:quantum_dot_two_site_superconductor}\\
    &\tilde{H}_\text{C} = \sum_{i=1}^2\sum_{\sigma} t\,d^\dag_{i\sigma}d_{i+1\sigma}+t\,d^\dag_{i+1\sigma}d_{i\sigma}+t_\text{so}\big(d^\dag_{i\uparrow}d_{i+1\downarrow}\notag\\
    &-d^\dag_{i\downarrow}d_{i+1\uparrow}+\text{h.c.}\big)+U^\text{nl}d^\dag_{1\sigma}d_{1\sigma}d^\dag_{2\sigma}d_{2\sigma}\label{eq:coupling_hamiltonian_two_dot_super},
\end{align}
where $d_{i\sigma},\,d_{i\sigma}^\dag$ are the fermionic annihilation and creation of a spin-$\sigma$ electron on the QD, with $\sigma\, \in \left\{\uparrow,\downarrow\right\}$, $n_{i\sigma} = d_{i\sigma}^\dag d_{i\sigma}$ is the QD fermionic $\sigma$-spin number operator, 
and $N_i = n_{i,\uparrow}+n_{i,\downarrow}$ is the total occupation number on the QD, where $i$ is the dot index of the $i$th-QD;
$\epsilon_i$ is the chemical potential of each dot, $V_\text{Z}$ is the {Zeeman energy} due to the external applied magnetic field, $t$ and $t_\text{so}$ are the {spin-conserving} and {spin-flipping} tunneling amplitude chosen real and positive, $\Delta_i$ is the superconductive pairing strength.
For simplicity, we assume that $\epsilon_1=\epsilon_2=\epsilon$ and $\Delta_1 = \Delta_2=\Delta$. Moreover, the magnetic field is chosen to be perpendicular to the spin-orbit interaction, which is the physical mechanism responsible for the spin-flipping tunneling, since having it parallel would suppress the spin-orbit effects~\cite{Stepanenko2012Spinorbit}.

To derive the effective low-energy description we use the projector method~\cite{FESHBACH1958357} in the limit where $\Delta,\,V_\text{Z}\gg\,t,\,t_\text{so}$ known as {Kitaev limit}.
Treating the coupling Hamiltonian as a perturbation, we diagonalize the dot Hamiltonian. Rewriting  $\tilde{H}_\text{QD}$ in the BdG form, we arrive at
\begin{align}
\tilde{H}_\text{QD} &=\frac{1}{2}\sum_{i=1}^2\Psi_i^\dag\mathcal{H}_i\Psi_i+E_0\\
\mathcal{H}_i &= 
\begin{pmatrix}
\epsilon+V_{\text{z}}&0&0 & \Delta \\
0 & \epsilon-V_{\text{z}}&-\Delta&0\\
0 &-\Delta&-\epsilon-V_{\text{z}}&0 \\
\Delta&0 &0& -\epsilon+V_{\text{z}}\\
\end{pmatrix}\\
\Psi_i&=
\begin{pmatrix}
d_{i\uparrow} \\
d_{i\downarrow}\\
d_{i\uparrow}^\dag\\
d_{i\downarrow}^\dag
\end{pmatrix}
\end{align}
where $E_0=2\left(\epsilon-\sqrt{\epsilon^2+\Delta^2}\right)$. Next, we diagonalize $\mathcal{H}_i$ and rewrite the dot Hamiltonian in the diagonal form 
    $\tilde{H}_\text{QD} = \sum_{i=1}^2 E_\alpha\, \alpha_i^\dag \alpha_i+E_\beta \beta_i^\dag \beta_i+E_0$, where $E_\alpha=\gamma-V_\text{Z}$ and $E_\beta=\gamma+V_\text{Z}$ are the eigenvalues associated with the fermionic operators $\alpha_i$ and $\beta_i$,
\begin{align}
    &\alpha_i = \frac{1}{\sqrt{2\gamma}} \left(\sqrt{\gamma-\epsilon}\,d^\dag_{i\downarrow}+\sqrt{\gamma+\epsilon}\,d_{i\uparrow}\right),\\
  &\beta_i = \frac{1}{\sqrt{2\gamma}} \left(\sqrt{\gamma-\epsilon}\,d^\dag_{i\uparrow}-\sqrt{\gamma+\epsilon}\,d_{i\downarrow}\right),
\end{align}
and $\gamma= \sqrt{\Delta^2+\epsilon^2}$. 
Performing the inverse of the relations of the operators $\alpha_i$ and $\beta_i$, we obtain
\begin{align}
    d^\dag_{i\downarrow} &= \frac{\sqrt{\gamma-\epsilon}\,\alpha_i^\dag-\sqrt{\gamma+\epsilon}\,\beta_i^\dag}{\sqrt{2\gamma}}\label{eq:inverse_two_dot_1}\\
    d_{i\uparrow} &= \frac{\sqrt{\gamma+\epsilon}\,\alpha_i^\dag+\sqrt{\gamma-\epsilon}\,\beta_i^\dag}{\sqrt{2\gamma}}.
    \label{eq:inverse_two_dot_2}
\end{align}
Introducing the inverse relations Eq.~\eqref{eq:inverse_two_dot_1} and ~\eqref{eq:inverse_two_dot_2}, into the coupling Hamiltonian Eq.~\eqref{eq:coupling_hamiltonian_two_dot_super}, we arrive at
\begin{align}
    \tilde{H}_\text{C} &= \sum_{i=1}^2 t_{\alpha\alpha}\alpha_i^\dag \alpha_{i+1} + \Delta_{\alpha\alpha}\alpha_i\alpha_{i+1}+t_{\alpha\beta}\alpha_i^\dag \beta_{i+1}\notag\\&+\Delta_{\alpha\beta}\alpha_i\beta_{i+1}-\Delta_{\alpha\beta}\alpha_{i+1}\beta_i+t_{\alpha\beta}\beta_i^\dag \alpha_{i+1}\notag\\&+\Delta_{\beta\beta}\beta_i\beta_{i+1}+t_{\beta\beta}\beta_i^\dag \beta_{i+1}+\text{h.c.},
\end{align}
where the new hopping parameters $t_{\alpha\alpha}$, $t_{\beta\beta}$ $t_{\alpha\beta}$ and superconducting pairings $\Delta_{\alpha\alpha}$, $\Delta_{\beta\beta}$, $\Delta_{\alpha\beta}$ read
\begin{align}
    t_{\alpha\alpha}=-t_{\beta\beta} &= -\frac{\epsilon t}{\gamma}\\
    t_{\beta\alpha}=t_{\alpha\beta} &=\frac{-t\Delta}{\gamma}\\
       \Delta_{\alpha\alpha}=-\Delta_{bb} &=\frac{-t_\text{so}\Delta}{\gamma}\\
       \Delta_{\alpha\beta} = \Delta_{\beta\alpha}&=\frac{t_\text{so}\epsilon}{\gamma}.
       \label{eq:parameters_effective_superconductor}
\end{align}
In the Kitaev limit, i.e. $V_\text{Z},\,\Delta\gg t_\text{so},\,t$, the energy $E_\beta\gg E_\alpha$, and we can derive an effective Hamiltonian $H_\text{eff} = PHP$, where the projector is given by $P = (1-\beta_1^\dag \beta_1)(1-\beta_2^\dag \beta_2)$. To capture the leading physics of the system, it is sufficient to stop at the first order in the Fesbach projection expansion~\cite{samuelson2024minimal}. Therefore, the effective Hamiltonian $H_\text{eff}$ becomes
\begin{figure}[b]
    \centering
    \includegraphics[width=\columnwidth]{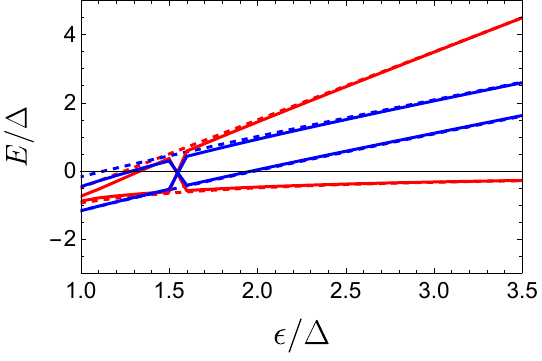}
    \caption{Many-body energy spectrum of two-quantum-dot chain with superconductivity as a function of $\epsilon/\Delta$.  The solid lines correspond to the even and odd parity eigenvalues of the effective Hamiltonian~\eqref{eq:effective_two_dot_hamiltonian}, while the dashed ones correspond to the even and odd parity eigenvalues of the two-quantum-dot chain with superconductivity given by Eqs.~\eqref{eq:quantum_dot_two_site_superconductor} and \eqref{eq:coupling_hamiltonian_two_dot_super}. The parameters chosen are $V_Z/\Delta=1.25$, $t/\Delta=0.5$, $t_\text{so}/\Delta=0.1$.}
    \label{fig:sectrumnocavity}
\end{figure}
\begin{align}
    H_\text{eff}^{1st} &=  \sum_{i=1}^2 E_\alpha^{1\text{st}} \alpha_i^\dag \alpha_i + \left(t_{\alpha\alpha}\alpha_i^\dag \alpha_{i+1}+\text{h.c.}\right) \notag\\
    &+ \left(\Delta_{\alpha\alpha}\alpha_i\alpha_{i+1}+\text{h.c.}\right)
    \label{eq:effective_two_dot_hamiltonian}
\end{align}
The second order in the expansion $\frac{PH_\text{C}QH_\text{C}P}{QHQ-E}$ produces only an energy shift of the on-site energies
\begin{equation}
    E_\alpha^{2\text{nd}}=\frac{2\left[(t_\text{so}\epsilon)^2-(t\,\Delta)^2\right]}{\gamma^2\,E_\beta}.
\end{equation}
Mapping $\alpha_1\rightarrow c_i$ and $\alpha_2 \rightarrow c_2$, we obtain the minimal Kitaev chain Hamiltonian, where the parameters $t$, $\Delta$ and $\mu$ of the original model now depend on the physical parameters of the system, see Eq.~\eqref{eq:parameters_effective_superconductor}. 
The poor man's MBSs emerge  in the system when the parameters are tuned to the sweet spot that for the effective Hamiltonian Eq.~\eqref{eq:effective_two_dot_hamiltonian} translates into the following conditions:
\begin{align}
    &E_\alpha^{1\text{st}}+ E_\alpha^{2\text{nd}}= 0, \label{eq:classical_sw_1}\\
    &t\,\epsilon= t_\text{so}\Delta.\label{eq:classical_sw_2}
\end{align}
We compare the many-body energy spectrum obtained with the effective Hamiltonian~\eqref{eq:effective_two_dot_hamiltonian}
to the spectrum of the microscopic model given by Eqs.~\eqref{eq:quantum_dot_two_site_superconductor} and \eqref{eq:coupling_hamiltonian_two_dot_super} in Fig.~\ref{fig:sectrumnocavity}. We find a relatively good agreement for $\epsilon > \sqrt{V_Z^2 - \Delta^2}$.

\section{Derivation of an effective light-matter Hamiltonian in the Kitaev limit} \label{KitaevLimit}
To derive the effective light-matter Hamiltonian, we can proceed in the same way as in Appendix~\ref{Kitaevlimitnocavity}. However, now $H_\text{QD}$ contains photonic operators $a, a^\dag$, therefore, we cannot directly proceed with its diagonalization. To overcome the problem, we rewrite the total Hamiltonian in the basis of the photon number states $\ket{n}$ and define the corresponding Hubbard operators $Y^{n,m}\equiv\ket{n}\bra{m}$~\cite{gomez2025majorana}.

The matrix element $\mathcal{M}_{mn}=\Braket{m|e^{ig\left(a+a^\dag\right)}|n}$  for $m\neq n$ is given by 
\begin{align}
\mathcal{M}_{mn} =  e^{-g^2/2} (ig)^{n-m}\sqrt{\frac{n!}{m!}}{}_1\tilde{F}_1(-m; n+1-m; g^2)
\end{align}
and it simplifies to 
 $\mathcal{M}_{nn} = e^{-g^2/2}L_{n}\left(g^2\right)$ for 
for $m = n$. Here, $L_{n}\left(g^2\right)$ is the $n$-th Laguerre polynomial and ${}_1\tilde{F}_1(-m; n+1-m; g^2)$ is the regularized confluent hypergeometric function.

We can identify two contributions for the $H_{\text{QD}}$, one diagonal in the photon subspace 
\begin{align}
H_{\text{QD}}^{D} &=\sum_n  \Bigg[\sum_{i=1}^2\left(\epsilon+V_\text{Z}\right)n_{i\uparrow}+\left(\epsilon-V_\text{Z}\right)n_{i\downarrow}\notag\\
&+\Delta e^{-g^2/2}L_n(g^2)\left(d^\dag_{1\uparrow}d^\dag_{1\downarrow}+\text{h.c.}\right)\notag\\
&+\Delta e^{-g^2/2}L_n(g^2)\left(d^\dag_{2\uparrow}d^\dag_{2\downarrow}+\text{h.c.}\right)\Bigg]Y^{n,n}
\label{equation:quantum_dot_hamiltonian_diagonal_photon}
\end{align}
and one off-diagonal
\begin{align}
H_\text{QD}^\text{OD}&=\sum_{m\neq n} \Big(\Delta_{m,n}(ig)\,d_{1,\uparrow}^\dagger d_{1,\downarrow}^\dagger\notag\\
&+\Delta_{m,n}(-ig)d_{2,\uparrow}^\dagger d_{2,\downarrow}^\dagger+\text{h.c.}\Big)Y^{m,n},
\end{align}
where $\Delta_{mn}(g)=\Delta e^{-g^2/2}(ig)^{n-m}\sqrt{\frac{n!}{m!}}{}_1\tilde{F}_1(-m;n+1-m;g^2)$. Next, we diagonalize $H_\text{QD}^\text{D}$ Eq. (\ref{equation:quantum_dot_hamiltonian_diagonal_photon}), where the eigenvalues $E_{\alpha} = \gamma(n)-V_\text{Z}$ and $E_{\beta} = \gamma(n)+V_\text{Z}$ now depend on the photon number $n$, and 
\begin{align}
   \alpha_i = \frac{1}{\sqrt{2\gamma(n)}} \left(\sqrt{\gamma(n)-\epsilon}\,d^\dag_{i\downarrow}+\sqrt{\gamma(n)+\epsilon}\,d_{i\uparrow}\right),\\
    \beta_i = \frac{1}{\sqrt{2\gamma(n)}} \left(\sqrt{\gamma(n)-\epsilon}\,d^\dag_{i\uparrow}-\sqrt{\gamma(n)+\epsilon}\,d_{i\downarrow}\right).
\end{align}
Here, $\gamma(n) $$=$$\sqrt{\epsilon^2+\Delta^2(n)}$, with $\Delta(n)=\Delta e^{-g^2/2}L_n(g^2)$. 
Consequently, substituting back the inverse relation of the $\alpha_i$ and $\beta_i$ in the total Hamiltonian, we can eliminate the high-energy electronic degree of freedom through the projector $P=~\left(1-\beta_1^\dag\beta_1\right)\left(1-\beta_2^\dag\beta_2\right)~\otimes~\mathbb{I}_\text{PH}$, where $\mathbb{I}_\text{PH}$ identity of the photonic Fock space.

Thus, the effective light-matter Hamiltonian in the Kitaev limit becomes
\begin{widetext}
\begin{align}
H_\text{QD}^\text{D}&=\sum_n\left[E_\alpha(n)\left(\alpha_1^\dag\alpha_1+\alpha_2^\dag\alpha_2\right)+2\left(\epsilon-\gamma(n)\right)\right]Y^{n,n}\\
    H_\text{QD}^\text{OD}&=\sum_{m\neq n}\frac{\Delta(n)}{\gamma(n)}\left[\Delta_{mn}(g)+\Delta_{mn}(-g)\right]-\frac{\Delta(n)}{2\gamma(n)}\left(\alpha_1^\dag\alpha_1+\alpha_2^\dag\alpha_2\right)\left[\Delta_{mn}(g)+\Delta_{mn}(-g)\right]Y^{m,n}\\
    H_\text{C,even}&=-\sum_{m,n}\frac{t_\text{so}}{2\gamma(n)}\left(\alpha_2\alpha_1+\alpha_1^\dag\alpha_2^\dag\right)\left(\Delta_{m,n}(g)+\Delta_{m,n}(-g)\right)Y^{m,n},\\
H_\text{C,odd}&=\sum_{m,n}\frac{t_{m,n}(g)}{2\gamma(n)}\left[\left(\gamma(n)-\epsilon\right)\alpha_1^\dag\alpha_2-\left(\gamma(n)+\epsilon\right)\alpha_2^\dag\alpha_1\right]+\frac{t_{m,n}(-g)}{2\gamma(n)}\left[\left(\gamma(n)-\epsilon\right)\alpha_2^\dag\alpha_1-\left(\gamma(n)+\epsilon\right)\alpha_1^\dag\alpha_2\right]Y^{n,m}.
\end{align}
\end{widetext}
Noticing that the Hamiltonian commutes with the fermionic parity operator, we have split the contribution to the $H_\text{C}$ in the even and odd sectors. Finally, we redefine $E_\alpha\rightarrow~E_\alpha~+~2\left[(t_\text{so}\,\epsilon)^2-(\Delta(n)\,t)^2\right]/\left[\gamma(n)^2\left(\gamma(n)+V_\text{Z}\right)\right]$ to include the contribution to the on-site energy coming from the second order in the projection method expansion.


\section{Derivation of an effective electronic Hamiltonian in a large detuning regime} \label{ProjectorsMethod}
To derive an effective electronic Hamiltonian from Eq.~\eqref{eq:hamiltonian_two_dot_final0},
we adiabatically eliminate the $n\pm1$ photonic subspace using the projector's method in the RWA approximation~\cite{FESHBACH1958357,gomez2025majorana}. We define the projectors $P = \ket{n}\bra{n}\otimes\mathbb{I}_\text{el}$ and $Q=\ket{n-1}\bra{n-1}+\ket{n+1}\bra{n+1}\otimes\mathbb{I}_\text{el}$, with $\mathbb{I}_\text{el}$ identity for the fermionic Fock space.
The terms $\left[\Delta_{m,n}(g)+\Delta_{m,n}(-g)\right]$ for $m=n\pm1$ sum up to zero, therefore $H_\text{QD}^\text{OD}$ has zero contribution.

We can then analyze separately the electronic even and odd sector, defined by:
\begin{align}
    &\text{Even states} \qquad \ket{00},\,\ket{11} = \alpha_1^\dag\alpha_2^\dag\ket{00}\\
    &\text{Odd states} \qquad \ket{10} = \alpha_1^\dag\ket{0},\,\ket{01} = \alpha_2^\dag\ket{0}.
\end{align}
In the even sector, the effective Hamiltonian reads
\begin{align}
    H_\text{even}^{D}&=-\sum_n\Big[ \frac{t_\text{so}\Delta(n)}{\gamma(n)}L_n(g^2)e^{-g^2/2}(\alpha_2\alpha_1+\text{h.c.})\notag\\
    &+n\omega_c\Big]Y^{n,n}+H^{D}_\text{QD},\label{eq:even_light_1}\\
 H_\text{even}^{OD}&=-\sum_{m\neq n}\frac{t_\text{so}}{2\gamma(n)}\left(\alpha_2\alpha_1+\text{h.c.}\right)\Big(\Delta_{m,n}(g)
 \notag\\
 &+\Delta_{m,n}(-g)\Big)Y^{m,n},\label{eq:even_light_2}
\end{align}
where $H_\text{even}^{D}$ is diagonal in the photon number basis and plays the role of $H_0$ in the projector method, and 
$H_\text{even}^{OD}$ is the perturbation corresponding to $H_1$.

We note that the even subspace is coupled to the cavity, see Eqs.~\eqref{eq:even_light_1} and \eqref{eq:even_light_2}, resulting in the renormalization of  the superconducting paring of the isolated QD Hamiltonian~\eqref{eq:quantum_dot_two_site_superconductor} as $\Delta \rightarrow \Delta_{m,n}(g)$. 

The terms $\left[\Delta_{m,n}(g)+\Delta_{m,n}(-g)\right]$ for $m=n\pm1$ sum up to zero. Therefore,  the relevant contribution to the adiabatic elimination of the photonic subspaces comes only from $PH_{\text{eff},\text{E}}^0P$. Consequently, the effective electronic Hamiltonian in the even subspace after the adiabatic elimination becomes
\begin{align}
\tilde{H}_\text{even}(n)&=-\left[\frac{t_\text{so}\Delta(n)}{\gamma(n)}L_n(g^2)e^{-g^2/2}(\alpha_2\alpha_1+\alpha_1^\dag\alpha_2^\dag)+n\omega_c\right]\notag\\&E_\alpha(n)\left(\alpha_1^\dag\alpha_1+\alpha_2^\dag\alpha_2\right)+2\left(\epsilon-\gamma(n)\right)
\label{eq:effective_even}
\end{align}

Continuing the analysis for the odd electronic many-body subspace, we obtain for the Hamiltonian in the photonic number basis 
\begin{align}
    H_\text{odd}^\text{D}&=-\sum_n \frac{\epsilon\,t e^{-g^2/2}}{\gamma(n)}L_n(g^2)\left(\alpha_1^\dag\alpha_2+\text{h.c.}\right)Y^{n,n}+H^{D}_\text{QD}\\
    H_\text{odd}^{OD}&=\sum_{m\neq n}\Big\{\frac{t_{m,n}(g)}{2\gamma(n)}\Big[\left(\gamma(n)-\epsilon\right)\alpha_1^\dag\alpha_2\notag\\
    &-\left(\gamma(n)+\epsilon\right)\alpha_2^\dag\alpha_1\Big]+\text{h.c.}\Big\}Y^{n,m}.
\end{align}
Here, $H_\text{odd}^\text{D}$ is the diagonal part in the photon number basis, 
$H_\text{odd}^{OD}$ is the off-diagonal part in the photon number basis. 
To apply the projector method, we identify $H_\text{odd}^\text{D}\equiv H_0$  as free Hamiltonian and $H_\text{odd}^{OD}\equiv H_1$ as perturbation.

At the first order in the projection $PH_\text{eff,O}^0P$, we obtain 
\begin{align}    
    \tilde{H}_\text{odd}^\text{1st}(n)&= n\omega_{\text{c}} +E_\alpha(n)(\alpha_1^\dag\alpha_1+\alpha_2^\dag\alpha_2)+2\left(\epsilon+\gamma(n)\right)\notag\\-&\frac{\epsilon\,t e^{-g^2/2}}{\gamma (n)}L_n(g^2)\left(\alpha_1^\dag\alpha_2+\alpha_2^\dag\alpha_2\right)
\label{eq:effective_odd_1}.
\end{align}
To analyze the second order, we specify the basis in the Fock space made by the tensor product of the photonic Fock space and the electronic Fock space. Since the projectors defined at the beginning are relative only to the photonic subspaces $n,\,n\pm1$, a valid basis choice, associated with the relative eigenenergies, is the following one:
\begin{align}
    \ket{\sigma_x^\pm,n} &= \ket{\sigma_x^\pm}\otimes \ket{n}\\
    \ket{\sigma_x^\pm,n\pm1} &= \ket{\sigma_x^\pm}\otimes \ket{n\pm1},
\end{align}
where $\ket{\sigma_x^\pm}=1/2 \left(\ket{1,0}\pm\ket{0,1}\right)$.
In this basis, the second order in the projection using RWA reduces to:
\begin{align}
 &-\sum_{a,b}\frac{PH_1Q H_1Q P}{E_b-E_a} =\notag\\  &\ket{\tilde{1}}\bra{\tilde{1}}\left(\frac{\left\vert{\Braket{\tilde{4}|H_\text{eff,O}^1|\tilde{1}}}^2\right\vert}{\tilde{E}_1-\tilde{E_4}}+\frac{\left\vert\Braket{\tilde{6}|H_\text{eff,O}^1|\tilde{1}}^2\right\vert}{\tilde{E}_1-\tilde{E}_6}\right) \notag\\ &+\ket{\tilde{2}}\bra{\tilde{2}}\left(\frac{\left\vert{\Braket{\tilde{3}|H_\text{eff,O}^1|\tilde{2}}}^2\right\vert}{\tilde{E}_2-\tilde{E}_3}+\frac{\left\vert{\Braket{\tilde{5}|H_\text{eff,O}^1|\tilde{2}}}^2\right\vert}{\tilde{E}_2-\tilde{E}_5}\right)\notag,
\end{align}
where we can notice that the non-zero matrix elements are only the ones with the same electronic part that differ of $\pm1$ in the number of photons.
 After a bit of algebra, the second-order Hamiltonian in the projection, written in terms of the electronic operators $\alpha_1,\,\alpha_2$, yields
\begin{align}
&\tilde{H}_\text{odd}^\text{2nd}(n)=(\alpha_1^\dag\alpha_1+\alpha_2^\dag\alpha_2)\left[\frac{n\tilde{\kappa}_n^2\omega_c}{\omega_c^2-\tilde{\omega}_n^2}-\frac{(n+1)\tilde{\kappa}^2_{n+1}\omega_c}{\omega_c^2-\tilde{\omega}_{n+1}^2}\right]+\notag
\\&\left(\alpha_1^\dag\alpha_2+\text{h.c.}\right)\left[\frac{n\tilde{\kappa}_n^2\tilde{\omega}_n}{\omega_c^2-\tilde{\omega}_n^2}-\frac{(n+1)\tilde{\kappa}^2_{n+1}\tilde{\omega}_{n+1}}{\omega_c^2-\tilde{\omega}_{n+1}^2}\right],
\label{eq:effective_odd_2}
\end{align}
where we define the following frequencies
\begin{align}
\tilde{\omega}_n&=\frac{t\epsilon}{\gamma(n)}e^{-g^2/2}\left[L_{n-1}(g^2)+L_n(g^2)\right],\\
\tilde{\kappa}_n&=t\,g\,e^{-g^2/2}\,{}_1F_1(1-n,2,g^2).
\end{align}

Adding the effective electronic Hamiltonian in the even subspace $\tilde{H}_{\text{even}}(n)$ (\ref{eq:effective_even}), the first order effective electronic Hamiltonian $\tilde{H}_{\text{odd}}^\text{1st}(n)$ (\ref{eq:effective_odd_1}) and the second order effective electronic Hamiltonian  $\tilde{H}_\text{odd}^\text{2nd}(n)$ (\ref{eq:effective_odd_2}) of the odd subspace, we arrive at
\begin{align}
    &\tilde{H}_\text{eff}(n) = \tilde{H}_\text{eff,E}(n)+\tilde{H}_\text{eff,O}^\text{1st}(n)+\tilde{H}_\text{eff,O}^\text{2nd}(n)\\
    &\tilde{H}_\text{eff}\left(n\right) =  \tilde{U}\left(n\right) \left(c_1^\dag c_2^\dag c_2c_1\right)-\tilde{\mu}\left(n\right)\left(c_1^\dag c_1+c_2^\dag c_2\right)+\tilde{C}(n)\notag\\&+\tilde{\Delta}\left(n\right)\left(c_1c_2+c_2^\dag c_1^\dag\right)-\tilde{t}\left(n\right)\left(c_1^\dag c_2+c_2^\dag c_1\right), 
    \label{eq:hamiltonian_two_dot_final}
\end{align}
where we write $\tilde{H}_\text{eff}(n)$ using the mapping $\alpha_1\rightarrow c_1$ and $\alpha_2\rightarrow c_2$.  The coefficients in Eq.~\eqref{eq:hamiltonian_two_dot_final} are given by
\begin{align}
    &\tilde{U}\left(n\right) = -2\left[\frac{n\tilde{\kappa}_n^2\omega_c}{\omega_c^2-\tilde{\omega}_n^2}-\frac{(n+1)\tilde{\kappa}^2_{n+1}\omega_c}{\omega_c^2-\tilde{\omega}_{n+1}^2}\right],\\
    &\tilde{\mu}\left(n\right) = -E_{\alpha}(n)-\frac{n\tilde{\kappa}_n^2\omega_c}{\omega_c^2+\tilde{\omega}_n^2}-\frac{(n+1)\tilde{\kappa}^2_{n+1}\omega_c}{\omega_c^2-\tilde{\omega}_{n+1}^2},\\
    &\tilde{C}(n)=n\omega_c+2\left(\epsilon-\gamma(n)\right),\\
&\tilde{\Delta}\left(n\right) = \frac{t_{so}\Delta(n) e^{-g^2/2}}{\gamma(n)}L_n\left(g^2\right),\\
&\tilde{t}\left(n\right)=-\frac{n\tilde{\kappa}_n^2\tilde{\omega}_n}{\omega_c^2-\tilde{\omega}_n^2}+\frac{(n+1)\tilde{\kappa}^2_{n+1}\tilde{\omega}_{n+1}}{\omega_c^2-\tilde{\omega}_{n+1}^2}+\frac{\epsilon t e^{-g^2/2}}{\gamma(n)}L_n\left(g^2\right).
\end{align}

The effective Hamiltonian~\eqref{eq:hamiltonian_two_dot_final} contains cavity-induced interaction term $\tilde{U}(n)$. 

\section{Derivation of an effective light-matter Hamiltonian in the limit of  $g\sqrt{n} = \text{const}$}\label{largeNlimit}

In this section, we provide details on the derivation of the effective Hamiltonian in the limit of $g\sqrt{n} = \text{const}$. To do so,  we follow the approach developed in Ref.~\cite{sentef2020quantum}. Starting with the total microscopic electron-photon Hamiltonian $H_{\text{total}}$ given by Eqs.~\eqref{eq:two_dot_ligt_micro_2} and \eqref{eq:two_dot_ligt_micro_3},
we note that $H_{\text{total}}$ is already in the Floquet form. However, to apply the Van Vleck high frequency expansion~\cite{Eckardt2015HFE}, we need to make the time dependence explicit. Therefore, we derive the Hamiltonian in the rotating frame by performing the unitary transformation $\mathcal{U}=e^{-i\omega_\text{c}a^\dag a\,t}$,

\begin{equation}
    H(t) = \mathcal{U}^\dagger H \mathcal{U}-i\, \mathcal{U}^\dagger \partial_t {\mathcal{U}},
\end{equation}
resulting in
    \begin{align}  
H_{\text{QD}}(t) &= \sum_{i=1}^2\left[\left(\epsilon+V_\text{Z}\right)n_{i\uparrow}+\left(\epsilon-V_\text{Z}\right)n_{i\downarrow}\right]\notag
\\&+\Delta \left(e^{iA}d^\dag_{1\uparrow}d^\dag_{1\downarrow}+ e^{-iA}d^\dag_{2\uparrow}d^\dag_{2\downarrow}+\text{h.c.}\right)\\
H_\text{C}(t) &= \sum_{\sigma\,\in\left\{\uparrow,\downarrow\right\}} t \left(e^{iA}d^\dag_{1\sigma}d_{2\sigma}+\text{h.c.}\right)\notag\\&+t_\text{so}\left[e^{iA}\left(d^\dag_{1\uparrow}d_{2\downarrow}-d^\dag_{1\downarrow}d_{2\uparrow}\right)+\text{h.c.}\right],
\end{align}
where $A=g(a^\dagger e^{i\omega_ct}+a e^{-i\omega_ct})$. 
Now it is straightforward to see that the total Hamiltonian is periodic, with a period $T=\frac{2\pi}{\omega_c}$. Therefore, the first order in high frequency expansion is given by the time average of the Hamiltonian, which in our case reduces to calculating  the time average of the Peierls phases. Using the Baker–Campbell–Hausdorff formula, we arrive at
\begin{equation}
    \frac{\omega_c}{2\pi}\int_0^\frac{\omega_c}{2\pi} dt e^{iA} = \int_0^1 dx e^{{iga^\dagger} e^{2i\pi x}}e^{{iga} e^{-2i\pi x}}e^{-\frac{1}{2}g^2},
\end{equation}
where $x=\omega_\text{c} t/(2\pi)$
 Next, we expand  the exponential
\begin{align}
 &e^{-\frac{g^2}{2}}\sum_{kk'}\int_0^1\frac{(ig)^{k+k'}}{k!k'!}{a^\dagger}^k{a^\dagger}^{k'}e^{2i\pi x(k-k')}dx\notag\\
 =&e^{-\frac{g^2}{2}}\sum_k\frac{(ig)^{2k}}{(k!)^2}{a^\dagger}^k{a^\dagger}^{k}.
\end{align}
 and then evaluate it in a photon number state basis $\ket{n}$:
\begin{equation}
    \braket{n|e^{iA}|n}  = e^{-g^2/2}\sum_{n=0}^n\frac{(ig)^{2k}}{(k!)^2}\frac{n!}{(n-k)!}.
\end{equation} 
Since we are interested in the limit for $n\rightarrow\infty$ we can do the approximation $\frac{n!}{(n-k)!}\approx n^k$ and arrive at the following form
\begin{equation}
    \braket{n|e^{iA}|n} \xrightarrow[n\to\infty]{} e^{-g^2/2} J_0\!\left(\lambda\right),
\end{equation}
where $J_0$ is the zero-order Bessel function of the first kind and $\lambda = 2g\sqrt{n}$.  Hence, both $t_\text{so}$, $t$, and $\Delta$ terms become renormalized as
\begin{align}
    &t\to tJ_0\!\left(\lambda\right)e^{-g^2/2}\\
    &\Delta\to \Delta J_0\!\left(\lambda\right)e^{-g^2/2}\\
    &t_\text{so}\to t_\text{so}J_0\!\left(\lambda\right)e^{-g^2/2},
\end{align}
which leads to the following time-independent electronic Hamiltonian
\begin{align}  
&H_{\text{QD}}= \sum_{i=1}^2\left(\epsilon+V_\text{Z}\right)n_{i\uparrow}+\left(\epsilon-V_\text{Z}\right)n_{i\downarrow}\notag
\\&+\Delta J_0\!\left(\lambda\right)e^{-g^2/2} \left(d^\dag_{i\uparrow}d^\dag_{i\downarrow}+\text{h.c.}\right)
\label{eq:renormalized_floquet_1}
\\
&H_\text{C} = \sum_{\sigma} t J_0\!\left(\lambda\right)e^{-g^2/2} \left(d^\dag_{1\sigma}d_{2\sigma}+\text{h.c.}\right)\notag\\&+t_\text{so} J_0\!\left(\lambda\right)e^{-g^2/2}\left(d^\dag_{1\uparrow}d_{2\downarrow}-d^\dag_{1\downarrow}d_{2\uparrow}+\text{h.c.}\right).
\label{eq:renormalized_floquet_2}
\end{align}

We diagonalize the resulting renormalized Hamiltonian given by Eqs.~\eqref{eq:renormalized_floquet_1} and \eqref{eq:renormalized_floquet_2}) by introducing the Bogoliubov operators
\begin{align}
    & \alpha_i = \frac{1}{\sqrt{2\gamma(\lambda)}} \left(\sqrt{\gamma(\lambda)-\epsilon}\,d^\dag_{i\downarrow}+\sqrt{\gamma(\lambda)+\epsilon}\,d_{i\uparrow}\right),\\
&\beta_i = \frac{1}{\sqrt{2\gamma(\lambda)}} \left(\sqrt{\gamma(\lambda)-\epsilon}\,d^\dag_{i\uparrow}-\sqrt{\gamma(\lambda)+\epsilon}\,d_{i\downarrow}\right),
\end{align}
and corresponding eigenvalues $E_{\alpha}(\lambda) = \gamma(\lambda)-V_\text{Z}$ and $E_{\beta}(\lambda) = \gamma(\lambda)+V_\text{Z}$. Here, $\gamma(\lambda)=\sqrt{\Delta^2e^{-g^2}J_0^2(\lambda)+\epsilon^2}$.

Then, as in Appendix~\ref{KitaevLimit}, we can eliminate the high-energy electronic degree of freedom through the projector $P=~\left(1-\beta_1^\dag\beta_1\right)\left(1-\beta_2^\dag\beta_2\right)$. For the effective electronic Hamiltonian in the limit of $\lambda=\text{const}$, we obtain
\begin{align}
&H_\text{QD}(\lambda)=E_\alpha(\lambda)\left(\alpha_1^\dag\alpha_1+\alpha_2^\dag\alpha_2\right)+2\left[\epsilon-\gamma(\lambda)\right],\\
&H_\text{even}(\lambda)=-\frac{t_\text{so}\Delta(\lambda) e^{-g^2/2}}{\gamma(\lambda)}L_n(g^2)(\alpha_2\alpha_1+\text{h.c.})\\
&H_\text{odd}(\lambda)=-\frac{\epsilon\,t e^{-g^2/2}}{\gamma(\lambda)}L_n(g^2)\left(\alpha_1^\dag\alpha_2+\alpha_2^\dag\alpha_2\right),
\end{align}
where $\Delta(\lambda)=\Delta J_0\!\left(\lambda\right)e^{-g^2/2}$ and 
$E_\alpha(\lambda)$$=$$ \gamma(\lambda)-V_{\text{Z}} +2 \left[(t_\text{so}\epsilon)^2-(t\,\Delta e^{-g^2/2}J_0(\lambda))^2\right]/\left[\gamma^2(\lambda)\left(\gamma(\lambda)+V_{\text{Z}}\right)\right]
$.
 
Finally, we can bring the above Hamiltonian in the same form used in Eq.~\eqref{eq:hamiltonian_two_dot_final}. Adopting the mapping $\alpha_1\rightarrow c_1$ and $\alpha_2\rightarrow c_2$, with $c_i(c_i^\dag)$ being spinless annihilation (creation) fermionic operator, we obtain
\begin{align}
    &\tilde{H}_\text{eff}\left(\lambda\right) = -\tilde{\mu}\left(\lambda\right)\left(c_1^\dag c_1+c_2^\dag c_2\right) +\tilde{C}(\lambda)\notag\\&+\tilde{\Delta}\left(\lambda\right)\left(c_1c_2+c_2^\dag c_1^\dag\right)-\tilde{t}\left(\lambda\right)\left(c_1^\dag c_2+c_2^\dag c_1\right).
    \label{eq:HeffA}
\end{align}
Here,
\begin{align}
&\tilde{\mu}\left(\lambda\right) = -E_{\alpha}(\lambda),\\
&\tilde{C}(\lambda)=2\left(\epsilon-\gamma(\lambda)\right),\\
&\tilde{\Delta}\left(\lambda\right) = \frac{t_{so}\Delta(\lambda) e^{-g^2/2}}{\gamma(\lambda)}L_n\left(g^2\right),\\
&\tilde{t}\left(\lambda\right)=\frac{\epsilon t}{\gamma(\lambda)}e^{-g^2/2}L_n\left(g^2\right).
\end{align}

Diagonalizing Eq.~\eqref{eq:HeffA}, one finds the energy spectrum of the effective Hamiltonian in the large photon regime. We note that for $\lambda\gg 1$ the zero-order Bessel function $J_0(\lambda) \approx 0$ is suppressed leading to a degenerate energy spectrum.


\begin{thebibliography}{108}%
	\makeatletter
	\providecommand \@ifxundefined [1]{%
		\@ifx{#1\undefined}
	}%
	\providecommand \@ifnum [1]{%
		\ifnum #1\expandafter \@firstoftwo
		\else \expandafter \@secondoftwo
		\fi
	}%
	\providecommand \@ifx [1]{%
		\ifx #1\expandafter \@firstoftwo
		\else \expandafter \@secondoftwo
		\fi
	}%
	\providecommand \natexlab [1]{#1}%
	\providecommand \enquote  [1]{``#1''}%
	\providecommand \bibnamefont  [1]{#1}%
	\providecommand \bibfnamefont [1]{#1}%
	\providecommand \citenamefont [1]{#1}%
	\providecommand \href@noop [0]{\@secondoftwo}%
	\providecommand \href [0]{\begingroup \@sanitize@url \@href}%
	\providecommand \@href[1]{\@@startlink{#1}\@@href}%
	\providecommand \@@href[1]{\endgroup#1\@@endlink}%
	\providecommand \@sanitize@url [0]{\catcode `\\12\catcode `\$12\catcode
		`\&12\catcode `\#12\catcode `\^12\catcode `\_12\catcode `\%12\relax}%
	\providecommand \@@startlink[1]{}%
	\providecommand \@@endlink[0]{}%
	\providecommand \url  [0]{\begingroup\@sanitize@url \@url }%
	\providecommand \@url [1]{\endgroup\@href {#1}{\urlprefix }}%
	\providecommand \urlprefix  [0]{URL }%
	\providecommand \Eprint [0]{\href }%
	\providecommand \doibase [0]{https://doi.org/}%
	\providecommand \selectlanguage [0]{\@gobble}%
	\providecommand \bibinfo  [0]{\@secondoftwo}%
	\providecommand \bibfield  [0]{\@secondoftwo}%
	\providecommand \translation [1]{[#1]}%
	\providecommand \BibitemOpen [0]{}%
	\providecommand \bibitemStop [0]{}%
	\providecommand \bibitemNoStop [0]{.\EOS\space}%
	\providecommand \EOS [0]{\spacefactor3000\relax}%
	\providecommand \BibitemShut  [1]{\csname bibitem#1\endcsname}%
	\let\auto@bib@innerbib\@empty
	\bibitem [{\citenamefont {G\'omez-Le\'on}\ \emph {et~al.}(2025)\citenamefont
		{G\'omez-Le\'on}, \citenamefont {Schir\`o},\ and\ \citenamefont
		{Dmytruk}}]{gomez2025majorana}%
	\BibitemOpen
	\bibfield  {author} {\bibinfo {author} {\bibfnamefont {A.}~\bibnamefont
			{G\'omez-Le\'on}}, \bibinfo {author} {\bibfnamefont {M.}~\bibnamefont
			{Schir\`o}},\ and\ \bibinfo {author} {\bibfnamefont {O.}~\bibnamefont
			{Dmytruk}},\ }\bibfield  {title} {\bibinfo {title} {Majorana bound states
			from cavity embedding in an interacting two-site {K}itaev chain},\ }\href
	{https://doi.org/10.1103/PhysRevB.111.155410} {\bibfield  {journal} {\bibinfo
			{journal} {Phys. Rev. B}\ }\textbf {\bibinfo {volume} {111}},\ \bibinfo
		{pages} {155410} (\bibinfo {year} {2025})}\BibitemShut {NoStop}%
	\bibitem [{\citenamefont {Kitaev}(2001)}]{kitaev2001unpaired}%
	\BibitemOpen
	\bibfield  {author} {\bibinfo {author} {\bibfnamefont {A.~Y.}\ \bibnamefont
			{Kitaev}},\ }\bibfield  {title} {\bibinfo {title} {Unpaired {M}ajorana
			fermions in quantum wires},\ }\href
	{https://doi.org/10.1070/1063-7869/44/10S/S29} {\bibfield  {journal}
		{\bibinfo  {journal} {Physics-Uspekhi}\ }\textbf {\bibinfo {volume} {44}},\
		\bibinfo {pages} {131} (\bibinfo {year} {2001})}\BibitemShut {NoStop}%
	\bibitem [{\citenamefont {Kitaev}(2003)}]{Kitaev2003quantumcomputationanyons}%
	\BibitemOpen
	\bibfield  {author} {\bibinfo {author} {\bibfnamefont {A.}~\bibnamefont
			{Kitaev}},\ }\bibfield  {title} {\bibinfo {title} {Fault-tolerant quantum
			computation by anyons},\ }\href
	{https://doi.org/https://doi.org/10.1016/S0003-4916(02)00018-0} {\bibfield
		{journal} {\bibinfo  {journal} {Annals of Physics}\ }\textbf {\bibinfo
			{volume} {303}},\ \bibinfo {pages} {2} (\bibinfo {year} {2003})}\BibitemShut
	{NoStop}%
	\bibitem [{\citenamefont {Oreg}\ \emph {et~al.}(2010)\citenamefont {Oreg},
		\citenamefont {Refael},\ and\ \citenamefont {von
			Oppen}}]{Oreg2010HelicalLiquids}%
	\BibitemOpen
	\bibfield  {author} {\bibinfo {author} {\bibfnamefont {Y.}~\bibnamefont
			{Oreg}}, \bibinfo {author} {\bibfnamefont {G.}~\bibnamefont {Refael}},\ and\
		\bibinfo {author} {\bibfnamefont {F.}~\bibnamefont {von Oppen}},\ }\bibfield
	{title} {\bibinfo {title} {Helical liquids and {M}ajorana bound states in
			quantum wires},\ }\href {https://doi.org/10.1103/PhysRevLett.105.177002}
	{\bibfield  {journal} {\bibinfo  {journal} {Phys. Rev. Lett.}\ }\textbf
		{\bibinfo {volume} {105}},\ \bibinfo {pages} {177002} (\bibinfo {year}
		{2010})}\BibitemShut {NoStop}%
	\bibitem [{\citenamefont {Lutchyn}\ \emph {et~al.}(2010)\citenamefont
		{Lutchyn}, \citenamefont {Sau},\ and\ \citenamefont
		{Das~Sarma}}]{Lutchyn2010Majorana}%
	\BibitemOpen
	\bibfield  {author} {\bibinfo {author} {\bibfnamefont {R.~M.}\ \bibnamefont
			{Lutchyn}}, \bibinfo {author} {\bibfnamefont {J.~D.}\ \bibnamefont {Sau}},\
		and\ \bibinfo {author} {\bibfnamefont {S.}~\bibnamefont {Das~Sarma}},\
	}\bibfield  {title} {\bibinfo {title} {Majorana fermions and a topological
			phase transition in semiconductor-superconductor heterostructures},\ }\href
	{https://doi.org/10.1103/PhysRevLett.105.077001} {\bibfield  {journal}
		{\bibinfo  {journal} {Phys. Rev. Lett.}\ }\textbf {\bibinfo {volume} {105}},\
		\bibinfo {pages} {077001} (\bibinfo {year} {2010})}\BibitemShut {NoStop}%
	\bibitem [{\citenamefont {Alicea}(2012)}]{Alicea2012newdirections}%
	\BibitemOpen
	\bibfield  {author} {\bibinfo {author} {\bibfnamefont {J.}~\bibnamefont
			{Alicea}},\ }\bibfield  {title} {\bibinfo {title} {New directions in the
			pursuit of {M}ajorana fermions in solid state systems},\ }\href
	{https://doi.org/10.1088/0034-4885/75/7/076501} {\bibfield  {journal}
		{\bibinfo  {journal} {Reports on Progress in Physics}\ }\textbf {\bibinfo
			{volume} {75}},\ \bibinfo {pages} {076501} (\bibinfo {year}
		{2012})}\BibitemShut {NoStop}%
	\bibitem [{\citenamefont {Mourik}\ \emph {et~al.}(2012)\citenamefont {Mourik},
		\citenamefont {Zuo}, \citenamefont {Frolov}, \citenamefont {Plissard},
		\citenamefont {Bakkers},\ and\ \citenamefont
		{Kouwenhoven}}]{Mourik2012Signatures}%
	\BibitemOpen
	\bibfield  {author} {\bibinfo {author} {\bibfnamefont {V.}~\bibnamefont
			{Mourik}}, \bibinfo {author} {\bibfnamefont {K.}~\bibnamefont {Zuo}},
		\bibinfo {author} {\bibfnamefont {S.~M.}\ \bibnamefont {Frolov}}, \bibinfo
		{author} {\bibfnamefont {S.~R.}\ \bibnamefont {Plissard}}, \bibinfo {author}
		{\bibfnamefont {E.~P. A.~M.}\ \bibnamefont {Bakkers}},\ and\ \bibinfo
		{author} {\bibfnamefont {L.~P.}\ \bibnamefont {Kouwenhoven}},\ }\bibfield
	{title} {\bibinfo {title} {Signatures of {M}ajorana fermions in hybrid
			superconductor-semiconductor nanowire devices},\ }\href
	{https://doi.org/10.1126/science.1222360} {\bibfield  {journal} {\bibinfo
			{journal} {Science}\ }\textbf {\bibinfo {volume} {336}},\ \bibinfo {pages}
		{1003} (\bibinfo {year} {2012})},\ \Eprint
	{https://arxiv.org/abs/https://www.science.org/doi/pdf/10.1126/science.1222360}
	{https://www.science.org/doi/pdf/10.1126/science.1222360} \BibitemShut
	{NoStop}%
	\bibitem [{\citenamefont {Aghaee}\ and\ \citenamefont
		{et~al.}(2023)}]{Microsoft2023}%
	\BibitemOpen
	\bibfield  {author} {\bibinfo {author} {\bibfnamefont {M.}~\bibnamefont
			{Aghaee}}\ and\ \bibinfo {author} {\bibnamefont {et~al.}} (\bibinfo
		{collaboration} {Microsoft Quantum}),\ }\bibfield  {title} {\bibinfo {title}
		{In{A}s-{A}l hybrid devices passing the topological gap protocol},\ }\href
	{https://doi.org/10.1103/PhysRevB.107.245423} {\bibfield  {journal} {\bibinfo
			{journal} {Phys. Rev. B}\ }\textbf {\bibinfo {volume} {107}},\ \bibinfo
		{pages} {245423} (\bibinfo {year} {2023})}\BibitemShut {NoStop}%
	\bibitem [{\citenamefont {{Microsoft Azure Quantum}}\ \emph
		{et~al.}(2025)\citenamefont {{Microsoft Azure Quantum}}, \citenamefont
		{Aghaee}, \citenamefont {Alcaraz~Ramirez}, \citenamefont {Alam},
		\citenamefont {Ali}, \citenamefont {Andrzejczuk}, \citenamefont {Antipov},
		\citenamefont {Astafev}, \citenamefont {Barzegar}, \citenamefont {Bauer}
		\emph {et~al.}}]{Microsoft2025}%
	\BibitemOpen
	\bibfield  {author} {\bibinfo {author} {\bibnamefont {{Microsoft Azure
					Quantum}}}, \bibinfo {author} {\bibfnamefont {M.}~\bibnamefont {Aghaee}},
		\bibinfo {author} {\bibfnamefont {A.}~\bibnamefont {Alcaraz~Ramirez}},
		\bibinfo {author} {\bibfnamefont {Z.}~\bibnamefont {Alam}}, \bibinfo {author}
		{\bibfnamefont {R.}~\bibnamefont {Ali}}, \bibinfo {author} {\bibfnamefont
			{M.}~\bibnamefont {Andrzejczuk}}, \bibinfo {author} {\bibfnamefont
			{A.}~\bibnamefont {Antipov}}, \bibinfo {author} {\bibfnamefont
			{M.}~\bibnamefont {Astafev}}, \bibinfo {author} {\bibfnamefont
			{A.}~\bibnamefont {Barzegar}}, \bibinfo {author} {\bibfnamefont
			{B.}~\bibnamefont {Bauer}}, \emph {et~al.},\ }\bibfield  {title} {\bibinfo
		{title} {Interferometric single-shot parity measurement in {I}n{A}s--{A}l
			hybrid devices},\ }\href {https://doi.org/10.1038/s41586-024-08445-2}
	{\bibfield  {journal} {\bibinfo  {journal} {Nature}\ }\textbf {\bibinfo
			{volume} {638}},\ \bibinfo {pages} {651} (\bibinfo {year}
		{2025})}\BibitemShut {NoStop}%
	\bibitem [{\citenamefont {Liu}\ \emph {et~al.}(2012)\citenamefont {Liu},
		\citenamefont {Potter}, \citenamefont {Law},\ and\ \citenamefont
		{Lee}}]{liu2012zero}%
	\BibitemOpen
	\bibfield  {author} {\bibinfo {author} {\bibfnamefont {J.}~\bibnamefont
			{Liu}}, \bibinfo {author} {\bibfnamefont {A.~C.}\ \bibnamefont {Potter}},
		\bibinfo {author} {\bibfnamefont {K.~T.}\ \bibnamefont {Law}},\ and\ \bibinfo
		{author} {\bibfnamefont {P.~A.}\ \bibnamefont {Lee}},\ }\bibfield  {title}
	{\bibinfo {title} {Zero-bias peaks in the tunneling conductance of
			spin-orbit-coupled superconducting wires with and without {M}ajorana
			end-states},\ }\href {https://doi.org/10.1103/PhysRevLett.109.267002}
	{\bibfield  {journal} {\bibinfo  {journal} {Phys. Rev. Lett.}\ }\textbf
		{\bibinfo {volume} {109}},\ \bibinfo {pages} {267002} (\bibinfo {year}
		{2012})}\BibitemShut {NoStop}%
	\bibitem [{\citenamefont {Kells}\ \emph {et~al.}(2012)\citenamefont {Kells},
		\citenamefont {Meidan},\ and\ \citenamefont {Brouwer}}]{kells2012near}%
	\BibitemOpen
	\bibfield  {author} {\bibinfo {author} {\bibfnamefont {G.}~\bibnamefont
			{Kells}}, \bibinfo {author} {\bibfnamefont {D.}~\bibnamefont {Meidan}},\ and\
		\bibinfo {author} {\bibfnamefont {P.~W.}\ \bibnamefont {Brouwer}},\
	}\bibfield  {title} {\bibinfo {title} {Near-zero-energy end states in
			topologically trivial spin-orbit coupled superconducting nanowires with a
			smooth confinement},\ }\href {https://doi.org/10.1103/PhysRevB.86.100503}
	{\bibfield  {journal} {\bibinfo  {journal} {Phys. Rev. B}\ }\textbf {\bibinfo
			{volume} {86}},\ \bibinfo {pages} {100503} (\bibinfo {year}
		{2012})}\BibitemShut {NoStop}%
	\bibitem [{\citenamefont {Prada}\ \emph {et~al.}(2012)\citenamefont {Prada},
		\citenamefont {San-Jose},\ and\ \citenamefont {Aguado}}]{prada2012transport}%
	\BibitemOpen
	\bibfield  {author} {\bibinfo {author} {\bibfnamefont {E.}~\bibnamefont
			{Prada}}, \bibinfo {author} {\bibfnamefont {P.}~\bibnamefont {San-Jose}},\
		and\ \bibinfo {author} {\bibfnamefont {R.}~\bibnamefont {Aguado}},\
	}\bibfield  {title} {\bibinfo {title} {Transport spectroscopy of {$NS$}
			nanowire junctions with {M}ajorana fermions},\ }\href
	{https://doi.org/10.1103/PhysRevB.86.180503} {\bibfield  {journal} {\bibinfo
			{journal} {Phys. Rev. B}\ }\textbf {\bibinfo {volume} {86}},\ \bibinfo
		{pages} {180503} (\bibinfo {year} {2012})}\BibitemShut {NoStop}%
	\bibitem [{\citenamefont {Liu}\ \emph {et~al.}(2017)\citenamefont {Liu},
		\citenamefont {Sau}, \citenamefont {Stanescu},\ and\ \citenamefont
		{Das~Sarma}}]{liu2017andreev}%
	\BibitemOpen
	\bibfield  {author} {\bibinfo {author} {\bibfnamefont {C.-X.}\ \bibnamefont
			{Liu}}, \bibinfo {author} {\bibfnamefont {J.~D.}\ \bibnamefont {Sau}},
		\bibinfo {author} {\bibfnamefont {T.~D.}\ \bibnamefont {Stanescu}},\ and\
		\bibinfo {author} {\bibfnamefont {S.}~\bibnamefont {Das~Sarma}},\ }\bibfield
	{title} {\bibinfo {title} {Andreev bound states versus {M}ajorana bound
			states in quantum dot-nanowire-superconductor hybrid structures: {T}rivial
			versus topological zero-bias conductance peaks},\ }\href
	{https://doi.org/10.1103/PhysRevB.96.075161} {\bibfield  {journal} {\bibinfo
			{journal} {Phys. Rev. B}\ }\textbf {\bibinfo {volume} {96}},\ \bibinfo
		{pages} {075161} (\bibinfo {year} {2017})}\BibitemShut {NoStop}%
	\bibitem [{\citenamefont {Reeg}\ \emph {et~al.}(2018)\citenamefont {Reeg},
		\citenamefont {Dmytruk}, \citenamefont {Chevallier}, \citenamefont {Loss},\
		and\ \citenamefont {Klinovaja}}]{Reeg2018ZeroAndreev}%
	\BibitemOpen
	\bibfield  {author} {\bibinfo {author} {\bibfnamefont {C.}~\bibnamefont
			{Reeg}}, \bibinfo {author} {\bibfnamefont {O.}~\bibnamefont {Dmytruk}},
		\bibinfo {author} {\bibfnamefont {D.}~\bibnamefont {Chevallier}}, \bibinfo
		{author} {\bibfnamefont {D.}~\bibnamefont {Loss}},\ and\ \bibinfo {author}
		{\bibfnamefont {J.}~\bibnamefont {Klinovaja}},\ }\bibfield  {title} {\bibinfo
		{title} {Zero-energy {A}ndreev bound states from quantum dots in proximitized
			{R}ashba nanowires},\ }\href {https://doi.org/10.1103/PhysRevB.98.245407}
	{\bibfield  {journal} {\bibinfo  {journal} {Phys. Rev. B}\ }\textbf {\bibinfo
			{volume} {98}},\ \bibinfo {pages} {245407} (\bibinfo {year}
		{2018})}\BibitemShut {NoStop}%
	\bibitem [{\citenamefont {Pe\~naranda}\ \emph {et~al.}(2018)\citenamefont
		{Pe\~naranda}, \citenamefont {Aguado}, \citenamefont {San-Jose},\ and\
		\citenamefont {Prada}}]{penaranda2018quantifying}%
	\BibitemOpen
	\bibfield  {author} {\bibinfo {author} {\bibfnamefont {F.}~\bibnamefont
			{Pe\~naranda}}, \bibinfo {author} {\bibfnamefont {R.}~\bibnamefont {Aguado}},
		\bibinfo {author} {\bibfnamefont {P.}~\bibnamefont {San-Jose}},\ and\
		\bibinfo {author} {\bibfnamefont {E.}~\bibnamefont {Prada}},\ }\bibfield
	{title} {\bibinfo {title} {Quantifying wave-function overlaps in
			inhomogeneous {M}ajorana nanowires},\ }\href
	{https://doi.org/10.1103/PhysRevB.98.235406} {\bibfield  {journal} {\bibinfo
			{journal} {Phys. Rev. B}\ }\textbf {\bibinfo {volume} {98}},\ \bibinfo
		{pages} {235406} (\bibinfo {year} {2018})}\BibitemShut {NoStop}%
	\bibitem [{\citenamefont {Prada}\ \emph {et~al.}(2020)\citenamefont {Prada},
		\citenamefont {San-Jose}, \citenamefont {de~Moor}, \citenamefont {Geresdi},
		\citenamefont {Lee}, \citenamefont {Klinovaja}, \citenamefont {Loss},
		\citenamefont {Nyg{\aa}rd}, \citenamefont {Aguado},\ and\ \citenamefont
		{Kouwenhoven}}]{prada2020}%
	\BibitemOpen
	\bibfield  {author} {\bibinfo {author} {\bibfnamefont {E.}~\bibnamefont
			{Prada}}, \bibinfo {author} {\bibfnamefont {P.}~\bibnamefont {San-Jose}},
		\bibinfo {author} {\bibfnamefont {M.~W.~A.}\ \bibnamefont {de~Moor}},
		\bibinfo {author} {\bibfnamefont {A.}~\bibnamefont {Geresdi}}, \bibinfo
		{author} {\bibfnamefont {E.~J.~H.}\ \bibnamefont {Lee}}, \bibinfo {author}
		{\bibfnamefont {J.}~\bibnamefont {Klinovaja}}, \bibinfo {author}
		{\bibfnamefont {D.}~\bibnamefont {Loss}}, \bibinfo {author} {\bibfnamefont
			{J.}~\bibnamefont {Nyg{\aa}rd}}, \bibinfo {author} {\bibfnamefont
			{R.}~\bibnamefont {Aguado}},\ and\ \bibinfo {author} {\bibfnamefont {L.~P.}\
			\bibnamefont {Kouwenhoven}},\ }\bibfield  {title} {\bibinfo {title} {From
			{A}ndreev to {M}ajorana bound states in hybrid superconductor--semiconductor
			nanowires},\ }\href {https://doi.org/10.1038/s42254-020-0228-y} {\bibfield
		{journal} {\bibinfo  {journal} {Nature Reviews Physics}\ }\textbf {\bibinfo
			{volume} {2}},\ \bibinfo {pages} {575} (\bibinfo {year} {2020})}\BibitemShut
	{NoStop}%
	\bibitem [{\citenamefont {Hess}\ \emph {et~al.}(2021)\citenamefont {Hess},
		\citenamefont {Legg}, \citenamefont {Loss},\ and\ \citenamefont
		{Klinovaja}}]{hess2021local}%
	\BibitemOpen
	\bibfield  {author} {\bibinfo {author} {\bibfnamefont {R.}~\bibnamefont
			{Hess}}, \bibinfo {author} {\bibfnamefont {H.~F.}\ \bibnamefont {Legg}},
		\bibinfo {author} {\bibfnamefont {D.}~\bibnamefont {Loss}},\ and\ \bibinfo
		{author} {\bibfnamefont {J.}~\bibnamefont {Klinovaja}},\ }\bibfield  {title}
	{\bibinfo {title} {Local and nonlocal quantum transport due to {A}ndreev
			bound states in finite {R}ashba nanowires with superconducting and normal
			sections},\ }\href {https://doi.org/10.1103/PhysRevB.104.075405} {\bibfield
		{journal} {\bibinfo  {journal} {Phys. Rev. B}\ }\textbf {\bibinfo {volume}
			{104}},\ \bibinfo {pages} {075405} (\bibinfo {year} {2021})}\BibitemShut
	{NoStop}%
	\bibitem [{\citenamefont {Hess}\ \emph {et~al.}(2023)\citenamefont {Hess},
		\citenamefont {Legg}, \citenamefont {Loss},\ and\ \citenamefont
		{Klinovaja}}]{hess2023trivial}%
	\BibitemOpen
	\bibfield  {author} {\bibinfo {author} {\bibfnamefont {R.}~\bibnamefont
			{Hess}}, \bibinfo {author} {\bibfnamefont {H.~F.}\ \bibnamefont {Legg}},
		\bibinfo {author} {\bibfnamefont {D.}~\bibnamefont {Loss}},\ and\ \bibinfo
		{author} {\bibfnamefont {J.}~\bibnamefont {Klinovaja}},\ }\bibfield  {title}
	{\bibinfo {title} {Trivial {A}ndreev band mimicking topological bulk gap
			reopening in the nonlocal conductance of long {R}ashba nanowires},\ }\href
	{https://doi.org/10.1103/PhysRevLett.130.207001} {\bibfield  {journal}
		{\bibinfo  {journal} {Phys. Rev. Lett.}\ }\textbf {\bibinfo {volume} {130}},\
		\bibinfo {pages} {207001} (\bibinfo {year} {2023})}\BibitemShut {NoStop}%
	\bibitem [{\citenamefont {Sahu}\ \emph {et~al.}(2023)\citenamefont {Sahu},
		\citenamefont {Khade},\ and\ \citenamefont {Gangadharaiah}}]{sahu2023effect}%
	\BibitemOpen
	\bibfield  {author} {\bibinfo {author} {\bibfnamefont {D.}~\bibnamefont
			{Sahu}}, \bibinfo {author} {\bibfnamefont {V.}~\bibnamefont {Khade}},\ and\
		\bibinfo {author} {\bibfnamefont {S.}~\bibnamefont {Gangadharaiah}},\
	}\bibfield  {title} {\bibinfo {title} {Effect of topological length on bound
			state signatures in a topological nanowire},\ }\href
	{https://doi.org/10.1103/PhysRevB.108.205426} {\bibfield  {journal} {\bibinfo
			{journal} {Phys. Rev. B}\ }\textbf {\bibinfo {volume} {108}},\ \bibinfo
		{pages} {205426} (\bibinfo {year} {2023})}\BibitemShut {NoStop}%
	\bibitem [{\citenamefont {Prem}\ \emph {et~al.}(2026)\citenamefont {Prem},
		\citenamefont {Dmytruk},\ and\ \citenamefont
		{Trif}}]{prem2026distinguishing}%
	\BibitemOpen
	\bibfield  {author} {\bibinfo {author} {\bibfnamefont {S.}~\bibnamefont
			{Prem}}, \bibinfo {author} {\bibfnamefont {O.}~\bibnamefont {Dmytruk}},\ and\
		\bibinfo {author} {\bibfnamefont {M.}~\bibnamefont {Trif}},\ }\bibfield
	{title} {\bibinfo {title} {Distinguishing {M}ajorana bound states from
			accidental zero-energy modes with a microwave cavity},\ }\href
	{https://doi.org/10.1103/q7l6-ytrr} {\bibfield  {journal} {\bibinfo
			{journal} {Phys. Rev. B}\ }\textbf {\bibinfo {volume} {113}},\ \bibinfo
		{pages} {085420} (\bibinfo {year} {2026})}\BibitemShut {NoStop}%
	\bibitem [{\citenamefont {Seoane~Souto}\ and\ \citenamefont
		{Aguado}(2024)}]{SeoaneSouto2024}%
	\BibitemOpen
	\bibfield  {author} {\bibinfo {author} {\bibfnamefont {R.}~\bibnamefont
			{Seoane~Souto}}\ and\ \bibinfo {author} {\bibfnamefont {R.}~\bibnamefont
			{Aguado}},\ }\bibinfo {title} {Subgap states in semiconductor-superconductor
		devices for quantum technologies: {A}ndreev qubits and minimal {M}ajorana
		chains},\ in\ \href {https://doi.org/10.1007/978-3-031-55657-9_3} {\emph
		{\bibinfo {booktitle} {New Trends and Platforms for Quantum Technologies}}},\
	\bibinfo {editor} {edited by\ \bibinfo {editor} {\bibfnamefont
			{R.}~\bibnamefont {Aguado}}, \bibinfo {editor} {\bibfnamefont
			{R.}~\bibnamefont {Citro}}, \bibinfo {editor} {\bibfnamefont
			{M.}~\bibnamefont {Lewenstein}},\ and\ \bibinfo {editor} {\bibfnamefont
			{M.}~\bibnamefont {Stern}}}\ (\bibinfo  {publisher} {Springer Nature
		Switzerland},\ \bibinfo {address} {Cham},\ \bibinfo {year} {2024})\ pp.\
	\bibinfo {pages} {133--223}\BibitemShut {NoStop}%
	\bibitem [{\citenamefont {Leijnse}\ and\ \citenamefont
		{Flensberg}(2012)}]{Leijnse2012Parityqubits}%
	\BibitemOpen
	\bibfield  {author} {\bibinfo {author} {\bibfnamefont {M.}~\bibnamefont
			{Leijnse}}\ and\ \bibinfo {author} {\bibfnamefont {K.}~\bibnamefont
			{Flensberg}},\ }\bibfield  {title} {\bibinfo {title} {Parity qubits and poor
			man's {M}ajorana bound states in double quantum dots},\ }\href
	{https://doi.org/10.1103/PhysRevB.86.134528} {\bibfield  {journal} {\bibinfo
			{journal} {Phys. Rev. B}\ }\textbf {\bibinfo {volume} {86}},\ \bibinfo
		{pages} {134528} (\bibinfo {year} {2012})}\BibitemShut {NoStop}%
	\bibitem [{\citenamefont {Sau}\ and\ \citenamefont
		{Sarma}(2012)}]{Sau2012RealizingrobustMajorana}%
	\BibitemOpen
	\bibfield  {author} {\bibinfo {author} {\bibfnamefont {J.~D.}\ \bibnamefont
			{Sau}}\ and\ \bibinfo {author} {\bibfnamefont {S.~D.}\ \bibnamefont
			{Sarma}},\ }\bibfield  {title} {\bibinfo {title} {Realizing a robust
			practical {M}ajorana chain in a quantum-dot-superconductor linear array},\
	}\href {https://doi.org/10.1038/ncomms1966} {\bibfield  {journal} {\bibinfo
			{journal} {Nature Communications}\ }\textbf {\bibinfo {volume} {3}},\
		\bibinfo {pages} {964} (\bibinfo {year} {2012})}\BibitemShut {NoStop}%
	\bibitem [{\citenamefont {Fulga}\ \emph {et~al.}(2013)\citenamefont {Fulga},
		\citenamefont {Haim}, \citenamefont {Akhmerov},\ and\ \citenamefont
		{Oreg}}]{Fulga2013AdaptiveMajorana}%
	\BibitemOpen
	\bibfield  {author} {\bibinfo {author} {\bibfnamefont {I.~C.}\ \bibnamefont
			{Fulga}}, \bibinfo {author} {\bibfnamefont {A.}~\bibnamefont {Haim}},
		\bibinfo {author} {\bibfnamefont {A.~R.}\ \bibnamefont {Akhmerov}},\ and\
		\bibinfo {author} {\bibfnamefont {Y.}~\bibnamefont {Oreg}},\ }\bibfield
	{title} {\bibinfo {title} {Adaptive tuning of {M}ajorana fermions in a
			quantum dot chain},\ }\href {https://doi.org/10.1088/1367-2630/15/4/045020}
	{\bibfield  {journal} {\bibinfo  {journal} {New Journal of Physics}\ }\textbf
		{\bibinfo {volume} {15}},\ \bibinfo {pages} {045020} (\bibinfo {year}
		{2013})}\BibitemShut {NoStop}%
	\bibitem [{\citenamefont {Tsintzis}\ \emph {et~al.}(2022)\citenamefont
		{Tsintzis}, \citenamefont {Souto},\ and\ \citenamefont
		{Leijnse}}]{Tsintzis2022}%
	\BibitemOpen
	\bibfield  {author} {\bibinfo {author} {\bibfnamefont {A.}~\bibnamefont
			{Tsintzis}}, \bibinfo {author} {\bibfnamefont {R.~S.}\ \bibnamefont
			{Souto}},\ and\ \bibinfo {author} {\bibfnamefont {M.}~\bibnamefont
			{Leijnse}},\ }\bibfield  {title} {\bibinfo {title} {Creating and detecting
			poor man's {M}ajorana bound states in interacting quantum dots},\ }\href
	{https://doi.org/10.1103/PhysRevB.106.L201404} {\bibfield  {journal}
		{\bibinfo  {journal} {Phys. Rev. B}\ }\textbf {\bibinfo {volume} {106}},\
		\bibinfo {pages} {L201404} (\bibinfo {year} {2022})}\BibitemShut {NoStop}%
	\bibitem [{\citenamefont {Liu}\ \emph {et~al.}(2022)\citenamefont {Liu},
		\citenamefont {Wang}, \citenamefont {Dvir},\ and\ \citenamefont
		{Wimmer}}]{LiuQDABS}%
	\BibitemOpen
	\bibfield  {author} {\bibinfo {author} {\bibfnamefont {C.-X.}\ \bibnamefont
			{Liu}}, \bibinfo {author} {\bibfnamefont {G.}~\bibnamefont {Wang}}, \bibinfo
		{author} {\bibfnamefont {T.}~\bibnamefont {Dvir}},\ and\ \bibinfo {author}
		{\bibfnamefont {M.}~\bibnamefont {Wimmer}},\ }\bibfield  {title} {\bibinfo
		{title} {Tunable superconducting coupling of quantum dots via {A}ndreev bound
			states in semiconductor-superconductor nanowires},\ }\href
	{https://doi.org/10.1103/PhysRevLett.129.267701} {\bibfield  {journal}
		{\bibinfo  {journal} {Phys. Rev. Lett.}\ }\textbf {\bibinfo {volume} {129}},\
		\bibinfo {pages} {267701} (\bibinfo {year} {2022})}\BibitemShut {NoStop}%
	\bibitem [{\citenamefont {Souto}\ \emph {et~al.}(2023)\citenamefont {Souto},
		\citenamefont {Tsintzis}, \citenamefont {Leijnse},\ and\ \citenamefont
		{Danon}}]{souto2023probing}%
	\BibitemOpen
	\bibfield  {author} {\bibinfo {author} {\bibfnamefont {R.~S.}\ \bibnamefont
			{Souto}}, \bibinfo {author} {\bibfnamefont {A.}~\bibnamefont {Tsintzis}},
		\bibinfo {author} {\bibfnamefont {M.}~\bibnamefont {Leijnse}},\ and\ \bibinfo
		{author} {\bibfnamefont {J.}~\bibnamefont {Danon}},\ }\bibfield  {title}
	{\bibinfo {title} {Probing {M}ajorana localization in minimal {K}itaev chains
			through a quantum dot},\ }\href
	{https://doi.org/10.1103/PhysRevResearch.5.043182} {\bibfield  {journal}
		{\bibinfo  {journal} {Phys. Rev. Res.}\ }\textbf {\bibinfo {volume} {5}},\
		\bibinfo {pages} {043182} (\bibinfo {year} {2023})}\BibitemShut {NoStop}%
	\bibitem [{\citenamefont {Samuelson}\ \emph {et~al.}(2024)\citenamefont
		{Samuelson}, \citenamefont {Svensson},\ and\ \citenamefont
		{Leijnse}}]{samuelson2024minimal}%
	\BibitemOpen
	\bibfield  {author} {\bibinfo {author} {\bibfnamefont {W.}~\bibnamefont
			{Samuelson}}, \bibinfo {author} {\bibfnamefont {V.}~\bibnamefont
			{Svensson}},\ and\ \bibinfo {author} {\bibfnamefont {M.}~\bibnamefont
			{Leijnse}},\ }\bibfield  {title} {\bibinfo {title} {Minimal quantum dot based
			{K}itaev chain with only local superconducting proximity effect},\ }\href
	{https://doi.org/10.1103/PhysRevB.109.035415} {\bibfield  {journal} {\bibinfo
			{journal} {Phys. Rev. B}\ }\textbf {\bibinfo {volume} {109}},\ \bibinfo
		{pages} {035415} (\bibinfo {year} {2024})}\BibitemShut {NoStop}%
	\bibitem [{\citenamefont {Svensson}\ and\ \citenamefont
		{Leijnse}(2024)}]{Svensson2024QuantumdotKitaev}%
	\BibitemOpen
	\bibfield  {author} {\bibinfo {author} {\bibfnamefont {V.}~\bibnamefont
			{Svensson}}\ and\ \bibinfo {author} {\bibfnamefont {M.}~\bibnamefont
			{Leijnse}},\ }\bibfield  {title} {\bibinfo {title} {Quantum dot based
			{K}itaev chains: {M}ajorana quality measures and scaling with increasing
			chain length},\ }\href {https://doi.org/10.1103/PhysRevB.110.155436}
	{\bibfield  {journal} {\bibinfo  {journal} {Phys. Rev. B}\ }\textbf {\bibinfo
			{volume} {110}},\ \bibinfo {pages} {155436} (\bibinfo {year}
		{2024})}\BibitemShut {NoStop}%
	\bibitem [{\citenamefont {Miles}\ \emph {et~al.}(2024)\citenamefont {Miles},
		\citenamefont {van Driel}, \citenamefont {Wimmer},\ and\ \citenamefont
		{Liu}}]{Miles2024dotAbs}%
	\BibitemOpen
	\bibfield  {author} {\bibinfo {author} {\bibfnamefont {S.}~\bibnamefont
			{Miles}}, \bibinfo {author} {\bibfnamefont {D.}~\bibnamefont {van Driel}},
		\bibinfo {author} {\bibfnamefont {M.}~\bibnamefont {Wimmer}},\ and\ \bibinfo
		{author} {\bibfnamefont {C.-X.}\ \bibnamefont {Liu}},\ }\bibfield  {title}
	{\bibinfo {title} {Kitaev chain in an alternating quantum dot-{A}ndreev bound
			state array},\ }\href {https://doi.org/10.1103/PhysRevB.110.024520}
	{\bibfield  {journal} {\bibinfo  {journal} {Phys. Rev. B}\ }\textbf {\bibinfo
			{volume} {110}},\ \bibinfo {pages} {024520} (\bibinfo {year}
		{2024})}\BibitemShut {NoStop}%
	\bibitem [{\citenamefont {Liu}\ \emph {et~al.}(2024{\natexlab{a}})\citenamefont
		{Liu}, \citenamefont {Bozkurt}, \citenamefont {Zatelli}, \citenamefont {ten
			Haaf}, \citenamefont {Dvir},\ and\ \citenamefont
		{Wimmer}}]{liu2024enhancing}%
	\BibitemOpen
	\bibfield  {author} {\bibinfo {author} {\bibfnamefont {C.-X.}\ \bibnamefont
			{Liu}}, \bibinfo {author} {\bibfnamefont {A.~M.}\ \bibnamefont {Bozkurt}},
		\bibinfo {author} {\bibfnamefont {F.}~\bibnamefont {Zatelli}}, \bibinfo
		{author} {\bibfnamefont {S.~L.}\ \bibnamefont {ten Haaf}}, \bibinfo {author}
		{\bibfnamefont {T.}~\bibnamefont {Dvir}},\ and\ \bibinfo {author}
		{\bibfnamefont {M.}~\bibnamefont {Wimmer}},\ }\bibfield  {title} {\bibinfo
		{title} {Enhancing the excitation gap of a quantum-dot-based {K}itaev
			chain},\ }\href {https://doi.org/https://doi.org/10.1038/s42005-024-01715-5}
	{\bibfield  {journal} {\bibinfo  {journal} {Communications Physics}\ }\textbf
		{\bibinfo {volume} {7}},\ \bibinfo {pages} {235} (\bibinfo {year}
		{2024}{\natexlab{a}})}\BibitemShut {NoStop}%
	\bibitem [{\citenamefont {Pino}\ \emph {et~al.}(2024)\citenamefont {Pino},
		\citenamefont {Souto},\ and\ \citenamefont {Aguado}}]{pino2024minimal}%
	\BibitemOpen
	\bibfield  {author} {\bibinfo {author} {\bibfnamefont {D.~M.}\ \bibnamefont
			{Pino}}, \bibinfo {author} {\bibfnamefont {R.~S.}\ \bibnamefont {Souto}},\
		and\ \bibinfo {author} {\bibfnamefont {R.}~\bibnamefont {Aguado}},\
	}\bibfield  {title} {\bibinfo {title} {Minimal {K}itaev-transmon qubit based
			on double quantum dots},\ }\href
	{https://doi.org/10.1103/PhysRevB.109.075101} {\bibfield  {journal} {\bibinfo
			{journal} {Phys. Rev. B}\ }\textbf {\bibinfo {volume} {109}},\ \bibinfo
		{pages} {075101} (\bibinfo {year} {2024})}\BibitemShut {NoStop}%
	\bibitem [{\citenamefont {Luna}\ \emph {et~al.}(2024)\citenamefont {Luna},
		\citenamefont {Bozkurt}, \citenamefont {Wimmer},\ and\ \citenamefont
		{Liu}}]{luna2024flux}%
	\BibitemOpen
	\bibfield  {author} {\bibinfo {author} {\bibfnamefont {J.~D.~T.}\
			\bibnamefont {Luna}}, \bibinfo {author} {\bibfnamefont {A.~M.}\ \bibnamefont
			{Bozkurt}}, \bibinfo {author} {\bibfnamefont {M.}~\bibnamefont {Wimmer}},\
		and\ \bibinfo {author} {\bibfnamefont {C.-X.}\ \bibnamefont {Liu}},\
	}\bibfield  {title} {\bibinfo {title} {{Flux-tunable {K}itaev chain in a
				quantum dot array}},\ }\href
	{https://doi.org/10.21468/SciPostPhysCore.7.3.065} {\bibfield  {journal}
		{\bibinfo  {journal} {SciPost Phys. Core}\ }\textbf {\bibinfo {volume} {7}},\
		\bibinfo {pages} {065} (\bibinfo {year} {2024})}\BibitemShut {NoStop}%
	\bibitem [{\citenamefont {Liu}\ \emph {et~al.}(2024{\natexlab{b}})\citenamefont
		{Liu}, \citenamefont {Zeng},\ and\ \citenamefont {Xu}}]{liu2024coupling}%
	\BibitemOpen
	\bibfield  {author} {\bibinfo {author} {\bibfnamefont {Z.-H.}\ \bibnamefont
			{Liu}}, \bibinfo {author} {\bibfnamefont {C.}~\bibnamefont {Zeng}},\ and\
		\bibinfo {author} {\bibfnamefont {H.~Q.}\ \bibnamefont {Xu}},\ }\bibfield
	{title} {\bibinfo {title} {Coupling of quantum-dot states via elastic
			cotunneling and crossed {A}ndreev reflection in a minimal {K}itaev chain},\
	}\href {https://doi.org/10.1103/PhysRevB.110.115302} {\bibfield  {journal}
		{\bibinfo  {journal} {Phys. Rev. B}\ }\textbf {\bibinfo {volume} {110}},\
		\bibinfo {pages} {115302} (\bibinfo {year} {2024}{\natexlab{b}})}\BibitemShut
	{NoStop}%
	\bibitem [{\citenamefont {Luethi}\ \emph {et~al.}(2024)\citenamefont {Luethi},
		\citenamefont {Legg}, \citenamefont {Loss},\ and\ \citenamefont
		{Klinovaja}}]{Luethi2024from}%
	\BibitemOpen
	\bibfield  {author} {\bibinfo {author} {\bibfnamefont {M.}~\bibnamefont
			{Luethi}}, \bibinfo {author} {\bibfnamefont {H.~F.}\ \bibnamefont {Legg}},
		\bibinfo {author} {\bibfnamefont {D.}~\bibnamefont {Loss}},\ and\ \bibinfo
		{author} {\bibfnamefont {J.}~\bibnamefont {Klinovaja}},\ }\bibfield  {title}
	{\bibinfo {title} {From perfect to imperfect poor man's {M}ajoranas in
			minimal {K}itaev chains},\ }\href
	{https://doi.org/10.1103/PhysRevB.110.245412} {\bibfield  {journal} {\bibinfo
			{journal} {Phys. Rev. B}\ }\textbf {\bibinfo {volume} {110}},\ \bibinfo
		{pages} {245412} (\bibinfo {year} {2024})}\BibitemShut {NoStop}%
	\bibitem [{\citenamefont {Luethi}\ \emph
		{et~al.}(2025{\natexlab{a}})\citenamefont {Luethi}, \citenamefont {Legg},
		\citenamefont {Loss},\ and\ \citenamefont {Klinovaja}}]{Luethi2025fate}%
	\BibitemOpen
	\bibfield  {author} {\bibinfo {author} {\bibfnamefont {M.}~\bibnamefont
			{Luethi}}, \bibinfo {author} {\bibfnamefont {H.~F.}\ \bibnamefont {Legg}},
		\bibinfo {author} {\bibfnamefont {D.}~\bibnamefont {Loss}},\ and\ \bibinfo
		{author} {\bibfnamefont {J.}~\bibnamefont {Klinovaja}},\ }\bibfield  {title}
	{\bibinfo {title} {Fate of poor man's {M}ajoranas in the long {K}itaev chain
			limit},\ }\href {https://doi.org/10.1103/PhysRevB.111.115419} {\bibfield
		{journal} {\bibinfo  {journal} {Phys. Rev. B}\ }\textbf {\bibinfo {volume}
			{111}},\ \bibinfo {pages} {115419} (\bibinfo {year}
		{2025}{\natexlab{a}})}\BibitemShut {NoStop}%
	\bibitem [{\citenamefont {Luethi}\ \emph
		{et~al.}(2025{\natexlab{b}})\citenamefont {Luethi}, \citenamefont {Legg},
		\citenamefont {Loss},\ and\ \citenamefont
		{Klinovaja}}]{Luethi2025properties}%
	\BibitemOpen
	\bibfield  {author} {\bibinfo {author} {\bibfnamefont {M.}~\bibnamefont
			{Luethi}}, \bibinfo {author} {\bibfnamefont {H.~F.}\ \bibnamefont {Legg}},
		\bibinfo {author} {\bibfnamefont {D.}~\bibnamefont {Loss}},\ and\ \bibinfo
		{author} {\bibfnamefont {J.}~\bibnamefont {Klinovaja}},\ }\bibfield  {title}
	{\bibinfo {title} {Properties and prevalence of false poor man's {M}ajoranas
			in two- and three-site artificial {K}itaev chains},\ }\href
	{https://doi.org/10.1103/bg4z-vx9m} {\bibfield  {journal} {\bibinfo
			{journal} {Phys. Rev. B}\ }\textbf {\bibinfo {volume} {112}},\ \bibinfo
		{pages} {205426} (\bibinfo {year} {2025}{\natexlab{b}})}\BibitemShut
	{NoStop}%
	\bibitem [{\citenamefont {Sanches}\ \emph {et~al.}(2025)\citenamefont
		{Sanches}, \citenamefont {Lustosa}, \citenamefont {Ricco}, \citenamefont
		{Sigurdsson}, \citenamefont {de~Souza}, \citenamefont {Figueira},
		\citenamefont {Marinho~Jr},\ and\ \citenamefont
		{Seridonio}}]{sanches2025spin}%
	\BibitemOpen
	\bibfield  {author} {\bibinfo {author} {\bibfnamefont {J.}~\bibnamefont
			{Sanches}}, \bibinfo {author} {\bibfnamefont {L.}~\bibnamefont {Lustosa}},
		\bibinfo {author} {\bibfnamefont {L.}~\bibnamefont {Ricco}}, \bibinfo
		{author} {\bibfnamefont {H.}~\bibnamefont {Sigurdsson}}, \bibinfo {author}
		{\bibfnamefont {M.}~\bibnamefont {de~Souza}}, \bibinfo {author}
		{\bibfnamefont {M.}~\bibnamefont {Figueira}}, \bibinfo {author}
		{\bibfnamefont {E.}~\bibnamefont {Marinho~Jr}},\ and\ \bibinfo {author}
		{\bibfnamefont {A.}~\bibnamefont {Seridonio}},\ }\bibfield  {title} {\bibinfo
		{title} {Spin-exchange induced spillover on poor man’s {M}ajoranas in
			minimal {K}itaev chains},\ }\href {https://doi.org/10.1088/1361-648X/adce6a}
	{\bibfield  {journal} {\bibinfo  {journal} {Journal of Physics: Condensed
				Matter}\ }\textbf {\bibinfo {volume} {37}},\ \bibinfo {pages} {205601}
		(\bibinfo {year} {2025})}\BibitemShut {NoStop}%
	\bibitem [{\citenamefont {Sanches}\ \emph {et~al.}(2026)\citenamefont
		{Sanches}, \citenamefont {Sobreira}, \citenamefont {Ricco}, \citenamefont
		{Figueira},\ and\ \citenamefont {Seridonio}}]{sanches2026revisiting}%
	\BibitemOpen
	\bibfield  {author} {\bibinfo {author} {\bibfnamefont {J.}~\bibnamefont
			{Sanches}}, \bibinfo {author} {\bibfnamefont {T.}~\bibnamefont {Sobreira}},
		\bibinfo {author} {\bibfnamefont {L.}~\bibnamefont {Ricco}}, \bibinfo
		{author} {\bibfnamefont {M.}~\bibnamefont {Figueira}},\ and\ \bibinfo
		{author} {\bibfnamefont {A.}~\bibnamefont {Seridonio}},\ }\bibfield  {title}
	{\bibinfo {title} {Revisiting the poor man’s {M}ajoranas: the
			spin--exchange induced spillover effect},\ }\href
	{https://doi.org/10.1088/1361-648X/ae2f13} {\bibfield  {journal} {\bibinfo
			{journal} {Journal of Physics: Condensed Matter}\ }\textbf {\bibinfo {volume}
			{38}},\ \bibinfo {pages} {023001} (\bibinfo {year} {2026})}\BibitemShut
	{NoStop}%
	\bibitem [{\citenamefont {Zhang}\ \emph {et~al.}(2026)\citenamefont {Zhang},
		\citenamefont {Yue}, \citenamefont {Qiao},\ and\ \citenamefont
		{Sun}}]{zhang2026sensitivedependencepoormans}%
	\BibitemOpen
	\bibfield  {author} {\bibinfo {author} {\bibfnamefont {Z.-L.}\ \bibnamefont
			{Zhang}}, \bibinfo {author} {\bibfnamefont {X.}~\bibnamefont {Yue}}, \bibinfo
		{author} {\bibfnamefont {G.-J.}\ \bibnamefont {Qiao}},\ and\ \bibinfo
		{author} {\bibfnamefont {C.~P.}\ \bibnamefont {Sun}},\ }\href
	{https://arxiv.org/abs/2604.12950} {\bibinfo {title} {Sensitive dependence of
			poor man's {M}ajorana modes on the length of superconductor}} (\bibinfo
	{year} {2026}),\ \Eprint {https://arxiv.org/abs/2604.12950} {arXiv:2604.12950
		[cond-mat.mes-hall]} \BibitemShut {NoStop}%
	\bibitem [{\citenamefont {Alvarado}\ \emph {et~al.}(2026)\citenamefont
		{Alvarado}, \citenamefont {Souto}, \citenamefont {Calderón},\ and\
		\citenamefont {Aguado}}]{alvarado2026optimalmajoranasmesoscopickitaev}%
	\BibitemOpen
	\bibfield  {author} {\bibinfo {author} {\bibfnamefont {M.}~\bibnamefont
			{Alvarado}}, \bibinfo {author} {\bibfnamefont {R.~S.}\ \bibnamefont {Souto}},
		\bibinfo {author} {\bibfnamefont {M.~J.}\ \bibnamefont {Calderón}},\ and\
		\bibinfo {author} {\bibfnamefont {R.}~\bibnamefont {Aguado}},\ }\href
	{https://arxiv.org/abs/2604.13945} {\bibinfo {title} {Optimal {M}ajoranas in
			mesoscopic {K}itaev chains}} (\bibinfo {year} {2026}),\ \Eprint
	{https://arxiv.org/abs/2604.13945} {arXiv:2604.13945 [cond-mat.mes-hall]}
	\BibitemShut {NoStop}%
	\bibitem [{\citenamefont {Dvir}\ \emph {et~al.}(2023)\citenamefont {Dvir},
		\citenamefont {Wang}, \citenamefont {van Loo}, \citenamefont {Liu},
		\citenamefont {Mazur}, \citenamefont {Bordin}, \citenamefont {ten Haaf},
		\citenamefont {Wang}, \citenamefont {van Driel}, \citenamefont {Zatelli},
		\citenamefont {Li}, \citenamefont {Malinowski}, \citenamefont {Gazibegovic},
		\citenamefont {Badawy}, \citenamefont {Bakkers}, \citenamefont {Wimmer},\
		and\ \citenamefont {Kouwenhoven}}]{Dvir2023exp2site}%
	\BibitemOpen
	\bibfield  {author} {\bibinfo {author} {\bibfnamefont {T.}~\bibnamefont
			{Dvir}}, \bibinfo {author} {\bibfnamefont {G.}~\bibnamefont {Wang}}, \bibinfo
		{author} {\bibfnamefont {N.}~\bibnamefont {van Loo}}, \bibinfo {author}
		{\bibfnamefont {C.-X.}\ \bibnamefont {Liu}}, \bibinfo {author} {\bibfnamefont
			{G.~P.}\ \bibnamefont {Mazur}}, \bibinfo {author} {\bibfnamefont
			{A.}~\bibnamefont {Bordin}}, \bibinfo {author} {\bibfnamefont {S.~L.~D.}\
			\bibnamefont {ten Haaf}}, \bibinfo {author} {\bibfnamefont {J.-Y.}\
			\bibnamefont {Wang}}, \bibinfo {author} {\bibfnamefont {D.}~\bibnamefont {van
				Driel}}, \bibinfo {author} {\bibfnamefont {F.}~\bibnamefont {Zatelli}},
		\bibinfo {author} {\bibfnamefont {X.}~\bibnamefont {Li}}, \bibinfo {author}
		{\bibfnamefont {F.~K.}\ \bibnamefont {Malinowski}}, \bibinfo {author}
		{\bibfnamefont {S.}~\bibnamefont {Gazibegovic}}, \bibinfo {author}
		{\bibfnamefont {G.}~\bibnamefont {Badawy}}, \bibinfo {author} {\bibfnamefont
			{E.~P. A.~M.}\ \bibnamefont {Bakkers}}, \bibinfo {author} {\bibfnamefont
			{M.}~\bibnamefont {Wimmer}},\ and\ \bibinfo {author} {\bibfnamefont {L.~P.}\
			\bibnamefont {Kouwenhoven}},\ }\bibfield  {title} {\bibinfo {title}
		{Realization of a minimal {K}itaev chain in coupled quantum dots},\ }\href
	{https://doi.org/10.1038/s41586-022-05585-1} {\bibfield  {journal} {\bibinfo
			{journal} {Nature}\ }\textbf {\bibinfo {volume} {614}},\ \bibinfo {pages}
		{445} (\bibinfo {year} {2023})}\BibitemShut {NoStop}%
	\bibitem [{\citenamefont {ten Haaf}\ \emph {et~al.}(2024)\citenamefont {ten
			Haaf}, \citenamefont {Wang}, \citenamefont {Bozkurt}, \citenamefont {Liu},
		\citenamefont {Kulesh}, \citenamefont {Kim}, \citenamefont {Xiao},
		\citenamefont {Thomas}, \citenamefont {Manfra}, \citenamefont {Dvir},
		\citenamefont {Wimmer},\ and\ \citenamefont {Goswami}}]{tenHaaf2024exp2site}%
	\BibitemOpen
	\bibfield  {author} {\bibinfo {author} {\bibfnamefont {S.~L.~D.}\
			\bibnamefont {ten Haaf}}, \bibinfo {author} {\bibfnamefont {Q.}~\bibnamefont
			{Wang}}, \bibinfo {author} {\bibfnamefont {A.~M.}\ \bibnamefont {Bozkurt}},
		\bibinfo {author} {\bibfnamefont {C.-X.}\ \bibnamefont {Liu}}, \bibinfo
		{author} {\bibfnamefont {I.}~\bibnamefont {Kulesh}}, \bibinfo {author}
		{\bibfnamefont {P.}~\bibnamefont {Kim}}, \bibinfo {author} {\bibfnamefont
			{D.}~\bibnamefont {Xiao}}, \bibinfo {author} {\bibfnamefont {C.}~\bibnamefont
			{Thomas}}, \bibinfo {author} {\bibfnamefont {M.~J.}\ \bibnamefont {Manfra}},
		\bibinfo {author} {\bibfnamefont {T.}~\bibnamefont {Dvir}}, \bibinfo {author}
		{\bibfnamefont {M.}~\bibnamefont {Wimmer}},\ and\ \bibinfo {author}
		{\bibfnamefont {S.}~\bibnamefont {Goswami}},\ }\bibfield  {title} {\bibinfo
		{title} {A two-site {K}itaev chain in a two-dimensional electron gas},\
	}\href {https://doi.org/10.1038/s41586-024-07434-9} {\bibfield  {journal}
		{\bibinfo  {journal} {Nature}\ }\textbf {\bibinfo {volume} {630}},\ \bibinfo
		{pages} {329} (\bibinfo {year} {2024})}\BibitemShut {NoStop}%
	\bibitem [{\citenamefont {Zatelli}\ \emph {et~al.}(2024)\citenamefont
		{Zatelli}, \citenamefont {van Driel}, \citenamefont {Xu}, \citenamefont
		{Wang}, \citenamefont {Liu}, \citenamefont {Bordin}, \citenamefont {Roovers},
		\citenamefont {Mazur}, \citenamefont {van Loo}, \citenamefont {Wolff},
		\citenamefont {Bozkurt}, \citenamefont {Badawy}, \citenamefont {Gazibegovic},
		\citenamefont {Bakkers}, \citenamefont {Wimmer}, \citenamefont
		{Kouwenhoven},\ and\ \citenamefont {Dvir}}]{Zatelli2024exp2site}%
	\BibitemOpen
	\bibfield  {author} {\bibinfo {author} {\bibfnamefont {F.}~\bibnamefont
			{Zatelli}}, \bibinfo {author} {\bibfnamefont {D.}~\bibnamefont {van Driel}},
		\bibinfo {author} {\bibfnamefont {D.}~\bibnamefont {Xu}}, \bibinfo {author}
		{\bibfnamefont {G.}~\bibnamefont {Wang}}, \bibinfo {author} {\bibfnamefont
			{C.-X.}\ \bibnamefont {Liu}}, \bibinfo {author} {\bibfnamefont
			{A.}~\bibnamefont {Bordin}}, \bibinfo {author} {\bibfnamefont
			{B.}~\bibnamefont {Roovers}}, \bibinfo {author} {\bibfnamefont {G.~P.}\
			\bibnamefont {Mazur}}, \bibinfo {author} {\bibfnamefont {N.}~\bibnamefont
			{van Loo}}, \bibinfo {author} {\bibfnamefont {J.~C.}\ \bibnamefont {Wolff}},
		\bibinfo {author} {\bibfnamefont {A.~M.}\ \bibnamefont {Bozkurt}}, \bibinfo
		{author} {\bibfnamefont {G.}~\bibnamefont {Badawy}}, \bibinfo {author}
		{\bibfnamefont {S.}~\bibnamefont {Gazibegovic}}, \bibinfo {author}
		{\bibfnamefont {E.~P. A.~M.}\ \bibnamefont {Bakkers}}, \bibinfo {author}
		{\bibfnamefont {M.}~\bibnamefont {Wimmer}}, \bibinfo {author} {\bibfnamefont
			{L.~P.}\ \bibnamefont {Kouwenhoven}},\ and\ \bibinfo {author} {\bibfnamefont
			{T.}~\bibnamefont {Dvir}},\ }\bibfield  {title} {\bibinfo {title} {Robust
			poor man's {M}ajorana zero modes using {Y}u-{S}hiba-{R}usinov states},\
	}\href {https://doi.org/10.1038/s41467-024-52066-2} {\bibfield  {journal}
		{\bibinfo  {journal} {Nature Communications}\ }\textbf {\bibinfo {volume}
			{15}},\ \bibinfo {pages} {7933} (\bibinfo {year} {2024})}\BibitemShut
	{NoStop}%
	\bibitem [{\citenamefont {van Loo}\ \emph {et~al.}(2026)\citenamefont {van
			Loo}, \citenamefont {Zatelli}, \citenamefont {Steffensen}, \citenamefont
		{Roovers}, \citenamefont {Wang}, \citenamefont {Van~Caekenberghe},
		\citenamefont {Bordin}, \citenamefont {van Driel}, \citenamefont {Zhang},
		\citenamefont {Huisman}, \citenamefont {Badawy}, \citenamefont {Bakkers},
		\citenamefont {Mazur}, \citenamefont {Aguado},\ and\ \citenamefont
		{Kouwenhoven}}]{vanLoo2026single}%
	\BibitemOpen
	\bibfield  {author} {\bibinfo {author} {\bibfnamefont {N.}~\bibnamefont {van
				Loo}}, \bibinfo {author} {\bibfnamefont {F.}~\bibnamefont {Zatelli}},
		\bibinfo {author} {\bibfnamefont {G.~O.}\ \bibnamefont {Steffensen}},
		\bibinfo {author} {\bibfnamefont {B.}~\bibnamefont {Roovers}}, \bibinfo
		{author} {\bibfnamefont {G.}~\bibnamefont {Wang}}, \bibinfo {author}
		{\bibfnamefont {T.}~\bibnamefont {Van~Caekenberghe}}, \bibinfo {author}
		{\bibfnamefont {A.}~\bibnamefont {Bordin}}, \bibinfo {author} {\bibfnamefont
			{D.}~\bibnamefont {van Driel}}, \bibinfo {author} {\bibfnamefont
			{Y.}~\bibnamefont {Zhang}}, \bibinfo {author} {\bibfnamefont {W.~D.}\
			\bibnamefont {Huisman}}, \bibinfo {author} {\bibfnamefont {G.}~\bibnamefont
			{Badawy}}, \bibinfo {author} {\bibfnamefont {E.~P. A.~M.}\ \bibnamefont
			{Bakkers}}, \bibinfo {author} {\bibfnamefont {G.~P.}\ \bibnamefont {Mazur}},
		\bibinfo {author} {\bibfnamefont {R.}~\bibnamefont {Aguado}},\ and\ \bibinfo
		{author} {\bibfnamefont {L.~P.}\ \bibnamefont {Kouwenhoven}},\ }\bibfield
	{title} {\bibinfo {title} {Single-shot parity readout of a minimal {K}itaev
			chain},\ }\href {https://doi.org/10.1038/s41586-025-09927-7} {\bibfield
		{journal} {\bibinfo  {journal} {Nature}\ }\textbf {\bibinfo {volume} {650}},\
		\bibinfo {pages} {334} (\bibinfo {year} {2026})}\BibitemShut {NoStop}%
	\bibitem [{\citenamefont {Bordin}\ \emph {et~al.}(2025)\citenamefont {Bordin},
		\citenamefont {Liu}, \citenamefont {Dvir}, \citenamefont {Zatelli},
		\citenamefont {ten Haaf}, \citenamefont {van Driel}, \citenamefont {Wang},
		\citenamefont {van Loo}, \citenamefont {Zhang}, \citenamefont {Wolff},
		\citenamefont {Van~Caekenberghe}, \citenamefont {Badawy}, \citenamefont
		{Gazibegovic}, \citenamefont {Bakkers}, \citenamefont {Wimmer}, \citenamefont
		{Kouwenhoven},\ and\ \citenamefont {Mazur}}]{Bordin2025exp3site}%
	\BibitemOpen
	\bibfield  {author} {\bibinfo {author} {\bibfnamefont {A.}~\bibnamefont
			{Bordin}}, \bibinfo {author} {\bibfnamefont {C.-X.}\ \bibnamefont {Liu}},
		\bibinfo {author} {\bibfnamefont {T.}~\bibnamefont {Dvir}}, \bibinfo {author}
		{\bibfnamefont {F.}~\bibnamefont {Zatelli}}, \bibinfo {author} {\bibfnamefont
			{S.~L.~D.}\ \bibnamefont {ten Haaf}}, \bibinfo {author} {\bibfnamefont
			{D.}~\bibnamefont {van Driel}}, \bibinfo {author} {\bibfnamefont
			{G.}~\bibnamefont {Wang}}, \bibinfo {author} {\bibfnamefont {N.}~\bibnamefont
			{van Loo}}, \bibinfo {author} {\bibfnamefont {Y.}~\bibnamefont {Zhang}},
		\bibinfo {author} {\bibfnamefont {J.~C.}\ \bibnamefont {Wolff}}, \bibinfo
		{author} {\bibfnamefont {T.}~\bibnamefont {Van~Caekenberghe}}, \bibinfo
		{author} {\bibfnamefont {G.}~\bibnamefont {Badawy}}, \bibinfo {author}
		{\bibfnamefont {S.}~\bibnamefont {Gazibegovic}}, \bibinfo {author}
		{\bibfnamefont {E.~P. A.~M.}\ \bibnamefont {Bakkers}}, \bibinfo {author}
		{\bibfnamefont {M.}~\bibnamefont {Wimmer}}, \bibinfo {author} {\bibfnamefont
			{L.~P.}\ \bibnamefont {Kouwenhoven}},\ and\ \bibinfo {author} {\bibfnamefont
			{G.~P.}\ \bibnamefont {Mazur}},\ }\bibfield  {title} {\bibinfo {title}
		{Enhanced {M}ajorana stability in a three-site {K}itaev chain},\ }\bibfield
	{journal} {\bibinfo  {journal} {Nature Nanotechnology}\ }\href
	{https://doi.org/10.1038/s41565-025-01894-4} {10.1038/s41565-025-01894-4}
	(\bibinfo {year} {2025})\BibitemShut {NoStop}%
	\bibitem [{\citenamefont {Ten~Haaf}\ \emph {et~al.}(2025)\citenamefont
		{Ten~Haaf}, \citenamefont {Zhang}, \citenamefont {Wang}, \citenamefont
		{Bordin}, \citenamefont {Liu}, \citenamefont {Kulesh}, \citenamefont
		{Sietses}, \citenamefont {Prosko}, \citenamefont {Xiao}, \citenamefont
		{Thomas} \emph {et~al.}}]{ten2025exp3site}%
	\BibitemOpen
	\bibfield  {author} {\bibinfo {author} {\bibfnamefont {S.~L.}\ \bibnamefont
			{Ten~Haaf}}, \bibinfo {author} {\bibfnamefont {Y.}~\bibnamefont {Zhang}},
		\bibinfo {author} {\bibfnamefont {Q.}~\bibnamefont {Wang}}, \bibinfo {author}
		{\bibfnamefont {A.}~\bibnamefont {Bordin}}, \bibinfo {author} {\bibfnamefont
			{C.-X.}\ \bibnamefont {Liu}}, \bibinfo {author} {\bibfnamefont
			{I.}~\bibnamefont {Kulesh}}, \bibinfo {author} {\bibfnamefont {V.~P.}\
			\bibnamefont {Sietses}}, \bibinfo {author} {\bibfnamefont {C.~G.}\
			\bibnamefont {Prosko}}, \bibinfo {author} {\bibfnamefont {D.}~\bibnamefont
			{Xiao}}, \bibinfo {author} {\bibfnamefont {C.}~\bibnamefont {Thomas}}, \emph
		{et~al.},\ }\bibfield  {title} {\bibinfo {title} {Observation of edge and
			bulk states in a three-site {K}itaev chain},\ }\href
	{https://doi.org/10.1038/s41586-025-08892-5} {\bibfield  {journal} {\bibinfo
			{journal} {Nature}\ ,\ \bibinfo {pages} {1}} (\bibinfo {year}
		{2025})}\BibitemShut {NoStop}%
	\bibitem [{\citenamefont {Garcia-Vidal}\ \emph {et~al.}(2021)\citenamefont
		{Garcia-Vidal}, \citenamefont {Ciuti},\ and\ \citenamefont
		{Ebbesen}}]{garciavidal2021manipulating}%
	\BibitemOpen
	\bibfield  {author} {\bibinfo {author} {\bibfnamefont {F.~J.}\ \bibnamefont
			{Garcia-Vidal}}, \bibinfo {author} {\bibfnamefont {C.}~\bibnamefont
			{Ciuti}},\ and\ \bibinfo {author} {\bibfnamefont {T.~W.}\ \bibnamefont
			{Ebbesen}},\ }\bibfield  {title} {\bibinfo {title} {Manipulating matter by
			strong coupling to vacuum fields},\ }\href
	{https://doi.org/10.1126/science.abd0336} {\bibfield  {journal} {\bibinfo
			{journal} {Science}\ }\textbf {\bibinfo {volume} {373}},\ \bibinfo {pages}
		{eabd0336} (\bibinfo {year} {2021})}\BibitemShut {NoStop}%
	\bibitem [{\citenamefont {Schlawin}\ \emph {et~al.}(2022)\citenamefont
		{Schlawin}, \citenamefont {Kennes},\ and\ \citenamefont
		{Sentef}}]{schlawin2022cavity}%
	\BibitemOpen
	\bibfield  {author} {\bibinfo {author} {\bibfnamefont {F.}~\bibnamefont
			{Schlawin}}, \bibinfo {author} {\bibfnamefont {D.~M.}\ \bibnamefont
			{Kennes}},\ and\ \bibinfo {author} {\bibfnamefont {M.~A.}\ \bibnamefont
			{Sentef}},\ }\bibfield  {title} {\bibinfo {title} {Cavity quantum
			materials},\ }\href {https://pubs.aip.org/aip/apr/article/9/1/011312/2835409}
	{\bibfield  {journal} {\bibinfo  {journal} {Applied Physics Reviews}\
		}\textbf {\bibinfo {volume} {9}} (\bibinfo {year} {2022})}\BibitemShut
	{NoStop}%
	\bibitem [{\citenamefont {Bretscher}\ \emph {et~al.}(2026)\citenamefont
		{Bretscher}, \citenamefont {Graziotto}, \citenamefont {Michael},
		\citenamefont {Montanaro}, \citenamefont {Lu}, \citenamefont {Grankin},
		\citenamefont {McIver}, \citenamefont {Faist}, \citenamefont {Fausti},
		\citenamefont {Eckstein}, \citenamefont {Ruggenthaler}, \citenamefont
		{Rubio}, \citenamefont {Basov}, \citenamefont {Hafezi}, \citenamefont
		{Claassen}, \citenamefont {Kennes},\ and\ \citenamefont
		{Sentef}}]{bretscher2026fluctuation}%
	\BibitemOpen
	\bibfield  {author} {\bibinfo {author} {\bibfnamefont {H.~M.}\ \bibnamefont
			{Bretscher}}, \bibinfo {author} {\bibfnamefont {L.}~\bibnamefont
			{Graziotto}}, \bibinfo {author} {\bibfnamefont {M.~H.}\ \bibnamefont
			{Michael}}, \bibinfo {author} {\bibfnamefont {A.}~\bibnamefont {Montanaro}},
		\bibinfo {author} {\bibfnamefont {I.-T.}\ \bibnamefont {Lu}}, \bibinfo
		{author} {\bibfnamefont {A.}~\bibnamefont {Grankin}}, \bibinfo {author}
		{\bibfnamefont {J.~W.}\ \bibnamefont {McIver}}, \bibinfo {author}
		{\bibfnamefont {J.}~\bibnamefont {Faist}}, \bibinfo {author} {\bibfnamefont
			{D.}~\bibnamefont {Fausti}}, \bibinfo {author} {\bibfnamefont
			{M.}~\bibnamefont {Eckstein}}, \bibinfo {author} {\bibfnamefont
			{M.}~\bibnamefont {Ruggenthaler}}, \bibinfo {author} {\bibfnamefont
			{A.}~\bibnamefont {Rubio}}, \bibinfo {author} {\bibfnamefont
			{D.}~\bibnamefont {Basov}}, \bibinfo {author} {\bibfnamefont
			{M.}~\bibnamefont {Hafezi}}, \bibinfo {author} {\bibfnamefont
			{M.}~\bibnamefont {Claassen}}, \bibinfo {author} {\bibfnamefont {D.~M.}\
			\bibnamefont {Kennes}},\ and\ \bibinfo {author} {\bibfnamefont {M.~A.}\
			\bibnamefont {Sentef}},\ }\href {https://arxiv.org/abs/2604.08666} {\bibinfo
		{title} {Fluctuation engineering in cavity quantum materials}} (\bibinfo
	{year} {2026}),\ \Eprint {https://arxiv.org/abs/2604.08666} {arXiv:2604.08666
		[cond-mat.mes-hall]} \BibitemShut {NoStop}%
	\bibitem [{\citenamefont {Appugliese}\ \emph {et~al.}(2022)\citenamefont
		{Appugliese}, \citenamefont {Enkner}, \citenamefont {Paravicini-Bagliani},
		\citenamefont {Beck}, \citenamefont {Reichl}, \citenamefont {Wegscheider},
		\citenamefont {Scalari}, \citenamefont {Ciuti},\ and\ \citenamefont
		{Faist}}]{appugliese2022breakdown}%
	\BibitemOpen
	\bibfield  {author} {\bibinfo {author} {\bibfnamefont {F.}~\bibnamefont
			{Appugliese}}, \bibinfo {author} {\bibfnamefont {J.}~\bibnamefont {Enkner}},
		\bibinfo {author} {\bibfnamefont {G.~L.}\ \bibnamefont
			{Paravicini-Bagliani}}, \bibinfo {author} {\bibfnamefont {M.}~\bibnamefont
			{Beck}}, \bibinfo {author} {\bibfnamefont {C.}~\bibnamefont {Reichl}},
		\bibinfo {author} {\bibfnamefont {W.}~\bibnamefont {Wegscheider}}, \bibinfo
		{author} {\bibfnamefont {G.}~\bibnamefont {Scalari}}, \bibinfo {author}
		{\bibfnamefont {C.}~\bibnamefont {Ciuti}},\ and\ \bibinfo {author}
		{\bibfnamefont {J.}~\bibnamefont {Faist}},\ }\bibfield  {title} {\bibinfo
		{title} {Breakdown of topological protection by cavity vacuum fields in the
			integer quantum {H}all effect},\ }\href
	{https://doi.org/10.1126/science.abl5818} {\bibfield  {journal} {\bibinfo
			{journal} {Science}\ }\textbf {\bibinfo {volume} {375}},\ \bibinfo {pages}
		{1030} (\bibinfo {year} {2022})}\BibitemShut {NoStop}%
	\bibitem [{\citenamefont {Enkner}\ \emph {et~al.}(2025)\citenamefont {Enkner},
		\citenamefont {Graziotto}, \citenamefont {Bori{\c c}i}, \citenamefont
		{Appugliese}, \citenamefont {Reichl}, \citenamefont {Scalari}, \citenamefont
		{Regnault}, \citenamefont {Wegscheider}, \citenamefont {Ciuti},\ and\
		\citenamefont {Faist}}]{enkner2025}%
	\BibitemOpen
	\bibfield  {author} {\bibinfo {author} {\bibfnamefont {J.}~\bibnamefont
			{Enkner}}, \bibinfo {author} {\bibfnamefont {L.}~\bibnamefont {Graziotto}},
		\bibinfo {author} {\bibfnamefont {D.}~\bibnamefont {Bori{\c c}i}}, \bibinfo
		{author} {\bibfnamefont {F.}~\bibnamefont {Appugliese}}, \bibinfo {author}
		{\bibfnamefont {C.}~\bibnamefont {Reichl}}, \bibinfo {author} {\bibfnamefont
			{G.}~\bibnamefont {Scalari}}, \bibinfo {author} {\bibfnamefont
			{N.}~\bibnamefont {Regnault}}, \bibinfo {author} {\bibfnamefont
			{W.}~\bibnamefont {Wegscheider}}, \bibinfo {author} {\bibfnamefont
			{C.}~\bibnamefont {Ciuti}},\ and\ \bibinfo {author} {\bibfnamefont
			{J.}~\bibnamefont {Faist}},\ }\bibfield  {title} {\bibinfo {title} {Tunable
			vacuum-field control of fractional and integer quantum {H}all phases},\
	}\href {https://doi.org/10.1038/s41586-025-08894-3} {\bibfield  {journal}
		{\bibinfo  {journal} {Nature}\ }\textbf {\bibinfo {volume} {641}},\ \bibinfo
		{pages} {884} (\bibinfo {year} {2025})}\BibitemShut {NoStop}%
	\bibitem [{\citenamefont {Jarc}\ \emph {et~al.}(2023)\citenamefont {Jarc},
		\citenamefont {Mathengattil}, \citenamefont {Montanaro}, \citenamefont
		{Giusti}, \citenamefont {Rigoni}, \citenamefont {Sergo}, \citenamefont
		{Fassioli}, \citenamefont {Winnerl}, \citenamefont {Dal~Zilio}, \citenamefont
		{Mihailovic}, \citenamefont {Prelov{\v s}ek}, \citenamefont {Eckstein},\ and\
		\citenamefont {Fausti}}]{jarcNature2023}%
	\BibitemOpen
	\bibfield  {author} {\bibinfo {author} {\bibfnamefont {G.}~\bibnamefont
			{Jarc}}, \bibinfo {author} {\bibfnamefont {S.~Y.}\ \bibnamefont
			{Mathengattil}}, \bibinfo {author} {\bibfnamefont {A.}~\bibnamefont
			{Montanaro}}, \bibinfo {author} {\bibfnamefont {F.}~\bibnamefont {Giusti}},
		\bibinfo {author} {\bibfnamefont {E.~M.}\ \bibnamefont {Rigoni}}, \bibinfo
		{author} {\bibfnamefont {R.}~\bibnamefont {Sergo}}, \bibinfo {author}
		{\bibfnamefont {F.}~\bibnamefont {Fassioli}}, \bibinfo {author}
		{\bibfnamefont {S.}~\bibnamefont {Winnerl}}, \bibinfo {author} {\bibfnamefont
			{S.}~\bibnamefont {Dal~Zilio}}, \bibinfo {author} {\bibfnamefont
			{D.}~\bibnamefont {Mihailovic}}, \bibinfo {author} {\bibfnamefont
			{P.}~\bibnamefont {Prelov{\v s}ek}}, \bibinfo {author} {\bibfnamefont
			{M.}~\bibnamefont {Eckstein}},\ and\ \bibinfo {author} {\bibfnamefont
			{D.}~\bibnamefont {Fausti}},\ }\bibfield  {title} {\bibinfo {title}
		{Cavity-mediated thermal control of metal-to-insulator transition in
			1{T}-{T}a{S}2},\ }\href {https://doi.org/10.1038/s41586-023-06596-2}
	{\bibfield  {journal} {\bibinfo  {journal} {Nature}\ }\textbf {\bibinfo
			{volume} {622}},\ \bibinfo {pages} {487} (\bibinfo {year}
		{2023})}\BibitemShut {NoStop}%
	\bibitem [{\citenamefont {Keren}\ \emph {et~al.}(2026)\citenamefont {Keren},
		\citenamefont {Webb}, \citenamefont {Zhang}, \citenamefont {Xu},
		\citenamefont {Sun}, \citenamefont {Kim}, \citenamefont {Shin}, \citenamefont
		{Zhang}, \citenamefont {Zhang}, \citenamefont {Pereira}, \citenamefont {Yao},
		\citenamefont {Okugawa}, \citenamefont {Michael}, \citenamefont
		{Vi{\~n}as~Bostr{\"o}m}, \citenamefont {Edgar}, \citenamefont {Wolf},
		\citenamefont {Julian}, \citenamefont {Prasankumar}, \citenamefont
		{Miyagawa}, \citenamefont {Kanoda}, \citenamefont {Gu}, \citenamefont
		{Cothrine}, \citenamefont {Mandrus}, \citenamefont {Buzzi}, \citenamefont
		{Cavalleri}, \citenamefont {Dean}, \citenamefont {Kennes}, \citenamefont
		{Millis}, \citenamefont {Li}, \citenamefont {Sentef}, \citenamefont {Rubio},
		\citenamefont {Pasupathy},\ and\ \citenamefont {Basov}}]{Keren2026cavity}%
	\BibitemOpen
	\bibfield  {author} {\bibinfo {author} {\bibfnamefont {I.}~\bibnamefont
			{Keren}}, \bibinfo {author} {\bibfnamefont {T.~A.}\ \bibnamefont {Webb}},
		\bibinfo {author} {\bibfnamefont {S.}~\bibnamefont {Zhang}}, \bibinfo
		{author} {\bibfnamefont {J.}~\bibnamefont {Xu}}, \bibinfo {author}
		{\bibfnamefont {D.}~\bibnamefont {Sun}}, \bibinfo {author} {\bibfnamefont
			{B.~S.~Y.}\ \bibnamefont {Kim}}, \bibinfo {author} {\bibfnamefont
			{D.}~\bibnamefont {Shin}}, \bibinfo {author} {\bibfnamefont {S.~S.}\
			\bibnamefont {Zhang}}, \bibinfo {author} {\bibfnamefont {J.}~\bibnamefont
			{Zhang}}, \bibinfo {author} {\bibfnamefont {G.}~\bibnamefont {Pereira}},
		\bibinfo {author} {\bibfnamefont {J.}~\bibnamefont {Yao}}, \bibinfo {author}
		{\bibfnamefont {T.}~\bibnamefont {Okugawa}}, \bibinfo {author} {\bibfnamefont
			{M.~H.}\ \bibnamefont {Michael}}, \bibinfo {author} {\bibfnamefont
			{E.}~\bibnamefont {Vi{\~n}as~Bostr{\"o}m}}, \bibinfo {author} {\bibfnamefont
			{J.~H.}\ \bibnamefont {Edgar}}, \bibinfo {author} {\bibfnamefont
			{S.}~\bibnamefont {Wolf}}, \bibinfo {author} {\bibfnamefont {M.}~\bibnamefont
			{Julian}}, \bibinfo {author} {\bibfnamefont {R.~P.}\ \bibnamefont
			{Prasankumar}}, \bibinfo {author} {\bibfnamefont {K.}~\bibnamefont
			{Miyagawa}}, \bibinfo {author} {\bibfnamefont {K.}~\bibnamefont {Kanoda}},
		\bibinfo {author} {\bibfnamefont {G.}~\bibnamefont {Gu}}, \bibinfo {author}
		{\bibfnamefont {M.}~\bibnamefont {Cothrine}}, \bibinfo {author}
		{\bibfnamefont {D.}~\bibnamefont {Mandrus}}, \bibinfo {author} {\bibfnamefont
			{M.}~\bibnamefont {Buzzi}}, \bibinfo {author} {\bibfnamefont
			{A.}~\bibnamefont {Cavalleri}}, \bibinfo {author} {\bibfnamefont {C.~R.}\
			\bibnamefont {Dean}}, \bibinfo {author} {\bibfnamefont {D.~M.}\ \bibnamefont
			{Kennes}}, \bibinfo {author} {\bibfnamefont {A.~J.}\ \bibnamefont {Millis}},
		\bibinfo {author} {\bibfnamefont {Q.}~\bibnamefont {Li}}, \bibinfo {author}
		{\bibfnamefont {M.~A.}\ \bibnamefont {Sentef}}, \bibinfo {author}
		{\bibfnamefont {A.}~\bibnamefont {Rubio}}, \bibinfo {author} {\bibfnamefont
			{A.~N.}\ \bibnamefont {Pasupathy}},\ and\ \bibinfo {author} {\bibfnamefont
			{D.~N.}\ \bibnamefont {Basov}},\ }\bibfield  {title} {\bibinfo {title}
		{Cavity-altered superconductivity},\ }\href
	{https://doi.org/10.1038/s41586-025-10062-6} {\bibfield  {journal} {\bibinfo
			{journal} {Nature}\ }\textbf {\bibinfo {volume} {650}},\ \bibinfo {pages}
		{864} (\bibinfo {year} {2026})}\BibitemShut {NoStop}%
	\bibitem [{\citenamefont {Trif}\ and\ \citenamefont
		{Tserkovnyak}(2012)}]{trif2012resonantly}%
	\BibitemOpen
	\bibfield  {author} {\bibinfo {author} {\bibfnamefont {M.}~\bibnamefont
			{Trif}}\ and\ \bibinfo {author} {\bibfnamefont {Y.}~\bibnamefont
			{Tserkovnyak}},\ }\bibfield  {title} {\bibinfo {title} {Resonantly tunable
			{M}ajorana polariton in a microwave cavity},\ }\href
	{https://doi.org/10.1103/PhysRevLett.109.257002} {\bibfield  {journal}
		{\bibinfo  {journal} {Phys. Rev. Lett.}\ }\textbf {\bibinfo {volume} {109}},\
		\bibinfo {pages} {257002} (\bibinfo {year} {2012})}\BibitemShut {NoStop}%
	\bibitem [{\citenamefont {Cottet}\ \emph {et~al.}(2013)\citenamefont {Cottet},
		\citenamefont {Kontos},\ and\ \citenamefont {Dou\ifmmode~\mbox{\c{c}}\else
			\c{c}\fi{}ot}}]{cottet2013squeezing}%
	\BibitemOpen
	\bibfield  {author} {\bibinfo {author} {\bibfnamefont {A.}~\bibnamefont
			{Cottet}}, \bibinfo {author} {\bibfnamefont {T.}~\bibnamefont {Kontos}},\
		and\ \bibinfo {author} {\bibfnamefont {B.}~\bibnamefont
			{Dou\ifmmode~\mbox{\c{c}}\else \c{c}\fi{}ot}},\ }\bibfield  {title} {\bibinfo
		{title} {Squeezing light with {M}ajorana fermions},\ }\href
	{https://doi.org/10.1103/PhysRevB.88.195415} {\bibfield  {journal} {\bibinfo
			{journal} {Phys. Rev. B}\ }\textbf {\bibinfo {volume} {88}},\ \bibinfo
		{pages} {195415} (\bibinfo {year} {2013})}\BibitemShut {NoStop}%
	\bibitem [{\citenamefont {Dmytruk}\ \emph {et~al.}(2015)\citenamefont
		{Dmytruk}, \citenamefont {Trif},\ and\ \citenamefont
		{Simon}}]{dmytruk2015cavity}%
	\BibitemOpen
	\bibfield  {author} {\bibinfo {author} {\bibfnamefont {O.}~\bibnamefont
			{Dmytruk}}, \bibinfo {author} {\bibfnamefont {M.}~\bibnamefont {Trif}},\ and\
		\bibinfo {author} {\bibfnamefont {P.}~\bibnamefont {Simon}},\ }\bibfield
	{title} {\bibinfo {title} {Cavity quantum electrodynamics with mesoscopic
			topological superconductors},\ }\href
	{https://doi.org/10.1103/PhysRevB.92.245432} {\bibfield  {journal} {\bibinfo
			{journal} {Phys. Rev. B}\ }\textbf {\bibinfo {volume} {92}},\ \bibinfo
		{pages} {245432} (\bibinfo {year} {2015})}\BibitemShut {NoStop}%
	\bibitem [{\citenamefont {Dmytruk}\ \emph {et~al.}(2016)\citenamefont
		{Dmytruk}, \citenamefont {Trif},\ and\ \citenamefont
		{Simon}}]{dmytruk2016josephson}%
	\BibitemOpen
	\bibfield  {author} {\bibinfo {author} {\bibfnamefont {O.}~\bibnamefont
			{Dmytruk}}, \bibinfo {author} {\bibfnamefont {M.}~\bibnamefont {Trif}},\ and\
		\bibinfo {author} {\bibfnamefont {P.}~\bibnamefont {Simon}},\ }\bibfield
	{title} {\bibinfo {title} {Josephson effect in topological superconducting
			rings coupled to a microwave cavity},\ }\href
	{https://doi.org/10.1103/PhysRevB.94.115423} {\bibfield  {journal} {\bibinfo
			{journal} {Phys. Rev. B}\ }\textbf {\bibinfo {volume} {94}},\ \bibinfo
		{pages} {115423} (\bibinfo {year} {2016})}\BibitemShut {NoStop}%
	\bibitem [{\citenamefont {Dartiailh}\ \emph {et~al.}(2017)\citenamefont
		{Dartiailh}, \citenamefont {Kontos}, \citenamefont
		{Dou\ifmmode~\mbox{\c{c}}\else \c{c}\fi{}ot},\ and\ \citenamefont
		{Cottet}}]{dartiailh2017direct}%
	\BibitemOpen
	\bibfield  {author} {\bibinfo {author} {\bibfnamefont {M.~C.}\ \bibnamefont
			{Dartiailh}}, \bibinfo {author} {\bibfnamefont {T.}~\bibnamefont {Kontos}},
		\bibinfo {author} {\bibfnamefont {B.}~\bibnamefont
			{Dou\ifmmode~\mbox{\c{c}}\else \c{c}\fi{}ot}},\ and\ \bibinfo {author}
		{\bibfnamefont {A.}~\bibnamefont {Cottet}},\ }\bibfield  {title} {\bibinfo
		{title} {Direct cavity detection of {M}ajorana pairs},\ }\href
	{https://doi.org/10.1103/PhysRevLett.118.126803} {\bibfield  {journal}
		{\bibinfo  {journal} {Phys. Rev. Lett.}\ }\textbf {\bibinfo {volume} {118}},\
		\bibinfo {pages} {126803} (\bibinfo {year} {2017})}\BibitemShut {NoStop}%
	\bibitem [{\citenamefont {Trif}\ and\ \citenamefont
		{Simon}(2019)}]{trif2019braiding}%
	\BibitemOpen
	\bibfield  {author} {\bibinfo {author} {\bibfnamefont {M.}~\bibnamefont
			{Trif}}\ and\ \bibinfo {author} {\bibfnamefont {P.}~\bibnamefont {Simon}},\
	}\bibfield  {title} {\bibinfo {title} {Braiding of {M}ajorana fermions in a
			cavity},\ }\href {https://doi.org/10.1103/PhysRevLett.122.236803} {\bibfield
		{journal} {\bibinfo  {journal} {Phys. Rev. Lett.}\ }\textbf {\bibinfo
			{volume} {122}},\ \bibinfo {pages} {236803} (\bibinfo {year}
		{2019})}\BibitemShut {NoStop}%
	\bibitem [{\citenamefont {M\'endez-C\'ordoba}\ \emph
		{et~al.}(2020)\citenamefont {M\'endez-C\'ordoba}, \citenamefont
		{Mendoza-Arenas}, \citenamefont {G\'omez-Ruiz}, \citenamefont
		{Rodr\'{\i}guez}, \citenamefont {Tejedor},\ and\ \citenamefont
		{Quiroga}}]{mendez2020renyi}%
	\BibitemOpen
	\bibfield  {author} {\bibinfo {author} {\bibfnamefont {F.~P.~M.}\
			\bibnamefont {M\'endez-C\'ordoba}}, \bibinfo {author} {\bibfnamefont {J.~J.}\
			\bibnamefont {Mendoza-Arenas}}, \bibinfo {author} {\bibfnamefont {F.~J.}\
			\bibnamefont {G\'omez-Ruiz}}, \bibinfo {author} {\bibfnamefont {F.~J.}\
			\bibnamefont {Rodr\'{\i}guez}}, \bibinfo {author} {\bibfnamefont
			{C.}~\bibnamefont {Tejedor}},\ and\ \bibinfo {author} {\bibfnamefont
			{L.}~\bibnamefont {Quiroga}},\ }\bibfield  {title} {\bibinfo {title} {R\'enyi
			entropy singularities as signatures of topological criticality in coupled
			photon-fermion systems},\ }\href
	{https://doi.org/10.1103/PhysRevResearch.2.043264} {\bibfield  {journal}
		{\bibinfo  {journal} {Phys. Rev. Res.}\ }\textbf {\bibinfo {volume} {2}},\
		\bibinfo {pages} {043264} (\bibinfo {year} {2020})}\BibitemShut {NoStop}%
	\bibitem [{\citenamefont {Contamin}\ \emph {et~al.}(2021)\citenamefont
		{Contamin}, \citenamefont {Delbecq}, \citenamefont {Dou{\c c}ot},
		\citenamefont {Cottet},\ and\ \citenamefont
		{Kontos}}]{contamin2021topological}%
	\BibitemOpen
	\bibfield  {author} {\bibinfo {author} {\bibfnamefont {L.~C.}\ \bibnamefont
			{Contamin}}, \bibinfo {author} {\bibfnamefont {M.~R.}\ \bibnamefont
			{Delbecq}}, \bibinfo {author} {\bibfnamefont {B.}~\bibnamefont {Dou{\c
					c}ot}}, \bibinfo {author} {\bibfnamefont {A.}~\bibnamefont {Cottet}},\ and\
		\bibinfo {author} {\bibfnamefont {T.}~\bibnamefont {Kontos}},\ }\bibfield
	{title} {\bibinfo {title} {Hybrid light-matter networks of {M}ajorana zero
			modes},\ }\href {https://doi.org/10.1038/s41534-021-00508-w} {\bibfield
		{journal} {\bibinfo  {journal} {npj Quantum Information}\ }\textbf {\bibinfo
			{volume} {7}},\ \bibinfo {pages} {171} (\bibinfo {year} {2021})}\BibitemShut
	{NoStop}%
	\bibitem [{\citenamefont {Dmytruk}\ and\ \citenamefont
		{Trif}(2023)}]{dmytruk2023microwave}%
	\BibitemOpen
	\bibfield  {author} {\bibinfo {author} {\bibfnamefont {O.}~\bibnamefont
			{Dmytruk}}\ and\ \bibinfo {author} {\bibfnamefont {M.}~\bibnamefont {Trif}},\
	}\bibfield  {title} {\bibinfo {title} {Microwave detection of gliding
			{M}ajorana zero modes in nanowires},\ }\href
	{https://doi.org/10.1103/PhysRevB.107.115418} {\bibfield  {journal} {\bibinfo
			{journal} {Phys. Rev. B}\ }\textbf {\bibinfo {volume} {107}},\ \bibinfo
		{pages} {115418} (\bibinfo {year} {2023})}\BibitemShut {NoStop}%
	\bibitem [{\citenamefont {Bacciconi}\ \emph {et~al.}(2024)\citenamefont
		{Bacciconi}, \citenamefont {Andolina},\ and\ \citenamefont
		{Mora}}]{bacciconi2023topological}%
	\BibitemOpen
	\bibfield  {author} {\bibinfo {author} {\bibfnamefont {Z.}~\bibnamefont
			{Bacciconi}}, \bibinfo {author} {\bibfnamefont {G.~M.}\ \bibnamefont
			{Andolina}},\ and\ \bibinfo {author} {\bibfnamefont {C.}~\bibnamefont
			{Mora}},\ }\bibfield  {title} {\bibinfo {title} {Topological protection of
			{M}ajorana polaritons in a cavity},\ }\href
	{https://doi.org/10.1103/PhysRevB.109.165434} {\bibfield  {journal} {\bibinfo
			{journal} {Phys. Rev. B}\ }\textbf {\bibinfo {volume} {109}},\ \bibinfo
		{pages} {165434} (\bibinfo {year} {2024})}\BibitemShut {NoStop}%
	\bibitem [{\citenamefont {Dmytruk}\ and\ \citenamefont
		{Schir\`o}(2024)}]{dmytruk2024hybrid}%
	\BibitemOpen
	\bibfield  {author} {\bibinfo {author} {\bibfnamefont {O.}~\bibnamefont
			{Dmytruk}}\ and\ \bibinfo {author} {\bibfnamefont {M.}~\bibnamefont
			{Schir\`o}},\ }\bibfield  {title} {\bibinfo {title} {Hybrid light-matter
			states in topological superconductors coupled to cavity photons},\ }\href
	{https://doi.org/10.1103/PhysRevB.110.075416} {\bibfield  {journal} {\bibinfo
			{journal} {Phys. Rev. B}\ }\textbf {\bibinfo {volume} {110}},\ \bibinfo
		{pages} {075416} (\bibinfo {year} {2024})}\BibitemShut {NoStop}%
	\bibitem [{\citenamefont {Becerra}\ and\ \citenamefont
		{Dmytruk}(2025)}]{becerra2025fermion}%
	\BibitemOpen
	\bibfield  {author} {\bibinfo {author} {\bibfnamefont {V.~F.}\ \bibnamefont
			{Becerra}}\ and\ \bibinfo {author} {\bibfnamefont {O.}~\bibnamefont
			{Dmytruk}},\ }\href {https://arxiv.org/abs/2506.06237} {\bibinfo {title}
		{Fermion parity switches imprinted in the photonic field of cavity embedded
			{K}itaev chains}} (\bibinfo {year} {2025}),\ \Eprint
	{https://arxiv.org/abs/2506.06237} {arXiv:2506.06237 [cond-mat.mes-hall]}
	\BibitemShut {NoStop}%
	\bibitem [{\citenamefont {Kobiałka}\ \emph {et~al.}(2026)\citenamefont
		{Kobiałka}, \citenamefont {Ghosh}, \citenamefont {Arouca},\ and\
		\citenamefont {Black-Schaffer}}]{kobialka2026topology}%
	\BibitemOpen
	\bibfield  {author} {\bibinfo {author} {\bibfnamefont {A.}~\bibnamefont
			{Kobiałka}}, \bibinfo {author} {\bibfnamefont {A.~K.}\ \bibnamefont
			{Ghosh}}, \bibinfo {author} {\bibfnamefont {R.}~\bibnamefont {Arouca}},\ and\
		\bibinfo {author} {\bibfnamefont {A.~M.}\ \bibnamefont {Black-Schaffer}},\
	}\href {https://arxiv.org/abs/2602.03553} {\bibinfo {title} {Topology and
			energy dependence of {M}ajorana bound states in a photonic cavity}} (\bibinfo
	{year} {2026}),\ \Eprint {https://arxiv.org/abs/2602.03553} {arXiv:2602.03553
		[cond-mat.mes-hall]} \BibitemShut {NoStop}%
	\bibitem [{\citenamefont {Sentef}\ \emph {et~al.}()\citenamefont {Sentef},
		\citenamefont {Ruggenthaler},\ and\ \citenamefont {Rubio}}]{sentef2018}%
	\BibitemOpen
	\bibfield  {author} {\bibinfo {author} {\bibfnamefont {M.~A.}\ \bibnamefont
			{Sentef}}, \bibinfo {author} {\bibfnamefont {M.}~\bibnamefont
			{Ruggenthaler}},\ and\ \bibinfo {author} {\bibfnamefont {A.}~\bibnamefont
			{Rubio}},\ }\bibfield  {title} {\bibinfo {title} {Cavity
			quantum-electrodynamical polaritonically enhanced electron-phonon coupling
			and its influence on superconductivity},\ }\href
	{https://doi.org/10.1126/sciadv.aau6969} {\bibfield  {journal} {\bibinfo
			{journal} {Science Advances}\ }\textbf {\bibinfo {volume} {4}},\ \bibinfo
		{pages} {eaau6969}}\BibitemShut {NoStop}%
	\bibitem [{\citenamefont {Schlawin}\ \emph {et~al.}(2019)\citenamefont
		{Schlawin}, \citenamefont {Cavalleri},\ and\ \citenamefont
		{Jaksch}}]{schlawin2019cavity}%
	\BibitemOpen
	\bibfield  {author} {\bibinfo {author} {\bibfnamefont {F.}~\bibnamefont
			{Schlawin}}, \bibinfo {author} {\bibfnamefont {A.}~\bibnamefont
			{Cavalleri}},\ and\ \bibinfo {author} {\bibfnamefont {D.}~\bibnamefont
			{Jaksch}},\ }\bibfield  {title} {\bibinfo {title} {Cavity-mediated
			electron-photon superconductivity},\ }\href
	{https://doi.org/10.1103/PhysRevLett.122.133602} {\bibfield  {journal}
		{\bibinfo  {journal} {Phys. Rev. Lett.}\ }\textbf {\bibinfo {volume} {122}},\
		\bibinfo {pages} {133602} (\bibinfo {year} {2019})}\BibitemShut {NoStop}%
	\bibitem [{\citenamefont {Curtis}\ \emph {et~al.}(2019)\citenamefont {Curtis},
		\citenamefont {Raines}, \citenamefont {Allocca}, \citenamefont {Hafezi},\
		and\ \citenamefont {Galitski}}]{curtis2019cavity}%
	\BibitemOpen
	\bibfield  {author} {\bibinfo {author} {\bibfnamefont {J.~B.}\ \bibnamefont
			{Curtis}}, \bibinfo {author} {\bibfnamefont {Z.~M.}\ \bibnamefont {Raines}},
		\bibinfo {author} {\bibfnamefont {A.~A.}\ \bibnamefont {Allocca}}, \bibinfo
		{author} {\bibfnamefont {M.}~\bibnamefont {Hafezi}},\ and\ \bibinfo {author}
		{\bibfnamefont {V.~M.}\ \bibnamefont {Galitski}},\ }\bibfield  {title}
	{\bibinfo {title} {Cavity quantum {E}liashberg enhancement of
			superconductivity},\ }\href {https://doi.org/10.1103/PhysRevLett.122.167002}
	{\bibfield  {journal} {\bibinfo  {journal} {Phys. Rev. Lett.}\ }\textbf
		{\bibinfo {volume} {122}},\ \bibinfo {pages} {167002} (\bibinfo {year}
		{2019})}\BibitemShut {NoStop}%
	\bibitem [{\citenamefont {Allocca}\ \emph {et~al.}(2019)\citenamefont
		{Allocca}, \citenamefont {Raines}, \citenamefont {Curtis},\ and\
		\citenamefont {Galitski}}]{alloca2019}%
	\BibitemOpen
	\bibfield  {author} {\bibinfo {author} {\bibfnamefont {A.~A.}\ \bibnamefont
			{Allocca}}, \bibinfo {author} {\bibfnamefont {Z.~M.}\ \bibnamefont {Raines}},
		\bibinfo {author} {\bibfnamefont {J.~B.}\ \bibnamefont {Curtis}},\ and\
		\bibinfo {author} {\bibfnamefont {V.~M.}\ \bibnamefont {Galitski}},\
	}\bibfield  {title} {\bibinfo {title} {Cavity superconductor-polaritons},\
	}\href {https://doi.org/10.1103/PhysRevB.99.020504} {\bibfield  {journal}
		{\bibinfo  {journal} {Phys. Rev. B}\ }\textbf {\bibinfo {volume} {99}},\
		\bibinfo {pages} {020504} (\bibinfo {year} {2019})}\BibitemShut {NoStop}%
	\bibitem [{\citenamefont {Kozin}\ \emph {et~al.}(2025)\citenamefont {Kozin},
		\citenamefont {Thingstad}, \citenamefont {Loss},\ and\ \citenamefont
		{Klinovaja}}]{kozin2025cavity}%
	\BibitemOpen
	\bibfield  {author} {\bibinfo {author} {\bibfnamefont {V.~K.}\ \bibnamefont
			{Kozin}}, \bibinfo {author} {\bibfnamefont {E.}~\bibnamefont {Thingstad}},
		\bibinfo {author} {\bibfnamefont {D.}~\bibnamefont {Loss}},\ and\ \bibinfo
		{author} {\bibfnamefont {J.}~\bibnamefont {Klinovaja}},\ }\bibfield  {title}
	{\bibinfo {title} {Cavity-enhanced superconductivity via band engineering},\
	}\href {https://doi.org/10.1103/PhysRevB.111.035410} {\bibfield  {journal}
		{\bibinfo  {journal} {Phys. Rev. B}\ }\textbf {\bibinfo {volume} {111}},\
		\bibinfo {pages} {035410} (\bibinfo {year} {2025})}\BibitemShut {NoStop}%
	\bibitem [{\citenamefont {Mazza}\ and\ \citenamefont
		{Georges}(2019)}]{mazza2019}%
	\BibitemOpen
	\bibfield  {author} {\bibinfo {author} {\bibfnamefont {G.}~\bibnamefont
			{Mazza}}\ and\ \bibinfo {author} {\bibfnamefont {A.}~\bibnamefont
			{Georges}},\ }\bibfield  {title} {\bibinfo {title} {Superradiant quantum
			materials},\ }\href {https://doi.org/10.1103/PhysRevLett.122.017401}
	{\bibfield  {journal} {\bibinfo  {journal} {Phys. Rev. Lett.}\ }\textbf
		{\bibinfo {volume} {122}},\ \bibinfo {pages} {017401} (\bibinfo {year}
		{2019})}\BibitemShut {NoStop}%
	\bibitem [{\citenamefont {Passetti}\ \emph {et~al.}(2023)\citenamefont
		{Passetti}, \citenamefont {Eckhardt}, \citenamefont {Sentef},\ and\
		\citenamefont {Kennes}}]{passetti2023cavity}%
	\BibitemOpen
	\bibfield  {author} {\bibinfo {author} {\bibfnamefont {G.}~\bibnamefont
			{Passetti}}, \bibinfo {author} {\bibfnamefont {C.~J.}\ \bibnamefont
			{Eckhardt}}, \bibinfo {author} {\bibfnamefont {M.~A.}\ \bibnamefont
			{Sentef}},\ and\ \bibinfo {author} {\bibfnamefont {D.~M.}\ \bibnamefont
			{Kennes}},\ }\bibfield  {title} {\bibinfo {title} {Cavity light-matter
			entanglement through quantum fluctuations},\ }\href
	{https://doi.org/10.1103/PhysRevLett.131.023601} {\bibfield  {journal}
		{\bibinfo  {journal} {Phys. Rev. Lett.}\ }\textbf {\bibinfo {volume} {131}},\
		\bibinfo {pages} {023601} (\bibinfo {year} {2023})}\BibitemShut {NoStop}%
	\bibitem [{\citenamefont {Kass}\ \emph {et~al.}(2024)\citenamefont {Kass},
		\citenamefont {Talkington}, \citenamefont {Srivastava},\ and\ \citenamefont
		{Claassen}}]{kass2024manybody}%
	\BibitemOpen
	\bibfield  {author} {\bibinfo {author} {\bibfnamefont {B.}~\bibnamefont
			{Kass}}, \bibinfo {author} {\bibfnamefont {S.}~\bibnamefont {Talkington}},
		\bibinfo {author} {\bibfnamefont {A.}~\bibnamefont {Srivastava}},\ and\
		\bibinfo {author} {\bibfnamefont {M.}~\bibnamefont {Claassen}},\ }\href
	{https://arxiv.org/abs/2411.08964} {\bibinfo {title} {Many-body photon
			blockade and quantum light generation from cavity quantum materials}}
	(\bibinfo {year} {2024}),\ \Eprint {https://arxiv.org/abs/2411.08964}
	{arXiv:2411.08964 [cond-mat.str-el]} \BibitemShut {NoStop}%
	\bibitem [{\citenamefont {Fadler}\ \emph {et~al.}(2024)\citenamefont {Fadler},
		\citenamefont {Schmidt}, \citenamefont {Li},\ and\ \citenamefont
		{Eckstein}}]{Fadler2024}%
	\BibitemOpen
	\bibfield  {author} {\bibinfo {author} {\bibfnamefont {P.}~\bibnamefont
			{Fadler}}, \bibinfo {author} {\bibfnamefont {K.~P.}\ \bibnamefont {Schmidt}},
		\bibinfo {author} {\bibfnamefont {J.}~\bibnamefont {Li}},\ and\ \bibinfo
		{author} {\bibfnamefont {M.}~\bibnamefont {Eckstein}},\ }\bibfield  {title}
	{\bibinfo {title} {Engineering photon-mediated long-range spin interactions
			in {M}ott insulators},\ }\href {https://doi.org/10.1103/PhysRevB.109.085149}
	{\bibfield  {journal} {\bibinfo  {journal} {Phys. Rev. B}\ }\textbf {\bibinfo
			{volume} {109}},\ \bibinfo {pages} {085149} (\bibinfo {year}
		{2024})}\BibitemShut {NoStop}%
	\bibitem [{\citenamefont {Nguyen}\ \emph {et~al.}(2023)\citenamefont {Nguyen},
		\citenamefont {Arwas}, \citenamefont {Lin}, \citenamefont {Yao},\ and\
		\citenamefont {Ciuti}}]{nguyen2023electron}%
	\BibitemOpen
	\bibfield  {author} {\bibinfo {author} {\bibfnamefont {D.-P.}\ \bibnamefont
			{Nguyen}}, \bibinfo {author} {\bibfnamefont {G.}~\bibnamefont {Arwas}},
		\bibinfo {author} {\bibfnamefont {Z.}~\bibnamefont {Lin}}, \bibinfo {author}
		{\bibfnamefont {W.}~\bibnamefont {Yao}},\ and\ \bibinfo {author}
		{\bibfnamefont {C.}~\bibnamefont {Ciuti}},\ }\bibfield  {title} {\bibinfo
		{title} {Electron-photon {C}hern number in cavity-embedded {2D} moir\'e
			materials},\ }\href {https://doi.org/10.1103/PhysRevLett.131.176602}
	{\bibfield  {journal} {\bibinfo  {journal} {Phys. Rev. Lett.}\ }\textbf
		{\bibinfo {volume} {131}},\ \bibinfo {pages} {176602} (\bibinfo {year}
		{2023})}\BibitemShut {NoStop}%
	\bibitem [{\citenamefont {Ciuti}(2021)}]{ciuti2021cavity}%
	\BibitemOpen
	\bibfield  {author} {\bibinfo {author} {\bibfnamefont {C.}~\bibnamefont
			{Ciuti}},\ }\bibfield  {title} {\bibinfo {title} {Cavity-mediated electron
			hopping in disordered quantum {H}all systems},\ }\href
	{https://doi.org/10.1103/PhysRevB.104.155307} {\bibfield  {journal} {\bibinfo
			{journal} {Phys. Rev. B}\ }\textbf {\bibinfo {volume} {104}},\ \bibinfo
		{pages} {155307} (\bibinfo {year} {2021})}\BibitemShut {NoStop}%
	\bibitem [{\citenamefont {Winter}\ and\ \citenamefont
		{Zilberberg}(2025)}]{winter2025fractional}%
	\BibitemOpen
	\bibfield  {author} {\bibinfo {author} {\bibfnamefont {L.}~\bibnamefont
			{Winter}}\ and\ \bibinfo {author} {\bibfnamefont {O.}~\bibnamefont
			{Zilberberg}},\ }\bibfield  {title} {\bibinfo {title} {Fractional quantum
			{H}all edge polaritons},\ }\href {https://doi.org/10.1103/qc6z-87c6}
	{\bibfield  {journal} {\bibinfo  {journal} {Phys. Rev. B}\ }\textbf {\bibinfo
			{volume} {112}},\ \bibinfo {pages} {L241105} (\bibinfo {year}
		{2025})}\BibitemShut {NoStop}%
	\bibitem [{\citenamefont {Bori\ifmmode~\mbox{\c{c}}\else \c{c}\fi{}i}\ \emph
		{et~al.}(2025)\citenamefont {Bori\ifmmode~\mbox{\c{c}}\else \c{c}\fi{}i},
		\citenamefont {Arwas},\ and\ \citenamefont {Ciuti}}]{borici2025}%
	\BibitemOpen
	\bibfield  {author} {\bibinfo {author} {\bibfnamefont {D.}~\bibnamefont
			{Bori\ifmmode~\mbox{\c{c}}\else \c{c}\fi{}i}}, \bibinfo {author}
		{\bibfnamefont {G.}~\bibnamefont {Arwas}},\ and\ \bibinfo {author}
		{\bibfnamefont {C.}~\bibnamefont {Ciuti}},\ }\bibfield  {title} {\bibinfo
		{title} {Cavity-modified quantum electron transport in multiterminal devices
			and interferometers},\ }\href {https://doi.org/10.1103/4l6d-gqkw} {\bibfield
		{journal} {\bibinfo  {journal} {Phys. Rev. B}\ }\textbf {\bibinfo {volume}
			{112}},\ \bibinfo {pages} {045301} (\bibinfo {year} {2025})}\BibitemShut
	{NoStop}%
	\bibitem [{\citenamefont {Dmytruk}\ and\ \citenamefont
		{Schir{\`o}}(2022)}]{dmytruk2022controlling}%
	\BibitemOpen
	\bibfield  {author} {\bibinfo {author} {\bibfnamefont {O.}~\bibnamefont
			{Dmytruk}}\ and\ \bibinfo {author} {\bibfnamefont {M.}~\bibnamefont
			{Schir{\`o}}},\ }\bibfield  {title} {\bibinfo {title} {Controlling
			topological phases of matter with quantum light},\ }\href
	{https://doi.org/10.1038/s42005-022-01049-0} {\bibfield  {journal} {\bibinfo
			{journal} {Communications Physics}\ }\textbf {\bibinfo {volume} {5}},\
		\bibinfo {pages} {271} (\bibinfo {year} {2022})}\BibitemShut {NoStop}%
	\bibitem [{\citenamefont {P{\'{e}}rez-Gonz{\'{a}}lez}\ \emph
		{et~al.}(2025)\citenamefont {P{\'{e}}rez-Gonz{\'{a}}lez}, \citenamefont
		{Platero},\ and\ \citenamefont {Gomez-Le{\'{o}}n}}]{perez2023light}%
	\BibitemOpen
	\bibfield  {author} {\bibinfo {author} {\bibfnamefont {B.}~\bibnamefont
			{P{\'{e}}rez-Gonz{\'{a}}lez}}, \bibinfo {author} {\bibfnamefont
			{G.}~\bibnamefont {Platero}},\ and\ \bibinfo {author} {\bibfnamefont
			{{\'{A}}.}~\bibnamefont {Gomez-Le{\'{o}}n}},\ }\bibfield  {title} {\bibinfo
		{title} {Light-matter correlations in {Q}uantum {F}loquet engineering of
			cavity quantum materials},\ }\href
	{https://doi.org/10.22331/q-2025-02-17-1633} {\bibfield  {journal} {\bibinfo
			{journal} {{Quantum}}\ }\textbf {\bibinfo {volume} {9}},\ \bibinfo {pages}
		{1633} (\bibinfo {year} {2025})}\BibitemShut {NoStop}%
	\bibitem [{\citenamefont {Vlasiuk}\ \emph {et~al.}(2023)\citenamefont
		{Vlasiuk}, \citenamefont {Kozin}, \citenamefont {Klinovaja}, \citenamefont
		{Loss}, \citenamefont {Iorsh},\ and\ \citenamefont
		{Tokatly}}]{vlasiuk2023cavity}%
	\BibitemOpen
	\bibfield  {author} {\bibinfo {author} {\bibfnamefont {E.}~\bibnamefont
			{Vlasiuk}}, \bibinfo {author} {\bibfnamefont {V.~K.}\ \bibnamefont {Kozin}},
		\bibinfo {author} {\bibfnamefont {J.}~\bibnamefont {Klinovaja}}, \bibinfo
		{author} {\bibfnamefont {D.}~\bibnamefont {Loss}}, \bibinfo {author}
		{\bibfnamefont {I.~V.}\ \bibnamefont {Iorsh}},\ and\ \bibinfo {author}
		{\bibfnamefont {I.~V.}\ \bibnamefont {Tokatly}},\ }\bibfield  {title}
	{\bibinfo {title} {Cavity-induced charge transfer in periodic systems:
			{L}ength-gauge formalism},\ }\href
	{https://doi.org/10.1103/PhysRevB.108.085410} {\bibfield  {journal} {\bibinfo
			{journal} {Phys. Rev. B}\ }\textbf {\bibinfo {volume} {108}},\ \bibinfo
		{pages} {085410} (\bibinfo {year} {2023})}\BibitemShut {NoStop}%
	\bibitem [{\citenamefont {Nguyen}\ \emph {et~al.}(2024)\citenamefont {Nguyen},
		\citenamefont {Arwas},\ and\ \citenamefont {Ciuti}}]{nguyen2024electron}%
	\BibitemOpen
	\bibfield  {author} {\bibinfo {author} {\bibfnamefont {D.-P.}\ \bibnamefont
			{Nguyen}}, \bibinfo {author} {\bibfnamefont {G.}~\bibnamefont {Arwas}},\ and\
		\bibinfo {author} {\bibfnamefont {C.}~\bibnamefont {Ciuti}},\ }\bibfield
	{title} {\bibinfo {title} {Electron conductance and many-body marker of a
			cavity-embedded topological one-dimensional chain},\ }\href
	{https://doi.org/10.1103/PhysRevB.110.195416} {\bibfield  {journal} {\bibinfo
			{journal} {Phys. Rev. B}\ }\textbf {\bibinfo {volume} {110}},\ \bibinfo
		{pages} {195416} (\bibinfo {year} {2024})}\BibitemShut {NoStop}%
	\bibitem [{\citenamefont {Shaffer}\ \emph {et~al.}(2024)\citenamefont
		{Shaffer}, \citenamefont {Claassen}, \citenamefont {Srivastava},\ and\
		\citenamefont {Santos}}]{shaffer2024entanglement}%
	\BibitemOpen
	\bibfield  {author} {\bibinfo {author} {\bibfnamefont {D.}~\bibnamefont
			{Shaffer}}, \bibinfo {author} {\bibfnamefont {M.}~\bibnamefont {Claassen}},
		\bibinfo {author} {\bibfnamefont {A.}~\bibnamefont {Srivastava}},\ and\
		\bibinfo {author} {\bibfnamefont {L.~H.}\ \bibnamefont {Santos}},\ }\bibfield
	{title} {\bibinfo {title} {Entanglement and topology in
			{S}u-{S}chrieffer-{H}eeger cavity quantum electrodynamics},\ }\href
	{https://doi.org/10.1103/PhysRevB.109.155160} {\bibfield  {journal} {\bibinfo
			{journal} {Phys. Rev. B}\ }\textbf {\bibinfo {volume} {109}},\ \bibinfo
		{pages} {155160} (\bibinfo {year} {2024})}\BibitemShut {NoStop}%
	\bibitem [{\citenamefont {Sueiro}\ \emph {et~al.}(2025)\citenamefont {Sueiro},
		\citenamefont {Andolina},\ and\ \citenamefont {Schirò}}]{Sueiro25}%
	\BibitemOpen
	\bibfield  {author} {\bibinfo {author} {\bibfnamefont {J.}~\bibnamefont
			{Sueiro}}, \bibinfo {author} {\bibfnamefont {G.~M.}\ \bibnamefont
			{Andolina}},\ and\ \bibinfo {author} {\bibfnamefont {M.}~\bibnamefont
			{Schirò}},\ }\href {https://arxiv.org/abs/2507.22715} {\bibinfo {title}
		{Floquet theory of lattice electrons coupled to an off-resonant cavity}}
	(\bibinfo {year} {2025}),\ \Eprint {https://arxiv.org/abs/2507.22715}
	{arXiv:2507.22715 [cond-mat.str-el]} \BibitemShut {NoStop}%
	\bibitem [{\citenamefont {Ritz-Zwilling}\ and\ \citenamefont
		{Dmytruk}(2026)}]{ritzzwilling2026}%
	\BibitemOpen
	\bibfield  {author} {\bibinfo {author} {\bibfnamefont {A.}~\bibnamefont
			{Ritz-Zwilling}}\ and\ \bibinfo {author} {\bibfnamefont {O.}~\bibnamefont
			{Dmytruk}},\ }\href {https://arxiv.org/abs/2604.13936} {\bibinfo {title}
		{Topological markers for a one-dimensional fermionic chain coupled to a
			single-mode cavity}} (\bibinfo {year} {2026}),\ \Eprint
	{https://arxiv.org/abs/2604.13936} {arXiv:2604.13936 [cond-mat.mes-hall]}
	\BibitemShut {NoStop}%
	\bibitem [{\citenamefont {Delbecq}\ \emph {et~al.}(2011)\citenamefont
		{Delbecq}, \citenamefont {Schmitt}, \citenamefont {Parmentier}, \citenamefont
		{Roch}, \citenamefont {Viennot}, \citenamefont {F\`eve}, \citenamefont
		{Huard}, \citenamefont {Mora}, \citenamefont {Cottet},\ and\ \citenamefont
		{Kontos}}]{delbecq2011coupling}%
	\BibitemOpen
	\bibfield  {author} {\bibinfo {author} {\bibfnamefont {M.~R.}\ \bibnamefont
			{Delbecq}}, \bibinfo {author} {\bibfnamefont {V.}~\bibnamefont {Schmitt}},
		\bibinfo {author} {\bibfnamefont {F.~D.}\ \bibnamefont {Parmentier}},
		\bibinfo {author} {\bibfnamefont {N.}~\bibnamefont {Roch}}, \bibinfo {author}
		{\bibfnamefont {J.~J.}\ \bibnamefont {Viennot}}, \bibinfo {author}
		{\bibfnamefont {G.}~\bibnamefont {F\`eve}}, \bibinfo {author} {\bibfnamefont
			{B.}~\bibnamefont {Huard}}, \bibinfo {author} {\bibfnamefont
			{C.}~\bibnamefont {Mora}}, \bibinfo {author} {\bibfnamefont {A.}~\bibnamefont
			{Cottet}},\ and\ \bibinfo {author} {\bibfnamefont {T.}~\bibnamefont
			{Kontos}},\ }\bibfield  {title} {\bibinfo {title} {Coupling a quantum dot,
			fermionic leads, and a microwave cavity on a chip},\ }\href
	{https://doi.org/10.1103/PhysRevLett.107.256804} {\bibfield  {journal}
		{\bibinfo  {journal} {Phys. Rev. Lett.}\ }\textbf {\bibinfo {volume} {107}},\
		\bibinfo {pages} {256804} (\bibinfo {year} {2011})}\BibitemShut {NoStop}%
	\bibitem [{\citenamefont {Petersson}\ \emph {et~al.}(2012)\citenamefont
		{Petersson}, \citenamefont {McFaul}, \citenamefont {Schroer}, \citenamefont
		{Jung}, \citenamefont {Taylor}, \citenamefont {Houck},\ and\ \citenamefont
		{Petta}}]{petersson2012coupling}%
	\BibitemOpen
	\bibfield  {author} {\bibinfo {author} {\bibfnamefont {K.~D.}\ \bibnamefont
			{Petersson}}, \bibinfo {author} {\bibfnamefont {L.~W.}\ \bibnamefont
			{McFaul}}, \bibinfo {author} {\bibfnamefont {M.~D.}\ \bibnamefont {Schroer}},
		\bibinfo {author} {\bibfnamefont {M.}~\bibnamefont {Jung}}, \bibinfo {author}
		{\bibfnamefont {J.~M.}\ \bibnamefont {Taylor}}, \bibinfo {author}
		{\bibfnamefont {A.~A.}\ \bibnamefont {Houck}},\ and\ \bibinfo {author}
		{\bibfnamefont {J.~R.}\ \bibnamefont {Petta}},\ }\bibfield  {title} {\bibinfo
		{title} {Circuit quantum electrodynamics with a spin qubit},\ }\href
	{https://doi.org/10.1038/nature11559} {\bibfield  {journal} {\bibinfo
			{journal} {Nature}\ }\textbf {\bibinfo {volume} {490}},\ \bibinfo {pages}
		{380} (\bibinfo {year} {2012})}\BibitemShut {NoStop}%
	\bibitem [{\citenamefont {Frey}\ \emph {et~al.}(2012)\citenamefont {Frey},
		\citenamefont {Leek}, \citenamefont {Beck}, \citenamefont {Blais},
		\citenamefont {Ihn}, \citenamefont {Ensslin},\ and\ \citenamefont
		{Wallraff}}]{frey2012dipole}%
	\BibitemOpen
	\bibfield  {author} {\bibinfo {author} {\bibfnamefont {T.}~\bibnamefont
			{Frey}}, \bibinfo {author} {\bibfnamefont {P.~J.}\ \bibnamefont {Leek}},
		\bibinfo {author} {\bibfnamefont {M.}~\bibnamefont {Beck}}, \bibinfo {author}
		{\bibfnamefont {A.}~\bibnamefont {Blais}}, \bibinfo {author} {\bibfnamefont
			{T.}~\bibnamefont {Ihn}}, \bibinfo {author} {\bibfnamefont {K.}~\bibnamefont
			{Ensslin}},\ and\ \bibinfo {author} {\bibfnamefont {A.}~\bibnamefont
			{Wallraff}},\ }\bibfield  {title} {\bibinfo {title} {Dipole coupling of a
			double quantum dot to a microwave resonator},\ }\href
	{https://doi.org/10.1103/PhysRevLett.108.046807} {\bibfield  {journal}
		{\bibinfo  {journal} {Phys. Rev. Lett.}\ }\textbf {\bibinfo {volume} {108}},\
		\bibinfo {pages} {046807} (\bibinfo {year} {2012})}\BibitemShut {NoStop}%
	\bibitem [{\citenamefont {Basset}\ \emph {et~al.}(2013)\citenamefont {Basset},
		\citenamefont {Jarausch}, \citenamefont {Stockklauser}, \citenamefont {Frey},
		\citenamefont {Reichl}, \citenamefont {Wegscheider}, \citenamefont {Ihn},
		\citenamefont {Ensslin},\ and\ \citenamefont {Wallraff}}]{basset2013single}%
	\BibitemOpen
	\bibfield  {author} {\bibinfo {author} {\bibfnamefont {J.}~\bibnamefont
			{Basset}}, \bibinfo {author} {\bibfnamefont {D.-D.}\ \bibnamefont
			{Jarausch}}, \bibinfo {author} {\bibfnamefont {A.}~\bibnamefont
			{Stockklauser}}, \bibinfo {author} {\bibfnamefont {T.}~\bibnamefont {Frey}},
		\bibinfo {author} {\bibfnamefont {C.}~\bibnamefont {Reichl}}, \bibinfo
		{author} {\bibfnamefont {W.}~\bibnamefont {Wegscheider}}, \bibinfo {author}
		{\bibfnamefont {T.~M.}\ \bibnamefont {Ihn}}, \bibinfo {author} {\bibfnamefont
			{K.}~\bibnamefont {Ensslin}},\ and\ \bibinfo {author} {\bibfnamefont
			{A.}~\bibnamefont {Wallraff}},\ }\bibfield  {title} {\bibinfo {title}
		{Single-electron double quantum dot dipole-coupled to a single photonic
			mode},\ }\href {https://doi.org/10.1103/PhysRevB.88.125312} {\bibfield
		{journal} {\bibinfo  {journal} {Phys. Rev. B}\ }\textbf {\bibinfo {volume}
			{88}},\ \bibinfo {pages} {125312} (\bibinfo {year} {2013})}\BibitemShut
	{NoStop}%
	\bibitem [{\citenamefont {Delbecq}\ \emph {et~al.}(2013)\citenamefont
		{Delbecq}, \citenamefont {Bruhat}, \citenamefont {Viennot}, \citenamefont
		{Datta}, \citenamefont {Cottet},\ and\ \citenamefont
		{Kontos}}]{delbecq2013engineering}%
	\BibitemOpen
	\bibfield  {author} {\bibinfo {author} {\bibfnamefont {M.~R.}\ \bibnamefont
			{Delbecq}}, \bibinfo {author} {\bibfnamefont {L.~E.}\ \bibnamefont {Bruhat}},
		\bibinfo {author} {\bibfnamefont {J.~J.}\ \bibnamefont {Viennot}}, \bibinfo
		{author} {\bibfnamefont {S.}~\bibnamefont {Datta}}, \bibinfo {author}
		{\bibfnamefont {A.}~\bibnamefont {Cottet}},\ and\ \bibinfo {author}
		{\bibfnamefont {T.}~\bibnamefont {Kontos}},\ }\bibfield  {title} {\bibinfo
		{title} {Photon-mediated interaction between distant quantum dot circuits},\
	}\href {https://doi.org/10.1038/ncomms2407} {\bibfield  {journal} {\bibinfo
			{journal} {Nature Communications}\ }\textbf {\bibinfo {volume} {4}},\
		\bibinfo {pages} {1400} (\bibinfo {year} {2013})}\BibitemShut {NoStop}%
	\bibitem [{\citenamefont {Viennot}\ \emph {et~al.}(2014)\citenamefont
		{Viennot}, \citenamefont {Delbecq}, \citenamefont {Dartiailh}, \citenamefont
		{Cottet},\ and\ \citenamefont {Kontos}}]{viennot2014out}%
	\BibitemOpen
	\bibfield  {author} {\bibinfo {author} {\bibfnamefont {J.~J.}\ \bibnamefont
			{Viennot}}, \bibinfo {author} {\bibfnamefont {M.~R.}\ \bibnamefont
			{Delbecq}}, \bibinfo {author} {\bibfnamefont {M.~C.}\ \bibnamefont
			{Dartiailh}}, \bibinfo {author} {\bibfnamefont {A.}~\bibnamefont {Cottet}},\
		and\ \bibinfo {author} {\bibfnamefont {T.}~\bibnamefont {Kontos}},\
	}\bibfield  {title} {\bibinfo {title} {Out-of-equilibrium charge dynamics in
			a hybrid circuit quantum electrodynamics architecture},\ }\href
	{https://doi.org/10.1103/PhysRevB.89.165404} {\bibfield  {journal} {\bibinfo
			{journal} {Phys. Rev. B}\ }\textbf {\bibinfo {volume} {89}},\ \bibinfo
		{pages} {165404} (\bibinfo {year} {2014})}\BibitemShut {NoStop}%
	\bibitem [{\citenamefont {Viennot}\ \emph {et~al.}(2015)\citenamefont
		{Viennot}, \citenamefont {Dartiailh}, \citenamefont {Cottet},\ and\
		\citenamefont {Kontos}}]{viennot2015coheret}%
	\BibitemOpen
	\bibfield  {author} {\bibinfo {author} {\bibfnamefont {J.~J.}\ \bibnamefont
			{Viennot}}, \bibinfo {author} {\bibfnamefont {M.~C.}\ \bibnamefont
			{Dartiailh}}, \bibinfo {author} {\bibfnamefont {A.}~\bibnamefont {Cottet}},\
		and\ \bibinfo {author} {\bibfnamefont {T.}~\bibnamefont {Kontos}},\
	}\bibfield  {title} {\bibinfo {title} {Coherent coupling of a single spin to
			microwave cavity photons},\ }\href {https://doi.org/10.1126/science.aaa3786}
	{\bibfield  {journal} {\bibinfo  {journal} {Science}\ }\textbf {\bibinfo
			{volume} {349}},\ \bibinfo {pages} {408} (\bibinfo {year} {2015})},\ \Eprint
	{https://arxiv.org/abs/https://www.science.org/doi/pdf/10.1126/science.aaa3786}
	{https://www.science.org/doi/pdf/10.1126/science.aaa3786} \BibitemShut
	{NoStop}%
	\bibitem [{\citenamefont {Stockklauser}\ \emph {et~al.}(2015)\citenamefont
		{Stockklauser}, \citenamefont {Maisi}, \citenamefont {Basset}, \citenamefont
		{Cujia}, \citenamefont {Reichl}, \citenamefont {Wegscheider}, \citenamefont
		{Ihn}, \citenamefont {Wallraff},\ and\ \citenamefont
		{Ensslin}}]{stockklauser2015microwave}%
	\BibitemOpen
	\bibfield  {author} {\bibinfo {author} {\bibfnamefont {A.}~\bibnamefont
			{Stockklauser}}, \bibinfo {author} {\bibfnamefont {V.~F.}\ \bibnamefont
			{Maisi}}, \bibinfo {author} {\bibfnamefont {J.}~\bibnamefont {Basset}},
		\bibinfo {author} {\bibfnamefont {K.}~\bibnamefont {Cujia}}, \bibinfo
		{author} {\bibfnamefont {C.}~\bibnamefont {Reichl}}, \bibinfo {author}
		{\bibfnamefont {W.}~\bibnamefont {Wegscheider}}, \bibinfo {author}
		{\bibfnamefont {T.}~\bibnamefont {Ihn}}, \bibinfo {author} {\bibfnamefont
			{A.}~\bibnamefont {Wallraff}},\ and\ \bibinfo {author} {\bibfnamefont
			{K.}~\bibnamefont {Ensslin}},\ }\bibfield  {title} {\bibinfo {title}
		{Microwave emission from hybridized states in a semiconductor charge qubit},\
	}\href {https://doi.org/10.1103/PhysRevLett.115.046802} {\bibfield  {journal}
		{\bibinfo  {journal} {Phys. Rev. Lett.}\ }\textbf {\bibinfo {volume} {115}},\
		\bibinfo {pages} {046802} (\bibinfo {year} {2015})}\BibitemShut {NoStop}%
	\bibitem [{\citenamefont {Mi}\ \emph {et~al.}(2017)\citenamefont {Mi},
		\citenamefont {Cady}, \citenamefont {Zajac}, \citenamefont {Deelman},\ and\
		\citenamefont {Petta}}]{mi2016strong}%
	\BibitemOpen
	\bibfield  {author} {\bibinfo {author} {\bibfnamefont {X.}~\bibnamefont
			{Mi}}, \bibinfo {author} {\bibfnamefont {J.~V.}\ \bibnamefont {Cady}},
		\bibinfo {author} {\bibfnamefont {D.~M.}\ \bibnamefont {Zajac}}, \bibinfo
		{author} {\bibfnamefont {P.~W.}\ \bibnamefont {Deelman}},\ and\ \bibinfo
		{author} {\bibfnamefont {J.~R.}\ \bibnamefont {Petta}},\ }\bibfield  {title}
	{\bibinfo {title} {Strong coupling of a single electron in silicon to a
			microwave photon},\ }\href {https://doi.org/10.1126/science.aal2469}
	{\bibfield  {journal} {\bibinfo  {journal} {Science}\ }\textbf {\bibinfo
			{volume} {355}},\ \bibinfo {pages} {156} (\bibinfo {year} {2017})},\ \Eprint
	{https://arxiv.org/abs/https://www.science.org/doi/pdf/10.1126/science.aal2469}
	{https://www.science.org/doi/pdf/10.1126/science.aal2469} \BibitemShut
	{NoStop}%
	\bibitem [{\citenamefont {Cottet}\ \emph {et~al.}(2017)\citenamefont {Cottet},
		\citenamefont {Dartiailh}, \citenamefont {Desjardins}, \citenamefont
		{Cubaynes}, \citenamefont {Contamin}, \citenamefont {Delbecq}, \citenamefont
		{Viennot}, \citenamefont {Bruhat}, \citenamefont {Douçot},\ and\
		\citenamefont {Kontos}}]{cottet2017cavity}%
	\BibitemOpen
	\bibfield  {author} {\bibinfo {author} {\bibfnamefont {A.}~\bibnamefont
			{Cottet}}, \bibinfo {author} {\bibfnamefont {M.~C.}\ \bibnamefont
			{Dartiailh}}, \bibinfo {author} {\bibfnamefont {M.~M.}\ \bibnamefont
			{Desjardins}}, \bibinfo {author} {\bibfnamefont {T.}~\bibnamefont
			{Cubaynes}}, \bibinfo {author} {\bibfnamefont {L.~C.}\ \bibnamefont
			{Contamin}}, \bibinfo {author} {\bibfnamefont {M.}~\bibnamefont {Delbecq}},
		\bibinfo {author} {\bibfnamefont {J.~J.}\ \bibnamefont {Viennot}}, \bibinfo
		{author} {\bibfnamefont {L.~E.}\ \bibnamefont {Bruhat}}, \bibinfo {author}
		{\bibfnamefont {B.}~\bibnamefont {Douçot}},\ and\ \bibinfo {author}
		{\bibfnamefont {T.}~\bibnamefont {Kontos}},\ }\bibfield  {title} {\bibinfo
		{title} {Cavity {QED} with hybrid nanocircuits: from atomic-like physics to
			condensed matter phenomena},\ }\href
	{https://doi.org/10.1088/1361-648X/aa7b4d} {\bibfield  {journal} {\bibinfo
			{journal} {Journal of Physics: Condensed Matter}\ }\textbf {\bibinfo {volume}
			{29}},\ \bibinfo {pages} {433002} (\bibinfo {year} {2017})}\BibitemShut
	{NoStop}%
	\bibitem [{\citenamefont {Mi}\ \emph {et~al.}(2018)\citenamefont {Mi},
		\citenamefont {Benito}, \citenamefont {Putz}, \citenamefont {Zajac},
		\citenamefont {Taylor}, \citenamefont {Burkard},\ and\ \citenamefont
		{Petta}}]{mi2018electron}%
	\BibitemOpen
	\bibfield  {author} {\bibinfo {author} {\bibfnamefont {X.}~\bibnamefont
			{Mi}}, \bibinfo {author} {\bibfnamefont {M.}~\bibnamefont {Benito}}, \bibinfo
		{author} {\bibfnamefont {S.}~\bibnamefont {Putz}}, \bibinfo {author}
		{\bibfnamefont {D.~M.}\ \bibnamefont {Zajac}}, \bibinfo {author}
		{\bibfnamefont {J.~M.}\ \bibnamefont {Taylor}}, \bibinfo {author}
		{\bibfnamefont {G.}~\bibnamefont {Burkard}},\ and\ \bibinfo {author}
		{\bibfnamefont {J.~R.}\ \bibnamefont {Petta}},\ }\bibfield  {title} {\bibinfo
		{title} {A coherent spin--photon interface in silicon},\ }\href
	{https://doi.org/10.1038/nature25769} {\bibfield  {journal} {\bibinfo
			{journal} {Nature}\ }\textbf {\bibinfo {volume} {555}},\ \bibinfo {pages}
		{599} (\bibinfo {year} {2018})}\BibitemShut {NoStop}%
	\bibitem [{\citenamefont {Samkharadze}\ \emph {et~al.}(2018)\citenamefont
		{Samkharadze}, \citenamefont {Zheng}, \citenamefont {Kalhor}, \citenamefont
		{Brousse}, \citenamefont {Sammak}, \citenamefont {Mendes}, \citenamefont
		{Blais}, \citenamefont {Scappucci},\ and\ \citenamefont
		{Vandersypen}}]{samkharadze2018strong}%
	\BibitemOpen
	\bibfield  {author} {\bibinfo {author} {\bibfnamefont {N.}~\bibnamefont
			{Samkharadze}}, \bibinfo {author} {\bibfnamefont {G.}~\bibnamefont {Zheng}},
		\bibinfo {author} {\bibfnamefont {N.}~\bibnamefont {Kalhor}}, \bibinfo
		{author} {\bibfnamefont {D.}~\bibnamefont {Brousse}}, \bibinfo {author}
		{\bibfnamefont {A.}~\bibnamefont {Sammak}}, \bibinfo {author} {\bibfnamefont
			{U.~C.}\ \bibnamefont {Mendes}}, \bibinfo {author} {\bibfnamefont
			{A.}~\bibnamefont {Blais}}, \bibinfo {author} {\bibfnamefont
			{G.}~\bibnamefont {Scappucci}},\ and\ \bibinfo {author} {\bibfnamefont
			{L.~M.~K.}\ \bibnamefont {Vandersypen}},\ }\bibfield  {title} {\bibinfo
		{title} {Strong spin-photon coupling in silicon},\ }\href
	{https://doi.org/10.1126/science.aar4054} {\bibfield  {journal} {\bibinfo
			{journal} {Science}\ }\textbf {\bibinfo {volume} {359}},\ \bibinfo {pages}
		{1123} (\bibinfo {year} {2018})},\ \Eprint
	{https://arxiv.org/abs/https://www.science.org/doi/pdf/10.1126/science.aar4054}
	{https://www.science.org/doi/pdf/10.1126/science.aar4054} \BibitemShut
	{NoStop}%
	\bibitem [{\citenamefont {Landig}\ \emph {et~al.}(2018)\citenamefont {Landig},
		\citenamefont {Koski}, \citenamefont {Scarlino}, \citenamefont {Mendes},
		\citenamefont {Blais}, \citenamefont {Reichl}, \citenamefont {Wegscheider},
		\citenamefont {Wallraff}, \citenamefont {Ensslin},\ and\ \citenamefont
		{Ihn}}]{landig2018electron}%
	\BibitemOpen
	\bibfield  {author} {\bibinfo {author} {\bibfnamefont {A.~J.}\ \bibnamefont
			{Landig}}, \bibinfo {author} {\bibfnamefont {J.~V.}\ \bibnamefont {Koski}},
		\bibinfo {author} {\bibfnamefont {P.}~\bibnamefont {Scarlino}}, \bibinfo
		{author} {\bibfnamefont {U.~C.}\ \bibnamefont {Mendes}}, \bibinfo {author}
		{\bibfnamefont {A.}~\bibnamefont {Blais}}, \bibinfo {author} {\bibfnamefont
			{C.}~\bibnamefont {Reichl}}, \bibinfo {author} {\bibfnamefont
			{W.}~\bibnamefont {Wegscheider}}, \bibinfo {author} {\bibfnamefont
			{A.}~\bibnamefont {Wallraff}}, \bibinfo {author} {\bibfnamefont
			{K.}~\bibnamefont {Ensslin}},\ and\ \bibinfo {author} {\bibfnamefont
			{T.}~\bibnamefont {Ihn}},\ }\bibfield  {title} {\bibinfo {title} {Coherent
			spin--photon coupling using a resonant exchange qubit},\ }\href
	{https://doi.org/10.1038/s41586-018-0365-y} {\bibfield  {journal} {\bibinfo
			{journal} {Nature}\ }\textbf {\bibinfo {volume} {560}},\ \bibinfo {pages}
		{179} (\bibinfo {year} {2018})}\BibitemShut {NoStop}%
	\bibitem [{\citenamefont {Scarlino}\ \emph {et~al.}(2022)\citenamefont
		{Scarlino}, \citenamefont {Ungerer}, \citenamefont {van Woerkom},
		\citenamefont {Mancini}, \citenamefont {Stano}, \citenamefont {M\"uller},
		\citenamefont {Landig}, \citenamefont {Koski}, \citenamefont {Reichl},
		\citenamefont {Wegscheider}, \citenamefont {Ihn}, \citenamefont {Ensslin},\
		and\ \citenamefont {Wallraff}}]{scarlino2022in}%
	\BibitemOpen
	\bibfield  {author} {\bibinfo {author} {\bibfnamefont {P.}~\bibnamefont
			{Scarlino}}, \bibinfo {author} {\bibfnamefont {J.~H.}\ \bibnamefont
			{Ungerer}}, \bibinfo {author} {\bibfnamefont {D.~J.}\ \bibnamefont {van
				Woerkom}}, \bibinfo {author} {\bibfnamefont {M.}~\bibnamefont {Mancini}},
		\bibinfo {author} {\bibfnamefont {P.}~\bibnamefont {Stano}}, \bibinfo
		{author} {\bibfnamefont {C.}~\bibnamefont {M\"uller}}, \bibinfo {author}
		{\bibfnamefont {A.~J.}\ \bibnamefont {Landig}}, \bibinfo {author}
		{\bibfnamefont {J.~V.}\ \bibnamefont {Koski}}, \bibinfo {author}
		{\bibfnamefont {C.}~\bibnamefont {Reichl}}, \bibinfo {author} {\bibfnamefont
			{W.}~\bibnamefont {Wegscheider}}, \bibinfo {author} {\bibfnamefont
			{T.}~\bibnamefont {Ihn}}, \bibinfo {author} {\bibfnamefont {K.}~\bibnamefont
			{Ensslin}},\ and\ \bibinfo {author} {\bibfnamefont {A.}~\bibnamefont
			{Wallraff}},\ }\bibfield  {title} {\bibinfo {title} {In situ tuning of the
			electric-dipole strength of a double-dot charge qubit: {C}harge-noise
			protection and ultrastrong coupling},\ }\href
	{https://doi.org/10.1103/PhysRevX.12.031004} {\bibfield  {journal} {\bibinfo
			{journal} {Phys. Rev. X}\ }\textbf {\bibinfo {volume} {12}},\ \bibinfo
		{pages} {031004} (\bibinfo {year} {2022})}\BibitemShut {NoStop}%
	\bibitem [{\citenamefont {Gu}\ \emph {et~al.}(2023)\citenamefont {Gu},
		\citenamefont {Kohler}, \citenamefont {Xu}, \citenamefont {Wu}, \citenamefont
		{Jiang}, \citenamefont {Ye}, \citenamefont {Lin}, \citenamefont {Wang},
		\citenamefont {Li}, \citenamefont {Cao},\ and\ \citenamefont
		{Guo}}]{gu2023probing}%
	\BibitemOpen
	\bibfield  {author} {\bibinfo {author} {\bibfnamefont {S.-S.}\ \bibnamefont
			{Gu}}, \bibinfo {author} {\bibfnamefont {S.}~\bibnamefont {Kohler}}, \bibinfo
		{author} {\bibfnamefont {Y.-Q.}\ \bibnamefont {Xu}}, \bibinfo {author}
		{\bibfnamefont {R.}~\bibnamefont {Wu}}, \bibinfo {author} {\bibfnamefont
			{S.-L.}\ \bibnamefont {Jiang}}, \bibinfo {author} {\bibfnamefont {S.-K.}\
			\bibnamefont {Ye}}, \bibinfo {author} {\bibfnamefont {T.}~\bibnamefont
			{Lin}}, \bibinfo {author} {\bibfnamefont {B.-C.}\ \bibnamefont {Wang}},
		\bibinfo {author} {\bibfnamefont {H.-O.}\ \bibnamefont {Li}}, \bibinfo
		{author} {\bibfnamefont {G.}~\bibnamefont {Cao}},\ and\ \bibinfo {author}
		{\bibfnamefont {G.-P.}\ \bibnamefont {Guo}},\ }\bibfield  {title} {\bibinfo
		{title} {Probing two driven double quantum dots strongly coupled to a
			cavity},\ }\href {https://doi.org/10.1103/PhysRevLett.130.233602} {\bibfield
		{journal} {\bibinfo  {journal} {Phys. Rev. Lett.}\ }\textbf {\bibinfo
			{volume} {130}},\ \bibinfo {pages} {233602} (\bibinfo {year}
		{2023})}\BibitemShut {NoStop}%
	\bibitem [{\citenamefont {Dmytruk}\ and\ \citenamefont
		{Schir\'o}(2021)}]{dmytruk2021gauge}%
	\BibitemOpen
	\bibfield  {author} {\bibinfo {author} {\bibfnamefont {O.}~\bibnamefont
			{Dmytruk}}\ and\ \bibinfo {author} {\bibfnamefont {M.}~\bibnamefont
			{Schir\'o}},\ }\bibfield  {title} {\bibinfo {title} {Gauge fixing for
			strongly correlated electrons coupled to quantum light},\ }\href
	{https://doi.org/10.1103/PhysRevB.103.075131} {\bibfield  {journal} {\bibinfo
			{journal} {Phys. Rev. B}\ }\textbf {\bibinfo {volume} {103}},\ \bibinfo
		{pages} {075131} (\bibinfo {year} {2021})}\BibitemShut {NoStop}%
	\bibitem [{\citenamefont {P{\'e}rez-Gonz{\'a}lez}\ \emph
		{et~al.}(2022)\citenamefont {P{\'e}rez-Gonz{\'a}lez}, \citenamefont
		{G{\'o}mez-Le{\'o}n},\ and\ \citenamefont {Platero}}]{perez2022topology}%
	\BibitemOpen
	\bibfield  {author} {\bibinfo {author} {\bibfnamefont {B.}~\bibnamefont
			{P{\'e}rez-Gonz{\'a}lez}}, \bibinfo {author} {\bibfnamefont
			{{\'A}.}~\bibnamefont {G{\'o}mez-Le{\'o}n}},\ and\ \bibinfo {author}
		{\bibfnamefont {G.}~\bibnamefont {Platero}},\ }\bibfield  {title} {\bibinfo
		{title} {Topology detection in cavity {QED}},\ }\href
	{https://pubs.rsc.org/en/content/articlehtml/2022/cp/d2cp01806c} {\bibfield
		{journal} {\bibinfo  {journal} {Physical Chemistry Chemical Physics}\
		}\textbf {\bibinfo {volume} {24}},\ \bibinfo {pages} {15860} (\bibinfo {year}
		{2022})}\BibitemShut {NoStop}%
	\bibitem [{\citenamefont {Feshbach}(1958)}]{FESHBACH1958357}%
	\BibitemOpen
	\bibfield  {author} {\bibinfo {author} {\bibfnamefont {H.}~\bibnamefont
			{Feshbach}},\ }\bibfield  {title} {\bibinfo {title} {Unified theory of
			nuclear reactions},\ }\href
	{https://doi.org/https://doi.org/10.1016/0003-4916(58)90007-1} {\bibfield
		{journal} {\bibinfo  {journal} {Annals of Physics}\ }\textbf {\bibinfo
			{volume} {5}},\ \bibinfo {pages} {357} (\bibinfo {year} {1958})}\BibitemShut
	{NoStop}%
	\bibitem [{\citenamefont {Sentef}\ \emph {et~al.}(2020)\citenamefont {Sentef},
		\citenamefont {Li}, \citenamefont {K\"unzel},\ and\ \citenamefont
		{Eckstein}}]{sentef2020quantum}%
	\BibitemOpen
	\bibfield  {author} {\bibinfo {author} {\bibfnamefont {M.~A.}\ \bibnamefont
			{Sentef}}, \bibinfo {author} {\bibfnamefont {J.}~\bibnamefont {Li}}, \bibinfo
		{author} {\bibfnamefont {F.}~\bibnamefont {K\"unzel}},\ and\ \bibinfo
		{author} {\bibfnamefont {M.}~\bibnamefont {Eckstein}},\ }\bibfield  {title}
	{\bibinfo {title} {Quantum to classical crossover of {F}loquet engineering in
			correlated quantum systems},\ }\href
	{https://doi.org/10.1103/PhysRevResearch.2.033033} {\bibfield  {journal}
		{\bibinfo  {journal} {Phys. Rev. Res.}\ }\textbf {\bibinfo {volume} {2}},\
		\bibinfo {pages} {033033} (\bibinfo {year} {2020})}\BibitemShut {NoStop}%
	\bibitem [{\citenamefont {Stepanenko}\ \emph {et~al.}(2012)\citenamefont
		{Stepanenko}, \citenamefont {Rudner}, \citenamefont {Halperin},\ and\
		\citenamefont {Loss}}]{Stepanenko2012Spinorbit}%
	\BibitemOpen
	\bibfield  {author} {\bibinfo {author} {\bibfnamefont {D.}~\bibnamefont
			{Stepanenko}}, \bibinfo {author} {\bibfnamefont {M.}~\bibnamefont {Rudner}},
		\bibinfo {author} {\bibfnamefont {B.~I.}\ \bibnamefont {Halperin}},\ and\
		\bibinfo {author} {\bibfnamefont {D.}~\bibnamefont {Loss}},\ }\bibfield
	{title} {\bibinfo {title} {Singlet-triplet splitting in double quantum dots
			due to spin-orbit and hyperfine interactions},\ }\href
	{https://doi.org/10.1103/PhysRevB.85.075416} {\bibfield  {journal} {\bibinfo
			{journal} {Phys. Rev. B}\ }\textbf {\bibinfo {volume} {85}},\ \bibinfo
		{pages} {075416} (\bibinfo {year} {2012})}\BibitemShut {NoStop}%
	\bibitem [{\citenamefont {Eckardt}\ and\ \citenamefont
		{Anisimovas}(2015)}]{Eckardt2015HFE}%
	\BibitemOpen
	\bibfield  {author} {\bibinfo {author} {\bibfnamefont {A.}~\bibnamefont
			{Eckardt}}\ and\ \bibinfo {author} {\bibfnamefont {E.}~\bibnamefont
			{Anisimovas}},\ }\bibfield  {title} {\bibinfo {title} {High-frequency
			approximation for periodically driven quantum systems from a {F}loquet-space
			perspective},\ }\href {https://doi.org/10.1088/1367-2630/17/9/093039}
	{\bibfield  {journal} {\bibinfo  {journal} {New Journal of Physics}\ }\textbf
		{\bibinfo {volume} {17}},\ \bibinfo {pages} {093039} (\bibinfo {year}
		{2015})}\BibitemShut {NoStop}%
\end{thebibliography}
\end{document}